\pdfoutput=1
\documentclass[11pt,twoside]{book}
\usepackage{fancyhdr}
\fancypagestyle{plain}{%
\fancyhf{} 
\fancyfoot[R]{\textsf{ILC Reference Design Report\hspace{0.5cm}I-\thepage}}}
\usepackage{epsfig}
\usepackage{rotate}
\usepackage[figuresright]{rotating}
\usepackage{lscape}
\usepackage{hyperref}
\usepackage{url}
\usepackage{graphicx}
\usepackage{color}
\usepackage{epstopdf}
\usepackage{epsf}
\usepackage{hhline}
\usepackage{mytrc_style}
\usepackage{ulem}
\usepackage{verbatim}

\newcommand{\vbdlspacing}{\\[-15pt]}
\newcommand{\vbabovecaption}{\vspace{-24pt}}

\newcommand{\vbabove}{\vspace{-12pt}}
\newcommand{\vbbelow}{\vspace{-12pt}}
\newcommand{\itemspace}{\vspace{-6pt}}

\begin{document}
\frontmatter


{\sffamily\bfseries
\begin{titlepage}
\begin{center}
~
 ~ \vskip 4cm

    {\Huge I}{\huge NTERNATIONAL} 
    {\Huge L}{\huge INEAR} 
    {\Huge C}{\huge OLLIDER}
    
  \vskip 1.2cm

    {\Huge R}{\huge EFERENCE}
    {\Huge D}{\huge ESIGN}
    {\Huge R}{\huge EPORT}

  \vskip 1.2cm




\vskip 3cm

{\huge ILC Global Design Effort and} \\
    
  \vskip 0.5cm

{\huge World Wide Study }

  \vskip 3cm

    {\huge AUGUST, 2007}

\end{center}
\end{titlepage}

\newpage\thispagestyle{empty}
~
 ~ \vskip 2cm

{\LARGE Volume 1:~~~EXECUTIVE SUMMARY}
 \vskip 0.5cm
{\Large Editors:} 
 \vskip 0.25cm
{\Large James~Brau, Yasuhiro~Okada, Nicholas~Walker}

  \vskip 1.5cm

{\LARGE Volume 2:~~~PHYSICS AT THE ILC}
 \vskip 0.5cm
{\Large Editors:} 
 \vskip 0.25cm
{\Large Abdelhak~Djouadi, Joseph~Lykken, Klaus~M{\"o}nig} 
 \vskip 0.25cm
{\Large Yasuhiro~Okada, Mark~Oreglia, Satoru~Yamashita}

  \vskip 1.5cm

{\LARGE Volume 3:~~~ACCELERATOR}
 \vskip 0.5cm
{\Large Editors:} 
 \vskip 0.25cm
{\Large Nan~Phinney, Nobukazu~Toge, Nicholas~Walker}

  \vskip 1.5cm

{\LARGE Volume 4:~~~DETECTORS}
 \vskip 0.5cm
{\Large Editors:} 
 \vskip 0.25cm
{\Large Ties~Behnke, Chris~Damerell, John~Jaros, Akiya~Miyamoto}

\newpage\thispagestyle{empty}
~
 ~ \vskip 5cm
{\Huge Volume 1:~~~EXECUTIVE SUMMARY}
 \vskip 1cm
{\LARGE Editors:} 
 \vskip 0.5cm
{\LARGE James~Brau, Yasuhiro~Okada, Nicholas~Walker}

\newpage\thispagestyle{empty}

}
\cleardoublepage\setcounter{page}{1}

\chapter*{List of Contributors} 

\begin{center}

\begin{center}

Gerald~Aarons$^{203}$,
Toshinori~Abe$^{290}$,
Jason~Abernathy$^{293}$,
Medina~Ablikim$^{87}$,
Halina~Abramowicz$^{216}$,
David~Adey$^{236}$,
Catherine~Adloff$^{128}$,
Chris~Adolphsen$^{203}$,
Konstantin~Afanaciev$^{11,47}$,
Ilya~Agapov$^{192,35}$
% ...

\end{center}

\clearpage

\chapter*{List of Institutions}

\begin{center}

{\sl $^{1}$ Abdus Salam International Centre for Theoretical Physics, Strada Costriera 11, 34014 Trieste, Italy}

{\sl $^{2}$ Academy, RPR, National Institute of Physics and Nuclear Engineering `Horia Hulubei' (IFIN-HH), Str. Atomistilor no. 407, P.O. Box MG-6, R-76900 Bucharest - Magurele, Romania}

{\sl $^{3}$ AGH University of Science and Technology Akademia Gorniczo-Hutnicza im. Stanislawa Staszica w Krakowie al. Mickiewicza 30 PL-30-059 Cracow, Poland}

{\sl $^{4}$ Albert-Ludwigs Universit{\"a}t Freiburg, Physikalisches Institut, Hermann-Herder Str. 3, D-79104 Freiburg, Germany}

{\sl $^{5}$ Aligarh Muslim University, Aligarh, Uttar Pradesh 202002, India}

{\sl $^{6}$ Amberg Engineering AG, Trockenloostr. 21, P.O.Box 27, 8105 Regensdorf-Watt, Switzerland}

{\sl $^{7}$ Angstromquelle Karlsruhe (ANKA), Forschungszentrum Karlsruhe, Hermann-von-Helmholtz-Platz 1, D-76344 Eggenstein-Leopoldshafen, Germany}

{\sl $^{8}$ Argonne National Laboratory (ANL), 9700 S. Cass Avenue, Argonne, IL 60439, USA}

{\sl $^{9}$ Baylor University, Department of Physics, 101 Bagby Avenue, Waco, TX 76706, USA}

{\sl $^{10}$ Beijing University, Department of Physics, Beijing, China 100871}

{\sl $^{11}$ Belarusian State University, National Scientific \& Educational Center, Particle \& HEP Physics, M. Bogdanovich St., 153, 240040 Minsk, Belarus}

{\sl $^{12}$ Benares Hindu University, Benares, Varanasi 221005, India}

{\sl $^{13}$ Bhabha Atomic Research Centre, Trombay, Mumbai 400085, India}

{\sl $^{14}$ Birla Institute of Technology and Science, EEE Dept., Pilani, Rajasthan, India}

{\sl $^{15}$ Bogazici University, Physics Department, 34342 Bebek / Istanbul, 80820 Istanbul, Turkey}

{\sl $^{16}$ Boston University, Department of Physics, 590 Commonwealth Avenue, Boston, MA 02215, USA}

{\sl $^{17}$ Brandenburg University of Technology, Postfach 101344, D-03013 Cottbus, Germany}

{\sl $^{18}$ Brno University of Technology, Anton\'insk\'a; 548/1, CZ 601 90 Brno, Czech Republic}

{\sl $^{19}$ Brookhaven National Laboratory (BNL), P.O.Box 5000, Upton, NY 11973-5000, USA}

{\sl $^{20}$ Brown University, Department of Physics, Box 1843, Providence, RI 02912, USA}

{\sl $^{21}$ Budkar Institute for Nuclear Physics (BINP), 630090 Novosibirsk, Russia}

{\sl $^{22}$ Calcutta University, Department of Physics, 92 A.P.C. Road, Kolkata 700009, India}

{\sl $^{23}$ California Institute of Technology, Physics, Mathematics and Astronomy (PMA), 1200 East California Blvd, Pasadena, CA 91125, USA}

{\sl $^{24}$ Carleton University, Department of Physics, 1125 Colonel By Drive, Ottawa, Ontario, Canada K1S 5B6}

{\sl $^{25}$ Carnegie Mellon University, Department of Physics, Wean Hall 7235, Pittsburgh, PA 15213, USA}

{\sl $^{26}$ CCLRC Daresbury Laboratory, Daresbury, Warrington, Cheshire WA4 4AD, UK }

{\sl $^{27}$ CCLRC Rutherford Appleton Laboratory, Chilton, Didcot, Oxton OX11 0QX, UK }

{\sl $^{28}$ CEA Saclay, DAPNIA, F-91191 Gif-sur-Yvette, France}

{\sl $^{29}$ CEA Saclay, Service de Physique Th{\'e}orique, CEA/DSM/SPhT, F-91191 Gif-sur-Yvette Cedex, France}

{\sl $^{30}$ Center for High Energy Physics (CHEP) / Kyungpook National University, 1370 Sankyuk-dong, Buk-gu, Daegu 702-701, Korea}

{\sl $^{31}$ Center for High Energy Physics (TUHEP), Tsinghua University, Beijing, China 100084}

{\sl $^{32}$ Centre de Physique Theorique, CNRS - Luminy, Universiti d'Aix - Marseille II, Campus of Luminy, Case 907, 13288 Marseille Cedex 9, France}

{\sl $^{33}$ Centro de Investigaciones Energ\'eticas, Medioambientales y Technol\'ogicas, CIEMAT, Avenia Complutense 22, E-28040 Madrid, Spain}

{\sl $^{34}$ Centro Nacional de Microelectr\'onica (CNM), Instituto de Microelectr\'onica de Barcelona (IMB), Campus UAB, 08193 Cerdanyola del Vall\`es (Bellaterra), Barcelona, Spain}

{\sl $^{35}$ CERN, CH-1211 Gen\`eve 23, Switzerland}

{\sl $^{36}$ Charles University, Institute of Particle \& Nuclear Physics, Faculty of Mathematics and Physics, V Holesovickach 2, CZ-18000 Praque 8, Czech Republic}

{\sl $^{37}$ Chonbuk National University, Physics Department, Chonju 561-756, Korea}

{\sl $^{38}$ Cockcroft Institute, Daresbury, Warrington WA4 4AD, UK }

{\sl $^{39}$ College of William and Mary, Department of Physics, Williamsburg, VA, 23187, USA}

{\sl $^{40}$ Colorado State University, Department of Physics, Fort Collins, CO 80523, USA}

{\sl $^{41}$ Columbia University, Department of Physics, New York, NY 10027-6902, USA}

{\sl $^{42}$ Concordia University, Department of Physics, 1455 De Maisonneuve Blvd. West, Montreal, Quebec, Canada H3G 1M8}

{\sl $^{43}$ Cornell University, Laboratory for Elementary-Particle Physics (LEPP), Ithaca, NY 14853, USA}

{\sl $^{44}$ Cukurova University, Department of Physics, Fen-Ed. Fakultesi 01330, Balcali, Turkey}

{\sl $^{45}$ D.~V. Efremov Research Institute, SINTEZ, 196641 St. Petersburg, Russia}

{\sl $^{46}$ Dartmouth College, Department of Physics and Astronomy, 6127 Wilder Laboratory, Hanover, NH 03755, USA}

{\sl $^{47}$ DESY-Hamburg site, Deutsches Elektronen-Synchrotoron in der Helmholtz-Gemeinschaft, Notkestrasse 85, 22607 Hamburg, Germany}

{\sl $^{48}$ DESY-Zeuthen site, Deutsches Elektronen-Synchrotoron in der Helmholtz-Gemeinschaft, Platanenallee 6, D-15738 Zeuthen, Germany}

{\sl $^{49}$ Durham University,  Department of Physics, Ogen Center for Fundamental Physics, South Rd., Durham DH1 3LE, UK}

{\sl $^{50}$ Ecole Polytechnique, Laboratoire Leprince-Ringuet (LLR), Route de Saclay, F-91128 Palaiseau Cedex, France}

{\sl $^{51}$ Ege University, Department of Physics, Faculty of Science, 35100 Izmir, Turkey}

{\sl $^{52}$ Enrico Fermi Institute, University of Chicago, 5640 S. Ellis Avenue, RI-183, Chicago, IL 60637, USA}

{\sl $^{53}$ Ewha Womans University, 11-1 Daehyun-Dong, Seodaemun-Gu, Seoul, 120-750, Korea}

{\sl $^{54}$ Fermi National Accelerator Laboratory (FNAL), P.O.Box 500, Batavia, IL 60510-0500, USA}

{\sl $^{55}$ Fujita Gakuen Health University, Department of Physics, Toyoake, Aichi 470-1192, Japan}

{\sl $^{56}$ Fukui University of Technology, 3-6-1 Gakuen, Fukui-shi, Fukui 910-8505, Japan}

{\sl $^{57}$ Fukui University, Department of Physics, 3-9-1 Bunkyo, Fukui-shi, Fukui 910-8507, Japan}

{\sl $^{58}$ Georg-August-Universit{\"a}t G{\"o}ttingen, II. Physikalisches Institut, Friedrich-Hund-Platz 1, 37077 G{\"o}ttingen, Germany}

{\sl $^{59}$ Global Design Effort}

{\sl $^{60}$ Gomel State University, Department of Physics, Ul. Sovietskaya 104, 246699 Gomel, Belarus}

{\sl $^{61}$ Guangxi University, College of Physics science and Engineering Technology, Nanning, China 530004}

{\sl $^{62}$ Hanoi University of Technology, 1 Dai Co Viet road, Hanoi, Vietnam}

{\sl $^{63}$ Hanson Professional Services, Inc., 1525 S. Sixth St., Springfield, IL 62703, USA}

{\sl $^{64}$ Harish-Chandra Research Institute, Chhatnag Road, Jhusi, Allahabad 211019, India}

{\sl $^{65}$ Helsinki Institute of Physics (HIP), P.O. Box 64, FIN-00014 University of Helsinki, Finland}

{\sl $^{66}$ Henan Normal University, College of Physics and Information Engineering, Xinxiang, China 453007}

{\sl $^{67}$ High Energy Accelerator Research Organization, KEK, 1-1 Oho, Tsukuba, Ibaraki 305-0801, Japan}

{\sl $^{68}$ Hiroshima University, Department of Physics, 1-3-1 Kagamiyama, Higashi-Hiroshima, Hiroshima 739-8526, Japan}

{\sl $^{69}$ Humboldt Universit{\"a}t zu Berlin, Fachbereich Physik, Institut f\"ur Elementarteilchenphysik, Newtonstr. 15, D-12489 Berlin, Germany}

{\sl $^{70}$ Hungarian Academy of Sciences, KFKI Research Institute for Particle and Nuclear Physics, P.O. Box 49, H-1525 Budapest, Hungary}

{\sl $^{71}$ Ibaraki University, College of Technology, Department of Physics, Nakanarusawa 4-12-1, Hitachi, Ibaraki 316-8511, Japan}

{\sl $^{72}$ Imperial College, Blackett Laboratory, Department of Physics, Prince Consort Road, London, SW7 2BW, UK}

{\sl $^{73}$ Indian Association for the Cultivation of Science, Department of Theoretical Physics and Centre for Theoretical Sciences, Kolkata 700032, India}

{\sl $^{74}$ Indian Institute of Science, Centre for High Energy Physics, Bangalore 560012, Karnataka, India}

{\sl $^{75}$ Indian Institute of Technology, Bombay, Powai, Mumbai 400076, India}

{\sl $^{76}$ Indian Institute of Technology, Guwahati, Guwahati, Assam 781039, India}

{\sl $^{77}$ Indian Institute of Technology, Kanpur, Department of Physics,  IIT Post Office, Kanpur 208016, India}

{\sl $^{78}$ Indiana University - Purdue University, Indianapolis, Department of Physics, 402 N. Blackford St., LD 154, Indianapolis, IN 46202, USA}

{\sl $^{79}$ Indiana University, Department of Physics, Swain Hall West 117, 727 E. 3rd St., Bloomington, IN 47405-7105, USA}

{\sl $^{80}$ Institucio Catalana de Recerca i Estudis, ICREA,  Passeig Lluis Companys, 23, Barcelona 08010, Spain}

{\sl $^{81}$ Institut de Physique Nucl\'eaire, F-91406 Orsay, France }

{\sl $^{82}$ Institut f\"ur Theorie Elektromagnetischer Felder (TEMF), Technische Universit\"at Darmstadt, Schlo{\ss}gartenstr. 8, D-64289 Darmstadt, Germany}

{\sl $^{83}$ Institut National de Physique Nucleaire et de Physique des Particules, 3, Rue Michel- Ange, 75794 Paris Cedex 16, France}

{\sl $^{84}$ Institut Pluridisciplinaire Hubert Curien, 23 Rue du Loess - BP28, 67037 Strasbourg Cedex 2, France}

{\sl $^{85}$ Institute for Chemical Research, Kyoto University, Gokasho, Uji, Kyoto 611-0011, Japan}

{\sl $^{86}$ Institute for Cosmic Ray Research, University of Tokyo, 5-1-5 Kashiwa-no-Ha, Kashiwa, Chiba 277-8582, Japan}

{\sl $^{87}$ Institute of High Energy Physics - IHEP, Chinese Academy of Sciences, P.O. Box 918, Beijing, China 100049}

{\sl $^{88}$ Institute of Mathematical Sciences, Taramani, C.I.T. Campus, Chennai 600113, India}

{\sl $^{89}$ Institute of Physics and Electronics, Vietnamese Academy of Science and Technology (VAST), 10 Dao-Tan, Ba-Dinh, Hanoi 10000, Vietnam}

{\sl $^{90}$ Institute of Physics, ASCR, Academy of Science of the Czech Republic, Division of Elementary Particle Physics, Na Slovance 2, CS-18221 Prague 8, Czech Republic}

{\sl $^{91}$ Institute of Physics, Pomorska 149/153, PL-90-236 Lodz, Poland}

{\sl $^{92}$ Institute of Theoretical and Experimetal Physics, B. Cheremushkinskawa, 25, RU-117259, Moscow, Russia}

{\sl $^{93}$ Institute of Theoretical Physics, Chinese Academy of Sciences, P.O.Box 2735, Beijing, China 100080}

{\sl $^{94}$ Instituto de Fisica Corpuscular (IFIC), Centro Mixto CSIC-UVEG, Edificio Investigacion Paterna, Apartado 22085, 46071 Valencia, Spain}

{\sl $^{95}$ Instituto de Fisica de Cantabria, (IFCA, CSIC-UC), Facultad de Ciencias, Avda. Los Castros s/n, 39005 Santander, Spain}

{\sl $^{96}$ Instituto Nazionale di Fisica Nucleare (INFN), Laboratorio LASA, Via Fratelli Cervi 201, 20090 Segrate, Italy}

{\sl $^{97}$ Instituto Nazionale di Fisica Nucleare (INFN), Sezione di Ferrara, via Paradiso 12, I-44100 Ferrara, Italy}

{\sl $^{98}$ Instituto Nazionale di Fisica Nucleare (INFN), Sezione di Firenze, Via G. Sansone 1, I-50019 Sesto Fiorentino (Firenze), Italy}

{\sl $^{99}$ Instituto Nazionale di Fisica Nucleare (INFN), Sezione di Lecce, via Arnesano, I-73100 Lecce, Italy}

{\sl $^{100}$ Instituto Nazionale di Fisica Nucleare (INFN), Sezione di Napoli, Complesso Universit{\'a} di Monte Sant'Angelo,via, I-80126 Naples, Italy}

{\sl $^{101}$ Instituto Nazionale di Fisica Nucleare (INFN), Sezione di Pavia, Via Bassi 6, I-27100 Pavia, Italy}

{\sl $^{102}$ Instituto Nazionale di Fisica Nucleare (INFN), Sezione di Pisa, Edificio C - Polo Fibonacci Largo B. Pontecorvo, 3, I-56127 Pisa, Italy}

{\sl $^{103}$ Instituto Nazionale di Fisica Nucleare (INFN), Sezione di Torino, c/o Universit{\'a}' di Torino facolt{\'a}' di Fisica, via P Giuria 1, 10125 Torino, Italy}

{\sl $^{104}$ Instituto Nazionale di Fisica Nucleare (INFN), Sezione di Trieste, Padriciano 99, I-34012 Trieste (Padriciano), Italy}

{\sl $^{105}$ Inter-University Accelerator Centre, Aruna Asaf Ali Marg, Post Box 10502, New Delhi 110067, India}

{\sl $^{106}$ International Center for Elementary Particle Physics, University of Tokyo, Hongo 7-3-1, Bunkyo District, Tokyo 113-0033, Japan}

{\sl $^{107}$ Iowa State University, Department of Physics, High Energy Physics Group, Ames, IA 50011, USA}

{\sl $^{108}$ Jagiellonian University, Institute of Physics, Ul. Reymonta 4, PL-30-059 Cracow, Poland}

{\sl $^{109}$ Jamia Millia Islamia, Centre for Theoretical Physics, Jamia Nagar, New Delhi 110025, India}

{\sl $^{110}$ Jamia Millia Islamia, Department of Physics, Jamia Nagar, New Delhi 110025, India}

{\sl $^{111}$ Japan Aerospace Exploration Agency, Sagamihara Campus, 3-1-1 Yoshinodai, Sagamihara, Kanagawa 220-8510 , Japan}

{\sl $^{112}$ Japan Atomic Energy Agency, 4-49 Muramatsu, Tokai-mura, Naka-gun, Ibaraki 319-1184, Japan}

{\sl $^{113}$ Johannes Gutenberg Universit{\"a}t Mainz, Institut f{\"u}r Physik, 55099 Mainz, Germany}

{\sl $^{114}$ Johns Hopkins University, Applied Physics Laboratory, 11100 Johns Hopkins RD., Laurel, MD 20723-6099, USA}

{\sl $^{115}$ Joint Institute for Nuclear Research (JINR), Joliot-Curie 6, 141980, Dubna, Moscow Region, Russia}

{\sl $^{116}$ Kansas State University, Department of Physics, 116 Cardwell Hall, Manhattan, KS 66506, USA}

{\sl $^{117}$ KCS Corp., 2-7-25 Muramatsukita, Tokai, Ibaraki 319-1108, Japan}

{\sl $^{118}$ Kharkov Institute of Physics and Technology, National Science Center, 1, Akademicheskaya St., Kharkov, 61108, Ukraine}

{\sl $^{119}$ Kinki University, Department of Physics, 3-4-1 Kowakae, Higashi-Osaka, Osaka 577-8502, Japan}

{\sl $^{120}$ Kobe University, Faculty of Science, 1-1 Rokkodai-cho, Nada-ku, Kobe, Hyogo 657-8501, Japan}

{\sl $^{121}$ Kogakuin University, Department of Physics, Shinjuku Campus, 1-24-2 Nishi-Shinjuku, Shinjuku-ku, Tokyo 163-8677, Japan}

{\sl $^{122}$ Konkuk University, 93-1 Mojin-dong, Kwanglin-gu, Seoul 143-701, Korea}

{\sl $^{123}$ Korea Advanced Institute of Science \& Technology, Department of Physics, 373-1 Kusong-dong, Yusong-gu, Taejon 305-701, Korea}

{\sl $^{124}$ Korea Institute for Advanced Study (KIAS), School of Physics, 207-43 Cheongryangri-dong, Dongdaemun-gu, Seoul 130-012, Korea}

{\sl $^{125}$ Korea University, Department of Physics, Seoul 136-701, Korea}

{\sl $^{126}$ Kyoto University, Department of Physics, Kitashirakawa-Oiwakecho, Sakyo-ku, Kyoto 606-8502, Japan}

{\sl $^{127}$ L.P.T.A., UMR 5207 CNRS-UM2, Universit{\'e} Montpellier II, Case Courrier 070, B{\^a}t. 13, place Eug{\`e}ne Bataillon, 34095 Montpellier Cedex 5, France}

{\sl $^{128}$ Laboratoire d'Annecy-le-Vieux de Physique des Particules (LAPP), Chemin du Bellevue, BP 110, F-74941 Annecy-le-Vieux Cedex, France}

{\sl $^{129}$ Laboratoire d'Annecy-le-Vieux de Physique Theorique (LAPTH), Chemin de Bellevue, BP 110, F-74941 Annecy-le-Vieux Cedex, France}

{\sl $^{130}$ Laboratoire de l'Acc\'el\'erateur Lin\'eaire (LAL), Universit\'e Paris-Sud 11, B\^atiment 200, 91898 Orsay, France}

{\sl $^{131}$ Laboratoire de Physique Corpusculaire de Clermont-Ferrand (LPC), Universit\'e Blaise Pascal, I.N.2.P.3./C.N.R.S., 24 avenue des Landais, 63177 Aubi\`ere Cedex, France}

{\sl $^{132}$ Laboratoire de Physique Subatomique et de Cosmologie (LPSC), Universit\'e Joseph Fourier (Grenoble 1), 53, ave. des Marthyrs, F-38026 Grenoble Cedex, France}

{\sl $^{133}$ Laboratoire de Physique Theorique, Universit\'e de Paris-Sud XI, Batiment 210, F-91405 Orsay Cedex, France}

{\sl $^{134}$ Laboratori Nazionali di Frascati, via E. Fermi, 40, C.P. 13, I-00044 Frascati, Italy}

{\sl $^{135}$ Laboratory of High Energy Physics and Cosmology, Department of Physics, Hanoi National University, 334 Nguyen Trai, Hanoi, Vietnam}

{\sl $^{136}$ Lancaster University, Physics Department, Lancaster LA1 4YB, UK}

{\sl $^{137}$ Lawrence Berkeley National Laboratory (LBNL), 1 Cyclotron Rd, Berkeley, CA 94720, USA}

{\sl $^{138}$ Lawrence Livermore National Laboratory (LLNL), Livermore, CA 94551, USA}

{\sl $^{139}$ Lebedev Physical Institute, Leninsky Prospect 53, RU-117924 Moscow, Russia}

{\sl $^{140}$ Liaoning Normal University, Department of Physics, Dalian, China 116029}

{\sl $^{141}$ Lomonosov Moscow State University, Skobeltsyn Institute of Nuclear Physics (MSU SINP), 1(2), Leninskie gory, GSP-1, Moscow 119991, Russia}

{\sl $^{142}$ Los Alamos National Laboratory (LANL), P.O.Box 1663, Los Alamos, NM 87545, USA}

{\sl $^{143}$ Louisiana Technical University, Department of Physics, Ruston, LA 71272, USA}

{\sl $^{144}$ Ludwig-Maximilians-Universit{\"a}t M{\"u}nchen, Department f{\"u}r Physik, Schellingstr. 4, D-80799 Munich, Germany}

{\sl $^{145}$ Lunds Universitet, Fysiska Institutionen, Avdelningen f{\"o}r Experimentell H{\"o}genergifysik, Box 118, 221 00 Lund, Sweden}

{\sl $^{146}$ Massachusetts Institute of Technology, Laboratory for Nuclear Science \& Center for Theoretical Physics, 77 Massachusetts Ave., NW16, Cambridge, MA 02139, USA}

{\sl $^{147}$ Max-Planck-Institut f{\"u}r Physik (Werner-Heisenberg-Institut), F{\"o}hringer Ring 6, 80805 M{\"u}nchen, Germany}

{\sl $^{148}$ McGill University, Department of Physics, Ernest Rutherford Physics Bldg., 3600 University Ave., Montreal, Quebec, H3A 2T8 Canada}

{\sl $^{149}$ Meiji Gakuin University, Department of Physics, 2-37 Shirokanedai 1-chome, Minato-ku, Tokyo 244-8539, Japan}

{\sl $^{150}$ Michigan State University, Department of Physics and Astronomy, East Lansing, MI 48824, USA}

{\sl $^{151}$ Middle East Technical University, Department of Physics, TR-06531 Ankara, Turkey}

{\sl $^{152}$ Mindanao Polytechnic State College, Lapasan, Cagayan de Oro City 9000, Phillipines}

{\sl $^{153}$ MSU-Iligan Institute of Technology, Department of Physics, Andres Bonifacio Avenue, 9200 Iligan City, Phillipines}

{\sl $^{154}$ Nagasaki Institute of Applied Science, 536 Abamachi, Nagasaki-Shi, Nagasaki 851-0193, Japan}

{\sl $^{155}$ Nagoya University, Fundamental Particle Physics Laboratory, Division of Particle and Astrophysical Sciences, Furo-cho, Chikusa-ku, Nagoya, Aichi 464-8602, Japan}

{\sl $^{156}$ Nanchang University, Department of Physics, Nanchang, China 330031}

{\sl $^{157}$ Nanjing University, Department of Physics, Nanjing, China 210093}

{\sl $^{158}$ Nankai University, Department of Physics, Tianjin, China 300071}

{\sl $^{159}$ National Central University, High Energy Group, Department of Physics, Chung-li, Taiwan 32001}

{\sl $^{160}$ National Institute for Nuclear \& High Energy Physics, PO Box 41882, 1009 DB Amsterdam, Netherlands}

{\sl $^{161}$ National Institute of Radiological Sciences, 4-9-1 Anagawa, Inaga, Chiba 263-8555, Japan}

{\sl $^{162}$ National Synchrotron Radiation Laboratory, University of Science and Technology of china, Hefei, Anhui, China 230029}

{\sl $^{163}$ National Synchrotron Research Center, 101 Hsin-Ann Rd., Hsinchu Science Part, Hsinchu, Taiwan 30076}

{\sl $^{164}$ National Taiwan University, Physics Department, Taipei, Taiwan 106}

{\sl $^{165}$ Niels Bohr Institute (NBI), University of Copenhagen, Blegdamsvej 17, DK-2100 Copenhagen, Denmark}

{\sl $^{166}$ Niigata University, Department of Physics, Ikarashi, Niigata 950-218, Japan}

{\sl $^{167}$ Nikken Sekkai Ltd., 2-18-3 Iidabashi, Chiyoda-Ku, Tokyo 102-8117, Japan}

{\sl $^{168}$ Nippon Dental University, 1-9-20 Fujimi, Chiyoda-Ku, Tokyo 102-8159, Japan}

{\sl $^{169}$ North Asia University, Akita 010-8515, Japan}

{\sl $^{170}$ North Eastern Hill University, Department of Physics, Shillong 793022, India}

{\sl $^{171}$ Northern Illinois University, Department of Physics, DeKalb, Illinois 60115-2825, USA}

{\sl $^{172}$ Northwestern University, Department of Physics and Astronomy, 2145 Sheridan Road., Evanston, IL 60208, USA}

{\sl $^{173}$ Novosibirsk State University (NGU), Department of Physics, Pirogov st. 2, 630090 Novosibirsk, Russia}

{\sl $^{174}$ Obninsk State Technical University for Nuclear Engineering (IATE), Obninsk, Russia}

{\sl $^{175}$ Ochanomizu University, Department of Physics, Faculty of Science, 1-1 Otsuka 2, Bunkyo-ku, Tokyo 112-8610, Japan}

{\sl $^{176}$ Osaka University, Laboratory of Nuclear Studies, 1-1 Machikaneyama, Toyonaka, Osaka 560-0043, Japan}

{\sl $^{177}$ {\"O}sterreichische Akademie der Wissenschaften, Institut f{\"u}r Hochenergiephysik, Nikolsdorfergasse 18, A-1050 Vienna, Austria}

{\sl $^{178}$ Panjab University, Chandigarh 160014, India}

{\sl $^{179}$ Pavel Sukhoi Gomel State Technical University, ICTP Affiliated Centre \& Laboratory for Physical Studies, October Avenue, 48, 246746, Gomel, Belarus}

{\sl $^{180}$ Pavel Sukhoi Gomel State Technical University, Physics Department, October Ave. 48, 246746 Gomel, Belarus}

{\sl $^{181}$ Physical Research Laboratory, Navrangpura, Ahmedabad 380 009, Gujarat, India}

{\sl $^{182}$ Pohang Accelerator Laboratory (PAL), San-31 Hyoja-dong, Nam-gu, Pohang, Gyeongbuk 790-784, Korea}

{\sl $^{183}$ Polish Academy of Sciences (PAS), Institute of Physics, Al. Lotnikow 32/46, PL-02-668 Warsaw, Poland}

{\sl $^{184}$ Primera Engineers Ltd., 100 S Wacker Drive, Suite 700, Chicago, IL 60606, USA}

{\sl $^{185}$ Princeton University, Department of Physics, P.O. Box 708, Princeton, NJ 08542-0708, USA}

{\sl $^{186}$ Purdue University, Department of Physics, West Lafayette, IN 47907, USA}

{\sl $^{187}$ Pusan National University, Department of Physics, Busan 609-735, Korea}

{\sl $^{188}$ R. W. Downing Inc., 6590 W. Box Canyon Dr., Tucson, AZ 85745, USA}

{\sl $^{189}$ Raja Ramanna Center for Advanced Technology, Indore 452013, India}

{\sl $^{190}$ Rheinisch-Westf{\"a}lische Technische Hochschule (RWTH), Physikalisches Institut, Physikzentrum, Sommerfeldstrasse 14, D-52056 Aachen, Germany}

{\sl $^{191}$ RIKEN, 2-1 Hirosawa, Wako, Saitama 351-0198, Japan}

{\sl $^{192}$ Royal Holloway, University of London (RHUL), Department of Physics, Egham, Surrey TW20 0EX, UK }

{\sl $^{193}$ Saga University, Department of Physics, 1 Honjo-machi, Saga-shi, Saga 840-8502, Japan}

{\sl $^{194}$ Saha Institute of Nuclear Physics, 1/AF Bidhan Nagar, Kolkata 700064, India}

{\sl $^{195}$ Salalah College of Technology (SCOT), Engineering Department, Post Box No. 608, Postal Code 211, Salalah, Sultanate of Oman}

{\sl $^{196}$ Saube Co., Hanabatake, Tsukuba, Ibaraki 300-3261, Japan}

{\sl $^{197}$ Seoul National University, San 56-1, Shinrim-dong, Kwanak-gu, Seoul 151-742, Korea}

{\sl $^{198}$ Shandong University, 27 Shanda Nanlu, Jinan, China 250100}

{\sl $^{199}$ Shanghai Institute of Applied Physics, Chinese Academy of Sciences, 2019 Jiaruo Rd., Jiading, Shanghai, China 201800}

{\sl $^{200}$ Shinshu University, 3-1-1, Asahi, Matsumoto, Nagano 390-8621, Japan}

{\sl $^{201}$ Sobolev Institute of Mathematics, Siberian Branch of the Russian Academy of Sciences, 4 Acad. Koptyug Avenue, 630090 Novosibirsk, Russia}

{\sl $^{202}$ Sokendai, The Graduate University for Advanced Studies, Shonan Village, Hayama, Kanagawa 240-0193, Japan}

{\sl $^{203}$ Stanford Linear Accelerator Center (SLAC), 2575 Sand Hill Road, Menlo Park, CA 94025, USA}

{\sl $^{204}$ State University of New York at Binghamton, Department of Physics, PO Box 6016, Binghamton, NY 13902, USA}

{\sl $^{205}$ State University of New York at Buffalo, Department of Physics \& Astronomy, 239 Franczak Hall, Buffalo, NY 14260, USA}

{\sl $^{206}$ State University of New York at Stony Brook, Department of Physics and Astronomy, Stony Brook, NY 11794-3800, USA}

{\sl $^{207}$ Sumitomo Heavy Industries, Ltd., Natsushima-cho, Yokosuka, Kanagawa 237-8555, Japan}

{\sl $^{208}$ Sungkyunkwan University (SKKU), Natural Science Campus 300, Physics Research Division, Chunchun-dong, Jangan-gu, Suwon, Kyunggi-do 440-746, Korea}

{\sl $^{209}$ Swiss Light Source (SLS), Paul Scherrer Institut (PSI), PSI West, CH-5232 Villigen PSI, Switzerland}

{\sl $^{210}$ Syracuse University, Department of Physics, 201 Physics Building, Syracuse, NY 13244-1130, USA}

{\sl $^{211}$ Tata Institute of Fundamental Research, School of Natural Sciences, Homi Bhabha Rd., Mumbai 400005, India}

{\sl $^{212}$ Technical Institute of Physics and Chemistry, Chinese Academy of Sciences, 2 North 1st St., Zhongguancun, Beijing, China 100080}

{\sl $^{213}$ Technical University of Lodz, Department of Microelectronics and Computer Science, al. Politechniki 11, 90-924 Lodz, Poland}

{\sl $^{214}$ Technische Universit{\"a}t Dresden, Institut f{\"u}r Kern- und Teilchenphysik, D-01069 Dresden, Germany}

{\sl $^{215}$ Technische Universit{\"a}t Dresden, Institut f{\"u}r Theoretische Physik,D-01062 Dresden, Germany}

{\sl $^{216}$ Tel-Aviv University, School of Physics and Astronomy, Ramat Aviv, Tel Aviv 69978, Israel}

{\sl $^{217}$ Texas A\&M University, Physics Department, College Station, 77843-4242 TX, USA}

{\sl $^{218}$ Texas Tech University, Department of Physics, Campus Box 41051, Lubbock, TX 79409-1051, USA}

{\sl $^{219}$ The Henryk Niewodniczanski Institute of Nuclear Physics (NINP), High Energy Physics Lab, ul. Radzikowskiego 152, PL-31342 Cracow, Poland}

{\sl $^{220}$ Thomas Jefferson National Accelerator Facility (TJNAF), 12000 Jefferson Avenue, Newport News, VA 23606, USA}

{\sl $^{221}$ Tohoku Gakuin University, Faculty of Technology, 1-13-1 Chuo, Tagajo, Miyagi 985-8537, Japan}

{\sl $^{222}$ Tohoku University, Department of Physics, Aoba District, Sendai, Miyagi 980-8578, Japan}

{\sl $^{223}$ Tokyo Management College, Computer Science Lab, Ichikawa, Chiba 272-0001, Japan}

{\sl $^{224}$ Tokyo University of Agriculture Technology, Department of Applied Physics, Naka-machi, Koganei, Tokyo 183-8488, Japan}

{\sl $^{225}$ Toyama University, Department of Physics, 3190 Gofuku, Toyama-shi 930-8588, Japan}

{\sl $^{226}$ TRIUMF, 4004 Wesbrook Mall, Vancouver, BC V6T 2A3, Canada}

{\sl $^{227}$ Tufts University, Department of Physics and Astronomy, Robinson Hall, Medford, MA 02155, USA}

{\sl $^{228}$ Universidad Aut\`onoma de Madrid (UAM), Facultad de Ciencias C-XI, Departamento de Fisica Teorica, Cantoblanco, Madrid 28049, Spain}

{\sl $^{229}$ Universitat Aut\`onoma de Barcelona, Institut de Fisica d'Altes Energies (IFAE), Campus UAB, Edifici Cn, E-08193 Bellaterra, Barcelona, Spain}

{\sl $^{230}$ University College of London (UCL), High Energy Physics Group, Physics and Astronomy Department, Gower Street, London WC1E 6BT, UK }

{\sl $^{231}$ University College, National University of Ireland (Dublin), Department of Experimental Physics, Science Buildings, Belfield, Dublin 4, Ireland}

{\sl $^{232}$ University de Barcelona, Facultat de F\'isica, Av. Diagonal, 647, Barcelona 08028, Spain}

{\sl $^{233}$ University of Abertay Dundee, Department of Physics, Bell St, Dundee, DD1 1HG, UK}

{\sl $^{234}$ University of Auckland, Department of Physics, Private Bag, Auckland 1, New Zealand}

{\sl $^{235}$ University of Bergen, Institute of Physics, Allegaten 55, N-5007 Bergen, Norway}

{\sl $^{236}$ University of Birmingham, School of Physics and Astronomy, Particle Physics Group, Edgbaston, Birmingham B15 2TT, UK}

{\sl $^{237}$ University of Bristol, H. H. Wills Physics Lab, Tyndall Ave., Bristol BS8 1TL, UK}

{\sl $^{238}$ University of British Columbia, Department of Physics and Astronomy, 6224 Agricultural Rd., Vancouver, BC V6T 1Z1, Canada}

{\sl $^{239}$ University of California Berkeley, Department of Physics, 366 Le Conte Hall, \#7300, Berkeley, CA 94720, USA}

{\sl $^{240}$ University of California Davis, Department of Physics, One Shields Avenue, Davis, CA 95616-8677, USA}

{\sl $^{241}$ University of California Irvine, Department of Physics and Astronomy, High Energy Group, 4129 Frederick Reines Hall, Irvine, CA 92697-4575 USA}

{\sl $^{242}$ University of California Riverside, Department of Physics, Riverside, CA 92521, USA}

{\sl $^{243}$ University of California Santa Barbara, Department of Physics, Broida Hall, Mail Code 9530, Santa Barbara, CA 93106-9530, USA}

{\sl $^{244}$ University of California Santa Cruz, Department of Astronomy and Astrophysics, 1156 High Street, Santa Cruz, CA 05060, USA}

{\sl $^{245}$ University of California Santa Cruz, Institute for Particle Physics, 1156 High Street, Santa Cruz, CA 95064, USA}

{\sl $^{246}$ University of Cambridge, Cavendish Laboratory, J J Thomson Avenue, Cambridge CB3 0HE, UK}

{\sl $^{247}$ University of Colorado at Boulder, Department of Physics, 390 UCB, University of Colorado, Boulder, CO 80309-0390, USA}

{\sl $^{248}$ University of Delhi, Department of Physics and Astrophysics, Delhi 110007, India}

{\sl $^{249}$ University of Delhi, S.G.T.B. Khalsa College, Delhi 110007, India}

{\sl $^{250}$ University of Dundee, Department of Physics, Nethergate, Dundee, DD1 4HN,  Scotland, UK}

{\sl $^{251}$ University of Edinburgh, School of Physics, James Clerk Maxwell Building, The King's Buildings, Mayfield Road, Edinburgh EH9 3JZ, UK}

{\sl $^{252}$ University of Essex, Department of Physics, Wivenhoe Park, Colchester CO4 3SQ, UK}

{\sl $^{253}$ University of Florida, Department of Physics, Gainesville, FL 32611, USA}

{\sl $^{254}$ University of Glasgow, Department of Physics \& Astronomy, University Avenue, Glasgow G12 8QQ, Scotland, UK}

{\sl $^{255}$ University of Hamburg, Physics Department, Institut f{\"u}r Experimentalphysik, Luruper Chaussee 149, 22761 Hamburg, Germany}

{\sl $^{256}$ University of Hawaii, Department of Physics and Astronomy, HEP, 2505 Correa Rd., WAT 232, Honolulu, HI 96822-2219, USA}

{\sl $^{257}$ University of Heidelberg, Kirchhoff Institute of Physics, Albert {\"U}berle Strasse 3-5, DE-69120 Heidelberg, Germany}

{\sl $^{258}$ University of Helsinki, Department of Physical Sciences, P.O. Box 64 (Vaino Auerin katu 11), FIN-00014, Helsinki, Finland}

{\sl $^{259}$ University of Hyogo, School of Science, Kouto 3-2-1, Kamigori, Ako, Hyogo 678-1297,  Japan}

{\sl $^{260}$ University of Illinois at Urbana-Champaign, Department of Phys., High Energy Physics, 441 Loomis Lab. of Physics1110 W. Green St., Urbana, IL 61801-3080, USA}

{\sl $^{261}$ University of Iowa, Department of Physics and Astronomy, 203 Van Allen Hall, Iowa City, IA 52242-1479, USA}

{\sl $^{262}$ University of Kansas, Department of Physics and Astronomy, Malott Hall, 1251 Wescoe Hall Drive, Room 1082, Lawrence, KS 66045-7582, USA}

{\sl $^{263}$ University of Liverpool, Department of Physics, Oliver Lodge Lab, Oxford St., Liverpool L69 7ZE, UK}

{\sl $^{264}$ University of Louisville, Department of Physics, Louisville, KY 40292, USA}

{\sl $^{265}$ University of Manchester, School of Physics and Astronomy, Schuster Lab, Manchester M13 9PL, UK}

{\sl $^{266}$ University of Maryland, Department of Physics and Astronomy, Physics Building (Bldg. 082), College Park, MD 20742, USA}

{\sl $^{267}$ University of Melbourne, School of Physics, Victoria 3010, Australia}

{\sl $^{268}$ University of Michigan, Department of Physics, 500 E. University Ave., Ann Arbor, MI 48109-1120, USA}

{\sl $^{269}$ University of Minnesota, 148 Tate Laboratory Of Physics, 116 Church St. S.E., Minneapolis, MN 55455, USA}

{\sl $^{270}$ University of Mississippi, Department of Physics and Astronomy, 108 Lewis Hall, PO Box 1848, Oxford, Mississippi 38677-1848, USA}

{\sl $^{271}$ University of Montenegro, Faculty of Sciences and Math., Department of Phys., P.O. Box 211, 81001 Podgorica, Serbia and Montenegro}

{\sl $^{272}$ University of New Mexico, New Mexico Center for Particle Physics, Department of Physics and Astronomy, 800 Yale Boulevard N.E., Albuquerque, NM 87131, USA}

{\sl $^{273}$ University of Notre Dame, Department of Physics, 225 Nieuwland Science Hall, Notre Dame, IN 46556, USA}

{\sl $^{274}$ University of Oklahoma, Department of Physics and Astronomy, Norman, OK 73071, USA}

{\sl $^{275}$ University of Oregon, Department of Physics, 1371 E. 13th Ave., Eugene, OR 97403, USA}

{\sl $^{276}$ University of Oxford, Particle Physics Department, Denys Wilkinson Bldg., Keble Road, Oxford OX1 3RH England, UK }

{\sl $^{277}$ University of Patras, Department of Physics, GR-26100 Patras, Greece}

{\sl $^{278}$ University of Pavia, Department of Nuclear and Theoretical Physics, via Bassi 6, I-27100 Pavia, Italy}

{\sl $^{279}$ University of Pennsylvania, Department of Physics and Astronomy, 209 South 33rd Street, Philadelphia, PA 19104-6396, USA}

{\sl $^{280}$ University of Puerto Rico at Mayaguez, Department of Physics, P.O. Box 9016, Mayaguez, 00681-9016 Puerto Rico}

{\sl $^{281}$ University of Regina, Department of Physics, Regina, Saskatchewan, S4S 0A2 Canada}

{\sl $^{282}$ University of Rochester, Department of Physics and Astronomy, Bausch \& Lomb Hall, P.O. Box 270171, 600 Wilson Boulevard, Rochester, NY 14627-0171 USA}

{\sl $^{283}$ University of Science and Technology of China, Department of Modern Physics (DMP), Jin Zhai Road 96, Hefei, China 230026}

{\sl $^{284}$ University of Silesia, Institute of Physics, Ul. Uniwersytecka 4, PL-40007 Katowice, Poland}

{\sl $^{285}$ University of Southampton, School of Physics and Astronomy, Highfield, Southampton S017 1BJ, England, UK}

{\sl $^{286}$ University of Strathclyde, Physics Department, John Anderson Building, 107 Rottenrow, Glasgow, G4 0NG, Scotland, UK}

{\sl $^{287}$ University of Sydney, Falkiner High Energy Physics Group, School of Physics, A28, Sydney, NSW 2006, Australia}

{\sl $^{288}$ University of Texas, Center for Accelerator Science and Technology, Arlington, TX 76019, USA}

{\sl $^{289}$ University of Tokushima, Institute of Theoretical Physics, Tokushima-shi 770-8502, Japan}

{\sl $^{290}$ University of Tokyo, Department of Physics, 7-3-1 Hongo, Bunkyo District, Tokyo 113-0033, Japan}

{\sl $^{291}$ University of Toronto, Department of Physics, 60 St. George St., Toronto M5S 1A7, Ontario, Canada}

{\sl $^{292}$ University of Tsukuba, Institute of Physics, 1-1-1 Ten'nodai, Tsukuba, Ibaraki 305-8571, Japan}

{\sl $^{293}$ University of Victoria, Department of Physics and Astronomy, P.O.Box 3055 Stn Csc, Victoria, BC V8W 3P6, Canada}

{\sl $^{294}$ University of Warsaw, Institute of Physics, Ul. Hoza 69, PL-00 681 Warsaw, Poland}

{\sl $^{295}$ University of Warsaw, Institute of Theoretical Physics, Ul. Hoza 69, PL-00 681 Warsaw, Poland}

{\sl $^{296}$ University of Washington, Department of Physics, PO Box 351560, Seattle, WA 98195-1560, USA}

{\sl $^{297}$ University of Wisconsin, Physics Department, Madison, WI 53706-1390, USA}

{\sl $^{298}$ University of Wuppertal, Gau{\ss}stra{\ss}e 20, D-42119 Wuppertal, Germany}

{\sl $^{299}$ Universit\'e Claude Bernard Lyon-I, Institut de Physique Nucl\'eaire de Lyon (IPNL), 4, rue Enrico Fermi, F-69622 Villeurbanne Cedex, France}

{\sl $^{300}$ Universit\'e de Gen\`eve, Section de Physique, 24, quai E. Ansermet, 1211 Gen\`eve 4, Switzerland}

{\sl $^{301}$ Universit\'e Louis Pasteur (Strasbourg I), UFR de Sciences Physiques, 3-5 Rue de l'Universit\'e, F-67084 Strasbourg Cedex, France}

{\sl $^{302}$ Universit\'e Pierre et Marie Curie (Paris VI-VII) (6-7) (UPMC), Laboratoire de Physique Nucl\'eaire et de Hautes Energies (LPNHE), 4 place Jussieu, Tour 33, Rez de chausse, 75252 Paris Cedex 05, France}

{\sl $^{303}$ Universit{\"a}t Bonn, Physikalisches Institut, Nu{\ss}allee 12, 53115 Bonn, Germany}

{\sl $^{304}$ Universit{\"a}t Karlsruhe, Institut f{\"u}r Physik, Postfach 6980, Kaiserstrasse 12, D-76128 Karlsruhe, Germany}

{\sl $^{305}$ Universit{\"a}t Rostock, Fachbereich Physik, Universit{\"a}tsplatz 3, D-18051 Rostock, Germany}

{\sl $^{306}$ Universit{\"a}t Siegen, Fachbereich f{\"u}r Physik, Emmy Noether Campus, Walter-Flex-Str.3, D-57068 Siegen, Germany}

{\sl $^{307}$ Universit{\`a} de Bergamo, Dipartimento di Fisica, via Salvecchio, 19, I-24100 Bergamo, Italy}

{\sl $^{308}$ Universit{\`a} degli Studi di Roma La Sapienza, Dipartimento di Fisica, Istituto Nazionale di Fisica Nucleare, Piazzale Aldo Moro 2, I-00185 Rome, Italy}

{\sl $^{309}$ Universit{\`a} degli Studi di Trieste, Dipartimento di Fisica, via A. Valerio 2, I-34127 Trieste, Italy}

{\sl $^{310}$ Universit{\`a} degli Studi di ``Roma Tre'', Dipartimento di Fisica ``Edoardo Amaldi'', Istituto Nazionale di Fisica Nucleare, Via della Vasca Navale 84, 00146 Roma, Italy}

{\sl $^{311}$ Universit{\`a} dell'Insubria in Como, Dipartimento di Scienze CC.FF.MM., via Vallegio 11, I-22100 Como, Italy}

{\sl $^{312}$ Universit{\`a} di Pisa, Departimento di Fisica 'Enrico Fermi', Largo Bruno Pontecorvo 3, I-56127 Pisa, Italy}

{\sl $^{313}$ Universit{\`a} di Salento, Dipartimento di Fisica, via Arnesano, C.P. 193, I-73100 Lecce, Italy}

{\sl $^{314}$ Universit{\`a} di Udine, Dipartimento di Fisica, via delle Scienze, 208, I-33100 Udine, Italy}

{\sl $^{315}$ Variable Energy Cyclotron Centre, 1/AF, Bidhan Nagar, Kolkata 700064, India}

{\sl $^{316}$ VINCA Institute of Nuclear Sciences, Laboratory of Physics, PO Box 522, YU-11001 Belgrade, Serbia and Montenegro}

{\sl $^{317}$ Vinh University, 182 Le Duan, Vinh City, Nghe An Province, Vietnam}

{\sl $^{318}$ Virginia Polytechnic Institute and State University, Physics Department, Blacksburg, VA 2406, USA}

{\sl $^{319}$ Visva-Bharati University, Department of Physics, Santiniketan 731235, India}

{\sl $^{320}$ Waseda University, Advanced Research Institute for Science and Engineering, Shinjuku, Tokyo 169-8555, Japan}

{\sl $^{321}$ Wayne State University, Department of Physics, Detroit, MI 48202, USA}

{\sl $^{322}$ Weizmann Institute of Science, Department of Particle Physics, P.O. Box 26, Rehovot 76100, Israel}

{\sl $^{323}$ Yale University, Department of Physics, New Haven, CT 06520, USA}

{\sl $^{324}$ Yonsei University, Department of Physics, 134 Sinchon-dong, Sudaemoon-gu, Seoul 120-749, Korea}

{\sl $^{325}$ Zhejiang University, College of Science, Department of Physics, Hangzhou, China 310027}

{\sl * deceased } 

\end{center}

\end{center}

\chapter*{Acknowledgements} 
We would like to acknowledge the support and guidance of the International Committee on Future Accelerators (ICFA), chaired by A. Wagner of DESY, and the International Linear Collider Steering Committee (ILCSC), chaired by S. Kurokawa of KEK, who established the ILC Global Design Effort, as well as the World Wide Study of the Physics and Detectors.
      
\medskip
We are grateful to the ILC Machine Advisory Committee (MAC), chaired by F. Willeke of DESY and the International ILC Cost Review Committee, chaired by L. Evans of CERN, for their advice on the ILC Reference Design. We also thank the consultants who particpated in the Conventional Facilities Review at CalTech and in the RDR Cost Review at SLAC. 

\medskip 
We would like to thank the directors of the institutions who have hosted ILC meetings: KEK, ANL/FNAL/SLAC/U. Colorado (Snowmass), INFN/Frascati, IIT/Bangalore, TRIUMF/U. British Columbia, U. Valencia, IHEP/Beijing and DESY.

\medskip 
We are grateful for the support of the Funding Agencies for Large Colliders (FALC), chaired by R. Petronzio of INFN, and we thank all of the international, regional and national funding agencies whose generous support has made the ILC 
Reference Design possible.

\medskip 
Each of the GDE regional teams in the Americas, Asia and Europe are grateful for the support of their local scientific societies, industrial forums, advisory committees and reviewers.

\cleardoublepage

\thispagestyle{myheadings}\markright{Executive Summary}
\tableofcontents %

\setcounter{secnumdepth}{0}





\mainmatter
\setcounter{page}{1} \setcounter{chapter}{0}


\newcommand{\picturefolder}{}

\clearpage
\setcounter{figure}{0}
\setcounter{table}{0}
\setcounter{secnumdepth}{4}
\setcounter{subsection}{0} 
\setcounter{subsubsection}{0}
\usecounter{subsubsection}
\chapter{Physics at a Terascale e$^+$e$^-$ Linear Collider}
\section{Questions about the Universe}
\begin{itemize}

\item {\it What is the universe?  How did it begin?} \itemspace 
\item {\it What are matter and energy?  What are space and time?} \itemspace 

\end{itemize}
These basic questions have been the subject of scientific theories
and experiments throughout human history.  The answers have
revolutionized the enlightened view of the world, transforming
society and advancing civilization. Universal laws and principles
govern everyday phenomena, some of them manifesting themselves only
at scales of time and distance far beyond everyday experience.
Particle physics experiments using particle accelerators transform
matter and energy, to reveal the basic workings of the universe.
Other experiments exploit naturally occurring particles, such as
solar neutrinos or cosmic rays, and astrophysical observations, to
provide additional insights.

The triumph of 20th century particle physics was the development
of the Standard Model. Experiments determined the
particle constituents of ordinary matter, and identified four forces
binding matter and transforming it from one form to another. 
This
success leads particle physicists to address even more fundamental
questions, and explore deeper mysteries in science. The scope of
these questions is illustrated by the summary from the report {\it
Quantum Universe}\cite{PhysCase:QuantumUniverse}:%
\begin{enumerate}%

\item {\it Are there undiscovered principles of nature?} \itemspace 
\item {\it How can we solve the mystery of dark energy?} \itemspace 
\item {\it Are there extra dimensions of space?} \itemspace %
\item {\it Do all the forces become one?} \itemspace 
\item {\it Why are there so many particles?} \itemspace %
\item {\it What is dark matter? How can we make it in the laboratory?} \itemspace 
\item {\it What are neutrinos telling us?} \itemspace 
\item {\it How did the universe begin?} \itemspace %
\item {\it What happened to the antimatter?} \itemspace %

\end{enumerate}

A worldwide particle physics program explores this fascinating
scientific landscape. The International Linear Collider
(ILC)\cite{PhysCase:ilc} is expected to play a central role in an
era of revolutionary advances\cite{PhysCase:hepap06} with
breakthrough impact on many of these fundamental questions.

\begin{figure}[htb]
   \begin{center} \vbabove
      \label{fig:terascale}
\includegraphics[width=\textwidth]{\picturefolder 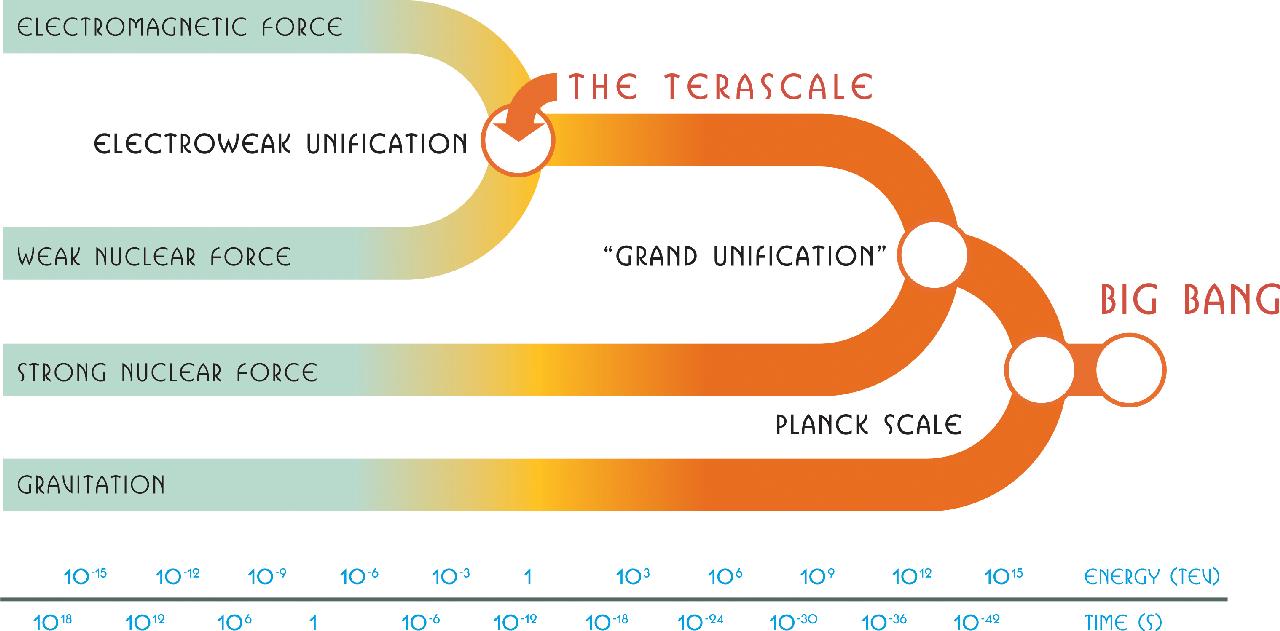}
\vbabovecaption
      \caption[The electromagnetic and weak nuclear forces unify at the
      Terascale.] {The electromagnetic and weak nuclear forces unify at the
      Terascale.The ILC will test unification at even high
      energy scales (from {\it Discovering the Quantum Universe}).}
   \end{center} \vbbelow
\end{figure}

The Standard Model includes a third component beyond particles and
forces that has not yet been verified, the Higgs mechanism that
gives mass to the particles. Many scientific opportunities for the
ILC involve the Higgs particle and related new phenomena at
Terascale energies. The Standard Model Higgs field permeates the
universe, giving mass to elementary particles, and breaking a
fundamental electroweak force into two, the electromagnetic and weak
forces (Figure~\ref{fig:terascale}).
But quantum effects should destabilize the Higgs of the
Standard Model, preventing its operation at Terascale energies. The
proposed antidotes for this quantum instability mostly involve
dramatic phenomena accessible to the ILC: new forces, a new
principle of nature called supersymmetry, or even extra dimensions
of space.

Thus the Higgs is central to a broad program of discovery. Is there
really a Higgs? Or are there other mechanisms that give mass to
particles and break the electroweak force? If there is a Higgs, does
it differ from the Standard Model? Is there more than one Higgs
particle? What new phenomena stabilize the Higgs at the Terascale?

\begin{figure}[htb]
   \begin{center} \vbabove 
      \label{fig:relicdensity}
\includegraphics[width=0.70\textwidth]{\picturefolder 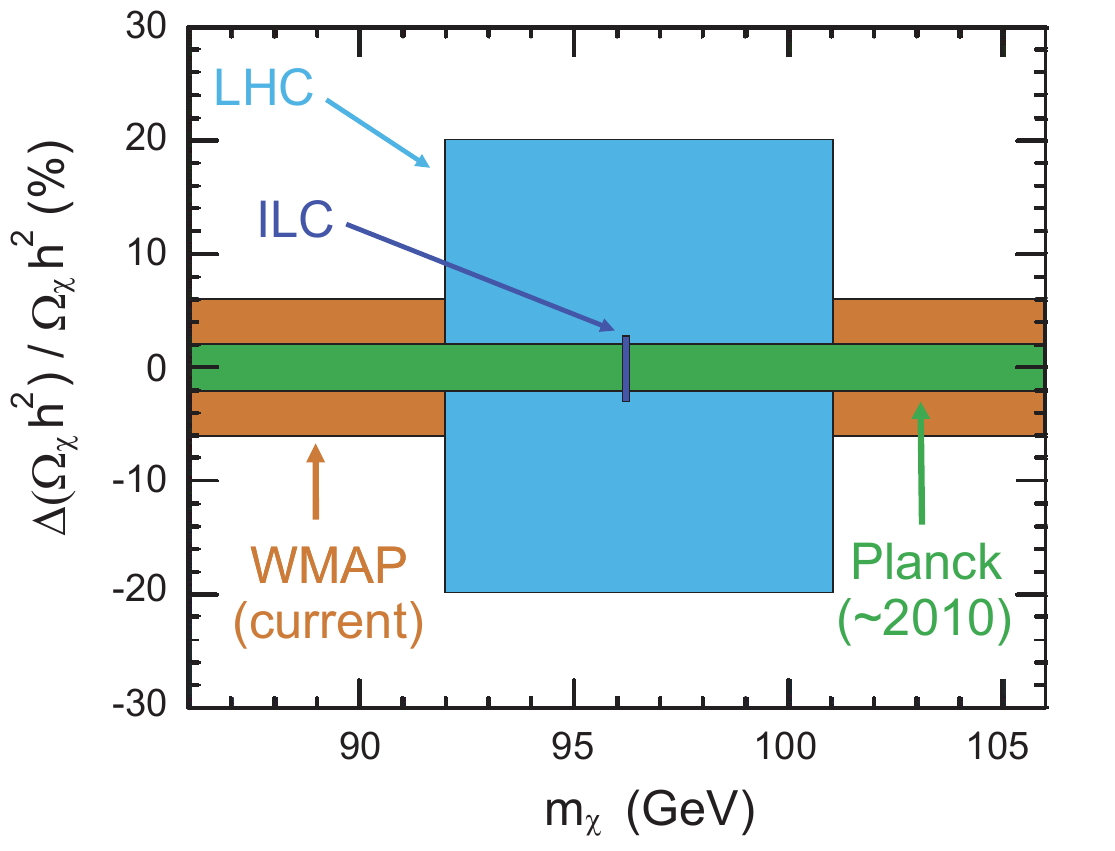}
\vbabovecaption
      \caption[Relic density and mass determinations for neutralino
      dark matter.] 
{Accuracy of
relic density $(\Omega_{\chi} h^2)$ and mass determinations for
neutralino dark matter. Comparison of the LHC and ILC data with that of the
WMAP and Planck satellites test neutralinos as the dark matter.
(ALCPG  Cosmology Subgroup,  from chapter 7, volume 2: Physics
at the ILC, ILC Reference Design Report) }
   \end{center} \vbbelow
\end{figure}

Astrophysical data show that dark matter dominates the matter content of the universe, and
cannot be explained by known particles. Dark matter may be comprised
of new weakly interacting particles with Terascale masses. If such
Terascale dark matter exists, experiments at the ILC should produce
and study such particles, raising important questions 
(Figure~\ref{fig:relicdensity}).
Do these new
particles have all the properties of the dark matter? Can they alone
account for all of the dark matter? How would they affect the
evolution of the universe? How do they connect to new principles or
forces of nature?

ILC experiments could test the idea that fundamental forces
originate from a single ``grand'' unified force, and search for
evidence of a related unified origin of matter involving
supersymmetry. They could distinguish among patterns of phenomena to
judge different unification models, providing a telescopic view of
the ultimate unification.

\section{The New Landscape of Particle Physics}

During the next few years, experiments at CERN's Large Hadron
Collider will have the first direct look at Terascale physics. While
those results are unpredictable \cite{PhysCase:atlas}, they could
considerably enhance the physics case for the ILC. Possible
discoveries include the Higgs particle, a recurrence of the Z boson(the Z$^{\prime}$), 
evidence for extra dimensions, or observation of
supersymmetry (SUSY) particles. Like the discovery of an uncharted continent, 
exploration of the Terascale could transform forever the geography of our universe. 
Equally compelling will be the interplay of LHC discoveries with other experiments and
observations. Particle physics should be entering a new era of
intellectual ferment and revolutionary advance.

If there is a Higgs boson, it is almost certain to be found at the
LHC and its mass measured by the ATLAS and CMS experiments. 
If there is a multiplet of Higgs bosons, there is a good chance the LHC 
experiments will see more than one. However it will be difficult for the LHC 
to measure the spin and parity of the Higgs particle and thus to 
establish its essential nature; the ILC can make these measurements accurately. 
If there is more than one decay channel of the Higgs, the LHC experiments will 
determine the ratio of branching fractions (roughly 7-30\%); the ILC will measure 
these couplings to quarks and vector bosons at the few percent level, and thus 
reveal whether the Higgs is the simple Standard Model object, or something more complex.

This first look at Terascale physics by the LHC can have three
possible outcomes. The first possibility is that a Higgs boson consistent 
with Standard Model properties has been found.  Then the ILC will be able to 
make a more complete and precise experimental analysis to verify if it is 
indeed Standard Model or something else. The second possibility is that a Higgs boson
is found with gross features at variance with the Standard Model.
Such discrepancies might be a Higgs mass significantly above Standard Model expectations, a large
deviation in the predicted pattern of Higgs decay, or the discovery
of multiple Higgs particles. The ILC measurements of couplings and quantum numbers 
will point to the new physics at work.  The third possibility is that no
Higgs boson is seen. In this case, the ILC precision measurements of top quark, 
Z and W boson properties will point the way to an alternate theory. 
In all cases, the ILC will be essential
to a full understanding of the Higgs and its relation to other new
fundamental phenomena.

The ATLAS and CMS experiments at LHC will have impressive
capabilities to discover new heavy particles. They could detect a
new Z$^{\prime}$ gauge boson as heavy as 5 TeV\cite{PhysCase:cousins}, or
squarks and gluinos of supersymmetry up to
2.5 TeV\cite{PhysCase:atlas}. New particles with mass up to a few TeV
associated with the existence of extra spatial dimensions could be
seen\cite{PhysCase:atlas}. The discovery of a Z$^{\prime}$ particle would
indicate a new fundamental force of nature. The question would be
to deduce the properties of this force, its origins, its relation to the other
forces in a unified framework, and its role in the earliest moments
of the universe. The ILC would play a definitive role in addressing
these questions.

If supersymmetry is responsible for stabilizing the electroweak unification 
at the Terascale and for providing a light Higgs
boson, signals of superpartner particles should be seen at the LHC.
But are the new heavy particles actually superpartners, with the right
spins and couplings? Is supersymmetry related to unification at a
higher energy scale? What was its role in our
cosmic origins? Definitive answers to these questions will require
precise measurements of the superpartner particles and the Higgs
particles. This will require the best possible results from the LHC
and the ILC in a combined analysis.

Supersymmetry illustrates the possible interplay between different
experiments and observations. Missing energy signatures at the LHC
may indicate a weakly interacting massive particle consistent with a
supersymmetric particle. Direct or indirect dark matter searches may
see a signal for weakly interacting exotic particles in our galactic
halo. Are these particles neutralinos, responsible for some or all of the
dark matter? Does the supersymmetry model preferred by collider data
predict the observed abundance of dark matter 
(Figure~\ref{fig:relicdensity}), or do assumptions
about the early history of the universe need to change? ILC
measurements will be mandatory for these analyses.

Alternative possible structures of the new physics include phenomena
containing extra dimensions, introducing connections between
Terascale physics and gravity. One possibility is that the weakness of 
gravity could be understood by the escape of the gravitons into the new 
large extra dimensions. Events with unbalanced momentum caused by the 
escaping gravitons could be seen at both the LHC and the ILC. The ILC 
could confirm this scenario by observing anomalous electron positron 
pair production caused by graviton exchange.

Another possible extra-dimensional model (warped extra-dimensions)
postulates two three-dimensional branes separated along one of the new dimensions. 
In this scenario, new resonances could appear at the colliders, and again 
pair production at the ILC would be critical to confirmation.  The measurement 
of the couplings to leptons at the ILC would reveal the nature of the new states.

In these differing scenarios, the ILC has a critical role to play in
resolving the confusing possible interpretations. In some scenarios
the new phenomena are effectively hidden from the LHC detectors, but
are revealed as small deviations in couplings that could be measured
at the ILC. In some cases the LHC experiments could
definitively identify the existence of extra dimensions. Then the
ILC would explore the size, shape, origins and impact of this
expanded universe. A powerful feature of the ILC is its capability
to explore new physics in a model independent way.

\section{Precision Requirements for ILC}

ILC has an unprecedented potential for precision measurements, with
new windows of exploration for physics beyond the Standard Model.
This implies new requirements on theoretical and experimental
accuracies. This in turn drives the need for more precise
theoretical calculations for standard, Higgs and supersymmetry
processes at the Terascale. There must be a corresponding effort to
eliminate all known instrumental limitations which could compromise
the precision of the measurements. These would include limits on the
accuracy of momentum resolution, jet reconstruction, or
reconstruction of short lived particles.

The ILC will search for invisible particles, candidates for the Dark
Matter. This requires that the detector be as hermetic as possible.
Machine backgrounds must be well controlled to reach the highest
precision. The luminosity and polarisation of the beams must also be
accurately known.

\section{Specifying Machine Parameters}

The accelerator described in Chapter~\ref{chapAcc} has been designed
to meet the basic parameters required for the planned physics
program \cite{PhysCase:ilcpars}.  The initial maximum center of mass energy
is $\sqrt{s}$ = 500 GeV.  Physics runs are possible for every energy
above $\sqrt{s}$ = 200 GeV and calibration runs with limited
luminosity are possible at $\sqrt{s}$ = 91 GeV.  The beam energy can
be changed in small steps for mass measurement threshold scans.

The total luminosity required is 500 fb$^{-1}$ within the first four
years of operation and 1000 fb$^{-1}$ during the first phase of
operation at 500 GeV.  The electron beam must have a polarisation
larger than 80\%. The positron source should be upgradable to
produce a beam with more than $\pm$50\%
polarisation\cite{PhysCase:gudi}. Beam energy and polarisation must
be stable and measurable at a level of about 0.1\%.

An e$^+$e$^-$  collider is uniquely capable of operation at a series
of energies near the threshold of a new physical process. This is an
extremely powerful tool for precision measurements of particle
masses and unambiguous particle spin determinations. In a broad
range of scenarios, including those with many new particles to
explore and thresholds to measure, it is possible to achieve
precision for all relevant observables in a reasonable time span.

All of the physics scenarios studied indicate that a $\sqrt{s}$ =
500 GeV collider can have a great impact on understanding the
physics of the Terascale. An energy upgrade up to $\sqrt{s}\sim$ 1
TeV opens the door to even greater discoveries. With modest
modifications, the ILC can also offer other options if required by
physics, although these are not all explicitly included in the RDR
design. For GigaZ, the ILC would run on the Z-resonance with high
luminosity and both beams polarised, producing 10$^9$ hadronic Z
decays in less than a year. The ILC could also run at the W-pair
production threshold for a high precision W-mass
measurement\cite{PhysCase:hawkings}.  Both linacs could accelerate
electrons for an e$^-$e$^-$ collider\cite{PhysCase:heusch},
measuring the mass of a particular supersymmetric particle, the
selectron, if it exists in the ILC energy range. Colliding electrons
with a very intense laser beam near the interaction point can
produce a high energy, high quality photon beam, resulting in an
e$^-\gamma$ or $\gamma\;\gamma$ collider\cite{PhysCase:ginzburg}.
After operating
below or at 500 GeV for a number of years, the ILC could be upgraded to 
higher energy or be modified for one of
the options. It would then operate for several years in the new 
configuration.

\clearpage
\setcounter{figure}{0}
\setcounter{table}{0}
\setcounter{secnumdepth}{4}
\setcounter{subsection}{0} 
\setcounter{subsubsection}{0}
\usecounter{subsubsection}
\chapter{The ILC Accelerator}\label{chapAcc}

The ILC is
based on 1.3 GHz superconducting radio-frequency (SCRF) accelerating
cavities. The use of the SCRF technology was recommended by
the International Technology Recommendation Panel (ITRP) in August
2004 \cite{itrp}, and shortly thereafter endorsed by the International
Committee for Future Accelerators (ICFA). In an unprecedented
milestone in high-energy physics, the many institutes around the world
involved in linear collider R\&D united in a common effort to produce a
global design for the ILC.  In November 2004, the 1st International
Linear Collider Workshop was held at KEK, Tsukuba, Japan. The workshop
was attended by some 200 physicists and engineers from around the world,
and paved the way for the 2nd ILC Workshop in August 2005, held at
Snowmass, Colorado, USA, where the ILC Global Design Effort (GDE) was
officially formed. The GDE membership reflects the global nature of
the collaboration, with accelerator experts from all three regions
(Americas, Asia and Europe). The first major goal of the GDE was to
define the basic parameters and layout of the machine -- the Baseline
Configuration. This was achieved at the first GDE meeting held at
INFN, Frascati, Italy in December 2005 with the creation of the
Baseline Configuration Document (BCD). During the next 14 months, the
BCD was used as the basis for the detailed design work and value
estimate culminating in the completion
of the second major milestone, the publication of the draft ILC
Reference Design Report (RDR).

The technical design and cost estimate for the ILC is based on two
decades of world-wide Linear Collider R\&D, beginning with the
construction and operation of the SLAC Linear Collider (SLC). The SLC
is acknowledged as a proof-of-principle machine for the linear
collider concept. The ILC SCRF linac technology was pioneered by the
TESLA collaboration\footnote{Now known as the TESLA Technology
Collaboration (TTC); see \url{http://tesla.desy.de}}, culminating in a
proposal for a 500 GeV center-of-mass linear collider in
2001~\cite{tdr}. The concurrent (competing) design work on a normal
conducting collider (NLC with X-band~\cite{nlc} and GLC with X- or
C-Band~\cite{glc}), has advanced the design concepts for the ILC
injectors, Damping Rings (DR) and Beam Delivery System (BDS), as well
as addressing overall operations, machine protection and availability
issues. The X- and C-band R\&D has led to concepts for RF power
sources that may eventually produce either cost and/or performance
benefits. Finally, the European XFEL \cite{euro-xfel} to be constructed at DESY,
Hamburg, Germany, will make use of the TESLA linac technology, and
represents a significant on-going R\&D effort of great
benefit for the ILC.

The current ILC baseline assumes an average accelerating gradient of
31.5 MV/m in the cavities to achieve a center-of-mass energy of 500
GeV. The high luminosity requires the use of high power and small
emittance beams. The choice of 1.3 GHz SCRF is well suited to the
requirements, primarily because the very low power loss in the SCRF
cavity walls allows the use of long RF pulses, relaxing the
requirements on the peak-power generation, and ultimately leading to
high wall-plug to beam transfer efficiency.

The primary cost drivers are the SCRF Main Linac technology and the
Conventional Facilities (including civil engineering). The choice of
gradient is a key cost and performance parameter, since it dictates
the length of the linacs, while the cavity quality factor 
($Q_0$) relates to the required cryogenic cooling power. The
achievement of 31.5 MV/m as the baseline average operational
accelerating gradient -- requiring a minimum performance of 35 MV/m
during cavity mass-production acceptance testing -- represents the
primary challenge to the global ILC R\&D

With the completion of the RDR, the GDE will begin an
engineering design study, closely coupled with a prioritized R\&D
program. The goal is to produce an Engineering Design Report (EDR) by
2010, presenting the matured technology, design and construction plan
for the ILC, allowing the world High Energy Physics community to seek
government-level project approvals, followed by start of construction
in 2012. When combined with the seven-year construction phase that is
assumed in studies presented in RDR, this timeline will allow
operations to begin in 2019. This is consistent with a technically
driven schedule for this international project.


\setcounter{secnumdepth}{4}
\setcounter{subsection}{0} 
\setcounter{subsubsection}{0}
\usecounter{subsubsection}

\section{Superconducting RF}

The primary cost driver for the ILC is the superconducting RF
technology used for the Main Linacs, bunch compressors and injector
linacs. In 1992, the TESLA Collaboration began R\&D on 1.3 GHz
technology with a goal of reducing the cost per MeV by a factor of 20
over the then state-of-the-art SCRF installation (CEBAF). This was
achieved by increasing the operating accelerating gradient by a factor
of five from ~5 MV/m to ~25 MV/m, and reducing the cost per meter of
the complete accelerating module by a factor of four for large-scale
production.

\begin{figure}[htb] \vbabove 
\begin{center}
  \includegraphics[width=0.95\textwidth]{\picturefolder 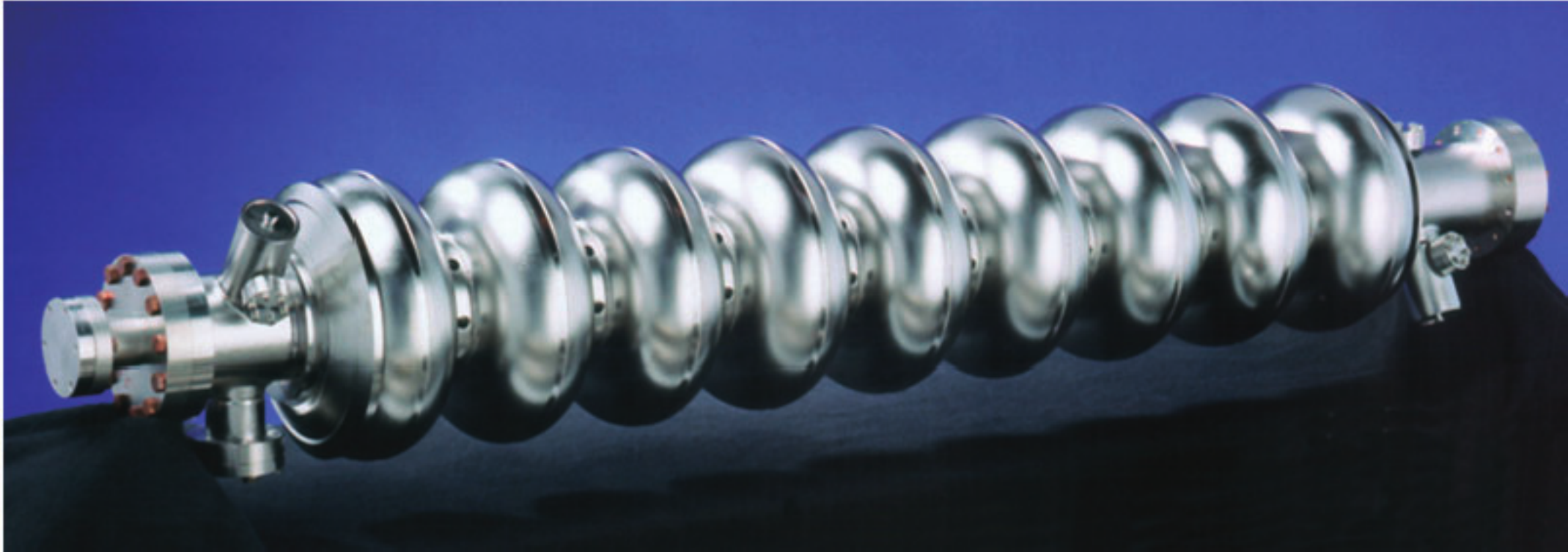}
  \vbabovecaption \caption{A TESLA nine-cell 1.3 GHz superconducting niobium cavity.}
  \label{fig:OVtesla9cell}
\end{center} \vbbelow 
\end{figure}

The TESLA cavity R\&D was based on extensive existing experience from
CEBAF (Jefferson Lab), CERN, Cornell University, KEK, Saclay and
Wuppertal. The basic element of the technology is a nine-cell 1.3 GHz
niobium cavity, shown in Figure~\ref{fig:OVtesla9cell}. Approximately
160 of these cavities have been fabricated by industry as part of the
on-going R\&D program at DESY; some 17,000 are needed for the ILC.

\begin{figure}[htb] \vbabove 
\begin{center}
  \includegraphics[width=0.95\textwidth]{\picturefolder 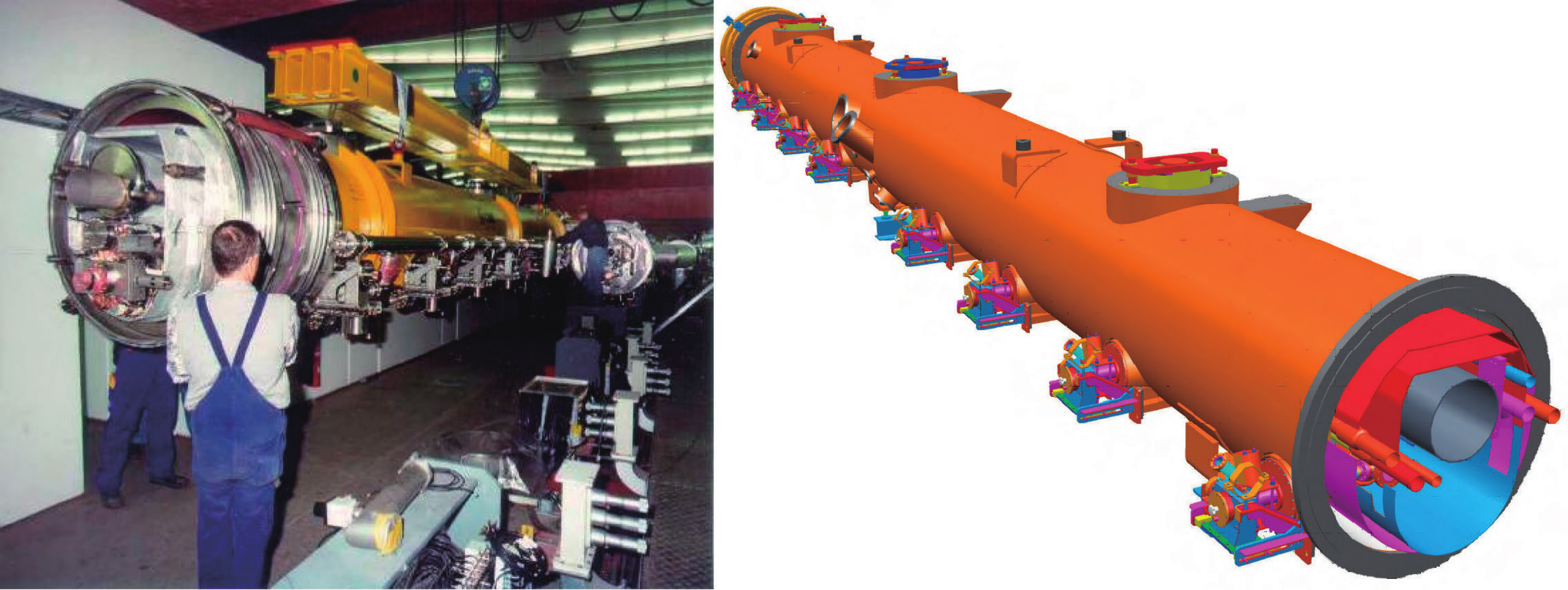}
  \vbabovecaption \caption [SCRF Cryomodules.]{ SCRF Cryomodules. Left: an 8 cavity TESLA cryomodule is
    installed into the FLASH linac at DESY. Right: design for the 4th
    generation ILC prototype cryomodule, due to be constructed at
    Fermilab National Laboratory.}
  \label{fig:OVcryomodule}
\end{center} \vbbelow 
\end{figure}

A single cavity is approximately 1 m long. The cavities must be
operated at 2 K to achieve their performance. Eight or nine cavities are
mounted together in a string and assembled into a common
low-temperature cryostat or {\it cryomodule}
(Figure~\ref{fig:OVcryomodule}), the design of which is already in the
third generation. Ten cryomodules have been produced to-date, five of
which are currently installed in the VUV free-electron laser
(FLASH)\footnote{Originally known as the TESLA Test Facility (TTF).}
at DESY, where they are routinely operated. DESY is currently
preparing for the construction of the European XFEL facility, which
will have a $\sim20$ GeV superconducting linac containing 116
cryomodules.

\begin{figure}[htb] \vbabove 
\begin{center}
  \includegraphics[width=0.95\textwidth]{\picturefolder 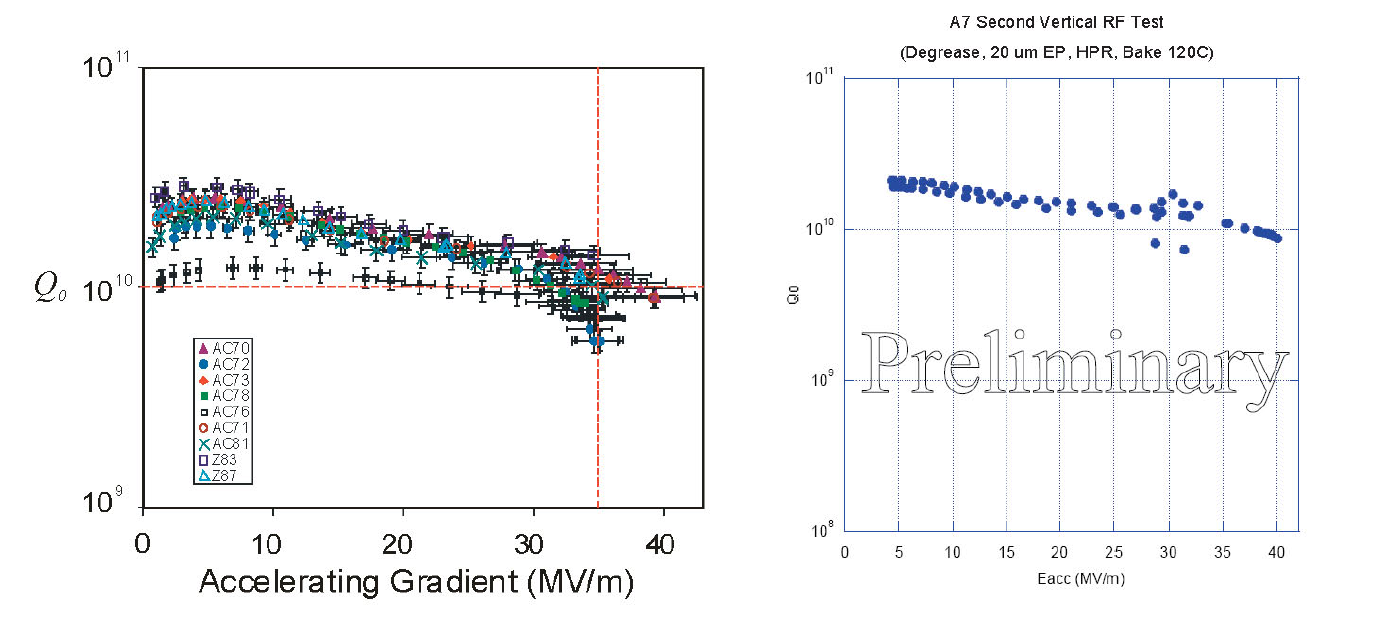}
 \vbabovecaption  \caption[High-performance nine-cell cavities.]{High-performance nine-cell cavities. Left: Examples of DESY
    nine-cell cavities achieving $\ge 35$ MV/m. Right: Recent result
    from Jefferson Lab of nine-cell cavity achieving ~40 MV/m.  }
  \label{fig:OV9cellperf}
\end{center}  \vbbelow 
\end{figure}

The ILC community has set an aggressive goal of routinely
achieving\footnote{Acceptance test.} 35 MV/m in nine-cell cavities,
with a minimum production yield of 80\%. Several cavities have already
achieved these and higher gradients (see
Figure~\ref{fig:OV9cellperf}), demonstrating proof of
principle. Records of over 50 MV/m have been achieved in single-cell
cavities at KEK and Cornell\cite{high-g}. However, it is still a challenge
to achieve the desired production yield for nine-cell cavities
at the mass-production levels ($\sim$17,000 cavities) required.

\begin{figure}[htb] \vbabove 
\begin{center}
  \includegraphics[width=\textwidth]{\picturefolder 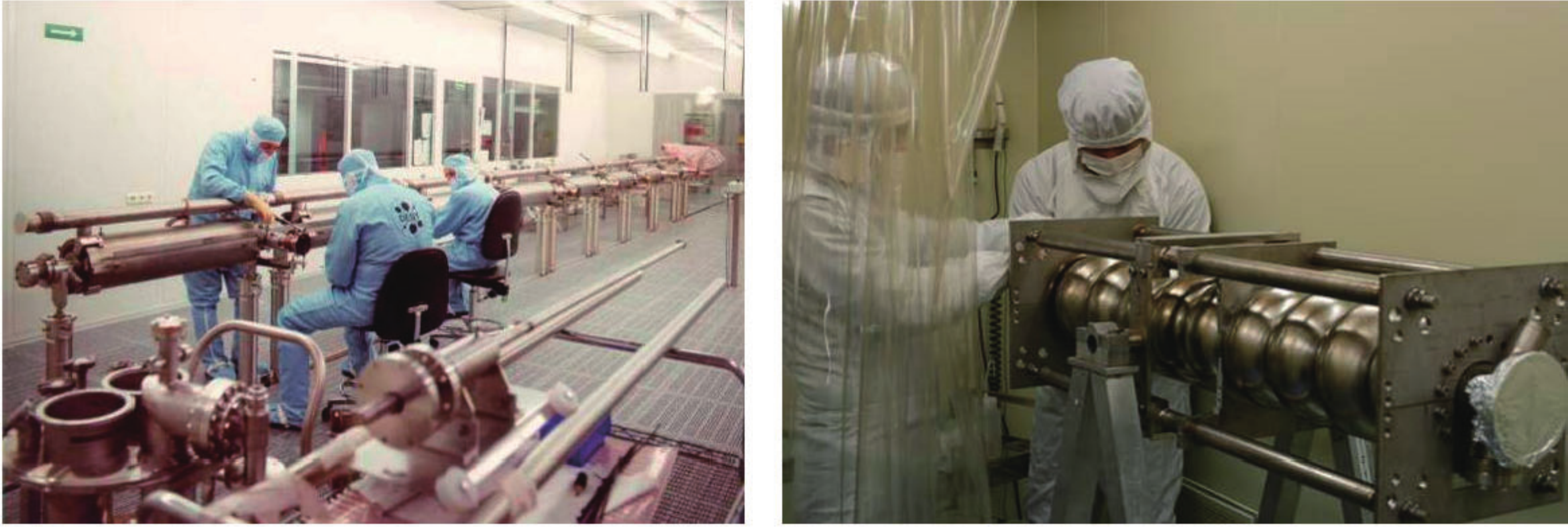}
  \vbabovecaption \caption[Clean room environments are mandatory.] {Clean room environments are mandatory. Left: the assembly
    of eight nine-cell TESLA cavities into a cryomodule string at
    DESY. Right: an ICHIRO nine-cell cavity is prepared for initial
    tests at the Superconducting RF Test Facility (STF) at KEK. }
  \label{fig:OVcleanrm}
\end{center} \vbbelow 
\end{figure}

The key to high-gradient performance is the ultra-clean and
defect-free inner surface of the cavity. Both cavity preparation and
assembly into cavity strings for the cryomodules must be performed in
clean-room environments (Figure~\ref{fig:OVcleanrm}).

\begin{figure}[htb] \vbabove 
\begin{center}
  \includegraphics[width=\textwidth]{\picturefolder 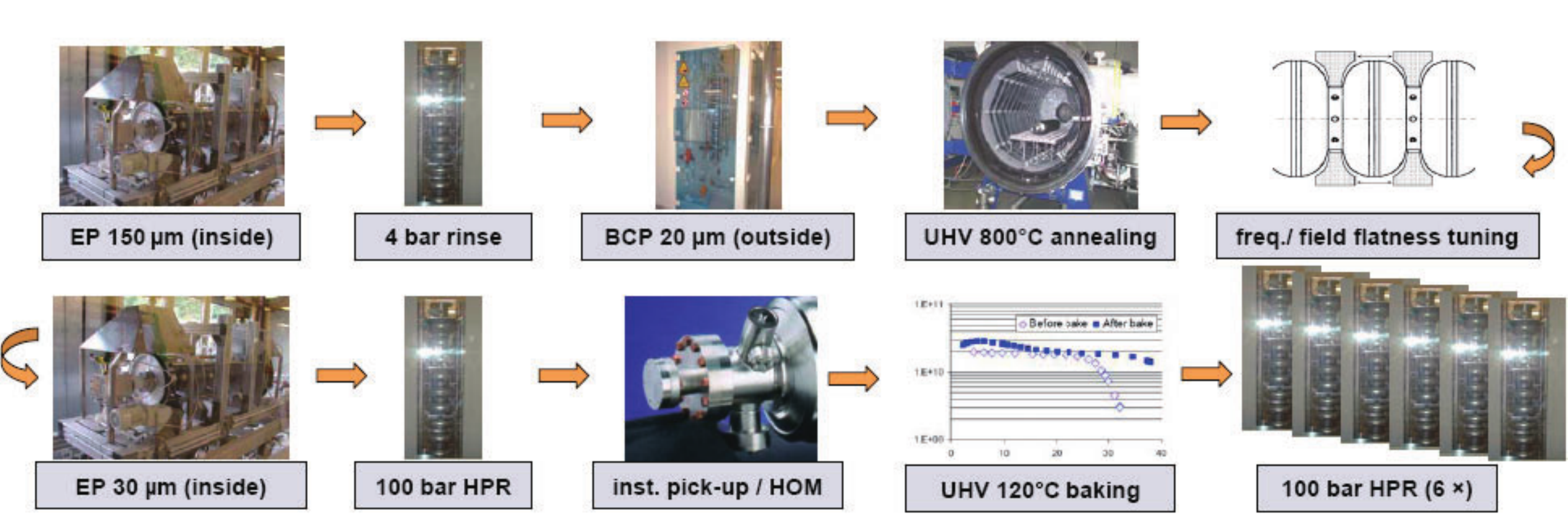}
  \vbabovecaption \caption[Birth of a nine-cell cavity.] {Birth of a nine-cell cavity: basic steps in surface
    treatment needed to achieve high-performance superconducting
    cavities. (EP = electropolishing; HPR = high-pressure rinsing.) }
  \label{fig:OVcavproc}
\end{center} \vbbelow 
\end{figure}

The best
cavities have been achieved using electropolishing, a common industry
practice which was first developed for use with superconducting
cavities by CERN and KEK. Over the last few years, research at
Cornell, DESY, KEK and Jefferson Lab has led to an agreed standard
procedure for cavity preparation, depicted in
Figure~\ref{fig:OVcavproc}. The focus of the R\&D is now to optimize
the process to guarantee the required yield. The ILC SCRF community
has developed an internationally agreed-upon plan to address the
priority issues.

The high-gradient SCRF R\&D required for ILC is expected to ramp-up
world-wide over the next years. The U.S. is currently investing in new
infrastructure for nine-cell cavity preparation and string and
cryomodule assembly. These efforts are centered at Fermilab (ILC Test
Accelerator, or ILCTA), together with ANL, Cornell University, SLAC
and Jefferson Lab. In Japan, KEK is developing the
Superconducting RF Test Facility (STF). In Europe, the focus of R\&D at
DESY has shifted to industrial preparation for construction of the
XFEL. There is continued R\&D to support the high-gradient program, as
well as other critical ILC-related R\&D such as high-power RF couplers
(LAL, Orsay, France) and cavity tuners (CEA Saclay, France; INFN
Milan, Italy).

The quest for high-gradient and affordable SCRF technology for
high-energy physics has revolutionized accelerator applications. In addition
to the recently completed Spallation Neutron Source (SNS) in Oak Ridge, Tennessee and the European XFEL under construction, many linac-based
projects utilizing
SCRF technology are being developed, including 4th-generation light
sources such as single-pass FELs and energy-recovery linacs. For the large
majority of new accelerator-based projects, SCRF has become the
technology of choice.


\setcounter{secnumdepth}{4}
\setcounter{subsection}{0} 
\setcounter{subsubsection}{0}
\usecounter{subsubsection}
\section{The ILC Baseline Design}\label{Ovr:baseline}

\begin{figure}[hb] \vbabove 
\begin{center}
  \includegraphics[width=8.5cm]{\picturefolder 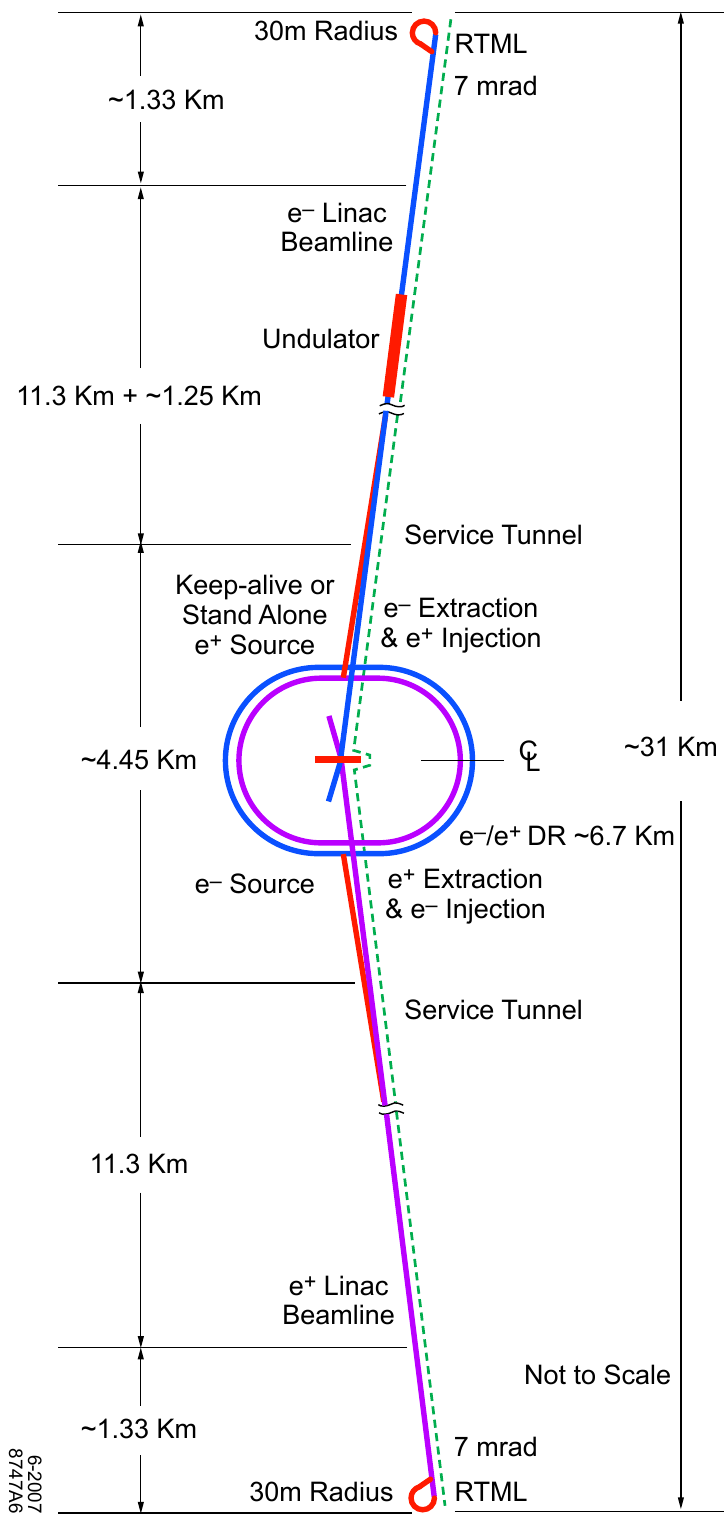}
  \vbabovecaption \caption{Schematic layout of the ILC complex for 500 GeV CM.}
  \label{fig:OVlayout}
\end{center} \vbbelow 
\end{figure}

The overall system design has been chosen to realize the physics
requirements with a maximum CM energy of 500 GeV and a peak luminosity
of $2\times10^{34}$ cm$^{-2}$s$^{-1}$. Figure~\ref{fig:OVlayout} shows
a schematic view of the overall layout of the ILC, indicating the
location of the major sub-systems:

\begin{itemize}

  \item a polarized electron source based on a photocathode DC gun; \itemspace 
  \item an undulator-based positron source, driven by the 150 GeV
    main electron beam; \itemspace 
  \item 5 GeV electron and positron damping rings (DR) with a
    circumference of 6.7 km, housed in a common tunnel at the center
    of the ILC complex; \itemspace 
  \item beam transport from the damping rings to the main linacs,
    followed by a two-stage bunch compressor system prior to injection into the
    main linac; \itemspace 
  \item two 11 km long main linacs, utilizing 1.3 GHz SCRF cavities,
    operating at an average gradient of 31.5 MV/m, with a pulse length
    of 1.6 ms; \itemspace 
  \item a 4.5 km long beam delivery system, which brings the two beams
    into collision with a 14 mrad crossing angle, at a single
    interaction point which can be shared by two detectors. \itemspace 

\end{itemize}

The total footprint is $\sim$31 km. The electron source, the damping
rings, and the positron auxiliary (`keep-alive') source are centrally
located around the interaction region (IR). The plane of the damping
rings is elevated by $\sim$10 m above that of the BDS to avoid
interference.

To upgrade the machine to $E_{\rm cms} =1$ TeV, the linacs and the beam
transport lines from the damping rings would be extended by another
$\sim 11$ km each. Certain components in the beam delivery system
would also need to be augmented or replaced.

\subsection{Beam Parameters}

\begin{table}[htb]
\caption[Basic design parameters for the ILC.]{Basic design parameters for the ILC ($^{a)}$ values at 500 GeV
  center-of-mass energy). \label{tab:BaseParams}}
\begin{center}
    \begin{tabular}{| l | l | l |} \hline 
    Parameter                   & Unit              &          \\ \hline
    & & \vbdlspacing \hline
    Center-of-mass energy range & GeV               & 200 - 500         \\ \hline
    Peak luminosity$^{a)}$      & cm$^{-2}$s$^{-1}$ & $2\times 10^{34}$ \\ \hline
    Average beam current in pulse& mA                & 9.0               \\ \hline
    Pulse rate                  & Hz                & 5.0               \\ \hline
    Pulse length (beam)         & ms                & $\sim1$           \\ \hline
    Number of bunches per pulse &                   & 1000 - 5400       \\ \hline
    Charge per bunch            & nC                & 1.6 - 3.2         \\ \hline
    Accelerating gradient$^{a)}$      & MV/m        & 31.5              \\ \hline
    RF pulse length             & ms                & 1.6               \\ \hline
    Beam power (per beam)$^{a)}$      & MW          & 10.8              \\ \hline
    Typical beam size at IP$^{a)}$ ($h\times v$) & nm & 640 $\times$ 5.7\\ \hline
    Total AC Power consumption$^{a)}$ &MW           & 230               \\
    \hline
    \end{tabular}
\end{center}
\end{table}

The nominal beam parameter set, corresponding to the design luminosity
of $2\times10^{34}$ cm$^{-2}$s$^{-1}$ at $E_{\rm cms} = 500$ GeV is given in Table~\ref{tab:BaseParams}. These parameters have been chosen to optimize between known accelerator physics and technology
challenges throughout the whole accelerator complex. Examples of such
challenges are:

\begin{itemize}

  \item beam instability and kicker hardware constraints in the
    damping rings; \itemspace 
  \item beam current, beam power, and pulse length limitations in the
    main linacs; \itemspace 
  \item emittance preservation requirements, in the main linacs and
    the BDS; \itemspace 
  \item background control and kink instability issues in the interaction region.
 \itemspace 

\end{itemize}

Nearly all high-energy physics accelerators have shown unanticipated
difficulties in reaching their design luminosity. The ILC design
specifies that each subsystem support a range of beam parameters. The
resulting flexibility in operating parameters will allow identified
problems in one area to be compensated for in another. The nominal IP
beam parameters and design ranges are presented in
Table~\ref{tab:OVparamrange}.

\begin{table}[htb] \vbabove 
\caption[Nominal and design range of beam parameters at the IP.] {Nominal and design range of beam parameters at the IP. The
  min. and max. columns do not represent consistent sets of
  parameters, but only indicate the span of the design range for each
  parameter. (Nominal vertical emittance assumes a 100\% emittance
  dilution budget from the damping ring to the IP.)}
\begin{center}
\begin{tabular}{| l | r | r | r | l |} \hline
             &  min  & nominal. & max. & unit \\ \hline & & & & \vbdlspacing \hline
Bunch population                            & 1       & 2      & 2    & $\times10^{10}$ \\ \hline
Number of bunches                           & 1260    & 2625   & 5340 &                 \\ \hline
Linac bunch interval                        & 180     & 369    & 500  & ns              \\ \hline
RMS bunch length                            & 200     & 300    & 500  & $\mu$m          \\ \hline
Normalized horizontal emittance at IP       & 10      & 10     & 12   & mm$\cdot$mrad   \\ \hline
Normalized vertical emittance at IP         & 0.02    & 0.04   & 0.08 & mm$\cdot$mrad   \\ \hline
Horizontal beta function at IP              & 10      & 20     & 20   & mm              \\ \hline
Vertical beta function at IP                & 0.2     & 0.4    & 0.6  & mm              \\ \hline
RMS horizontal beam size at IP              & 474     & 640    & 640  & nm              \\ \hline
RMS vertical beam size at IP                & 3.5     & 5.7    & 9.9  & nm              \\ \hline
Vertical disruption parameter               & 14      & 19.4   & 26.1 &                 \\ \hline
Fractional RMS energy loss to beamstrahlung &   1.7   &  2.4  &   5.5 &  \% \\
\hline
\end{tabular}
\label{tab:OVparamrange}
\end{center} \vbbelow 
\end{table}

\subsection{Electron Source}

\subparagraph{Functional Requirements} ~\\
The ILC polarized electron source must:
\begin{itemize}
  \item generate the required bunch train of polarized electrons
       ($>80\%$ polarization); \itemspace 
  \item capture and accelerate the beam to 5 GeV; \itemspace 
  \item transport the beam to the electron damping ring with minimal
       beam loss, and perform an energy compression and spin rotation
       prior to injection. \itemspace 
\end{itemize}

\begin{figure}[htb] \vbabove  
\begin{center}
  \includegraphics[width=0.95\textwidth]{\picturefolder 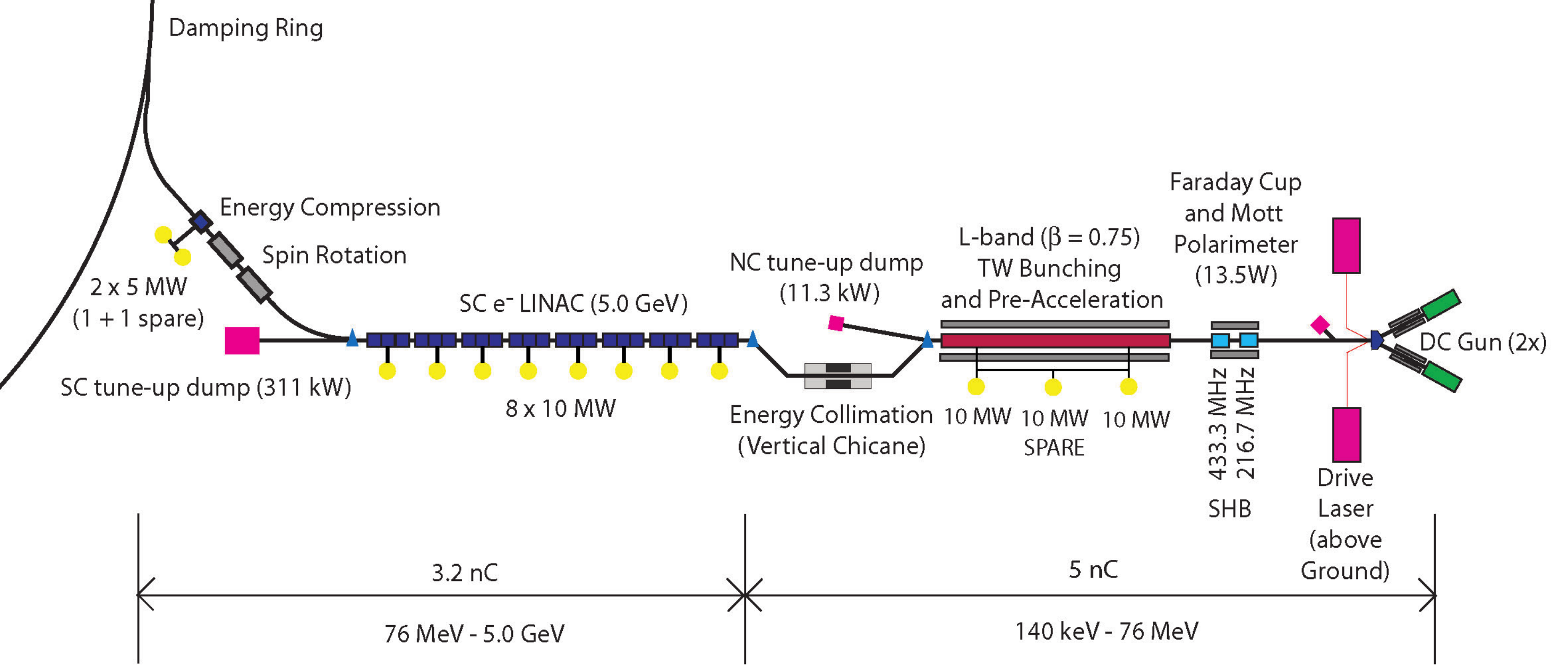}
 \vbabovecaption  \caption{Schematic View of the Polarized Electron Source.}
  \label{fig:OVPES}
\end{center} \vbbelow 
\end{figure}

\subparagraph{System Description} ~\\
The polarized electron source is located on the positron linac side of
the damping rings. The beam is produced by a laser illuminating a
photocathode in a DC gun. Two independent laser and gun systems
provide redundancy. Normal-conducting structures are used for bunching
and pre-acceleration to 76 MeV, after which the beam is accelerated to
5 GeV in a superconducting linac. Before injection into the damping
ring, superconducting solenoids rotate the spin vector into the
vertical, and a separate superconducting RF structure is used for
energy compression. The layout of the polarized electron source is
shown in Figure~\ref{fig:OVPES}.

\subparagraph{Challenges} ~\\
The SLC polarized electron source already meets the requirements for
polarization, charge and lifetime. The primary challenge for the ILC
electron source is the ~1 ms long bunch train, which demands a laser
system beyond that used at any existing accelerator.

\subsection{Positron Source}

\subparagraph{Functional requirements} ~\\
The positron source must perform several critical functions:

\begin{itemize}

  \item generate a high-power multi-MeV photon production drive beam
    in a suitably short-period, high K-value helical undulator; \itemspace 
  \item produce the needed positron bunches in a metal target that can
    reliably deal with the beam power and induced radioactivity; \itemspace 
  \item capture and accelerate the beam to 5 GeV \itemspace ;
  \item transport the beam to the positron damping ring with minimal
    beam loss, and perform an energy compression and spin rotation
    prior to injection. \itemspace 
\end{itemize}

\begin{figure}[htb] \vbabove 
\begin{center}
  \includegraphics[width=0.95\textwidth]{\picturefolder 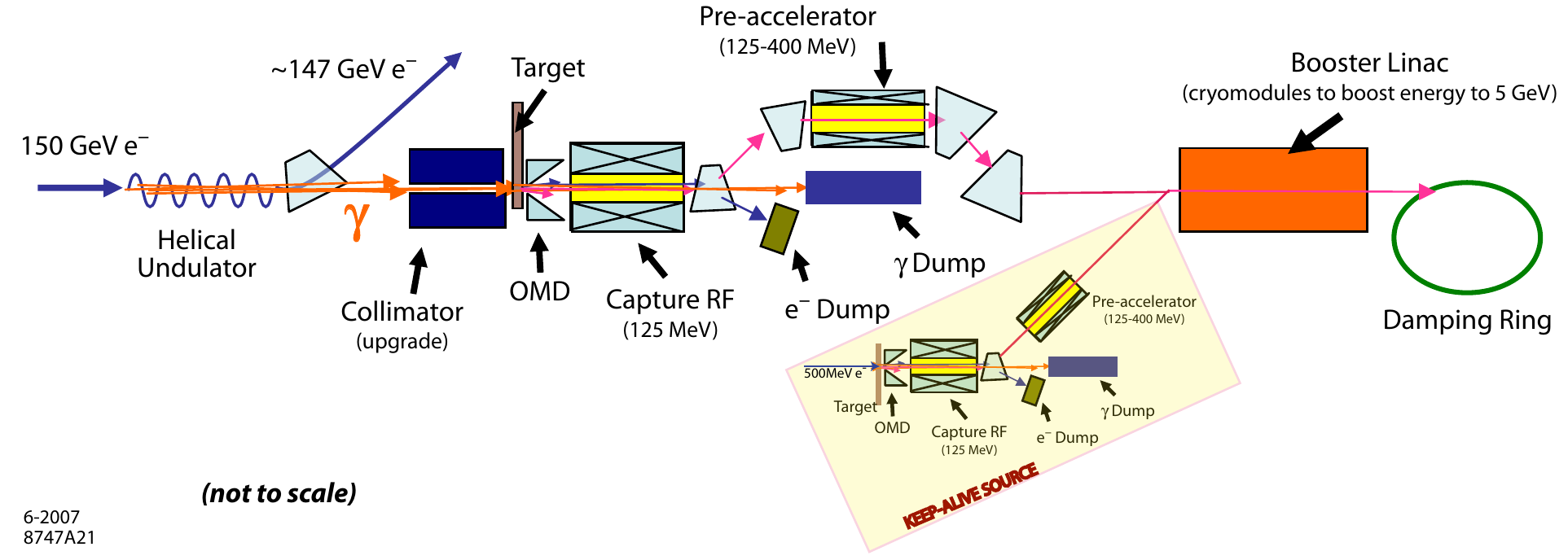}
 \vbabovecaption  \caption{Overall Layout of the Positron Source.}
  \label{fig:OVposi}
\end{center} \vbbelow 
\end{figure}

\subparagraph{System Description} ~\\
The major elements of the ILC positron source are shown in
Figure~\ref{fig:OVposi}. The source uses photoproduction to generate
positrons. After acceleration to 150 GeV, the electron beam is
diverted into an offset beamline, transported through a 150-meter
helical undulator, and returned to the electron linac. The high-energy
($\sim$10 MeV) photons from the undulator are directed onto a rotating
0.4 radiation-length Ti-alloy target $\sim$500 meters downstream,
producing a beam of electron and positron pairs. This beam is then
matched using an optical-matching device into a normal conducting (NC)
L-band RF and solenoidal-focusing capture system and accelerated to
125 MeV. The electrons and remaining photons are separated from the
positrons and dumped. The positrons are accelerated to 400 MeV in a NC
L-band linac with solenoidal focusing. The beam is transported ~5 km
through the rest of the electron main linac tunnel, brought to the
central injector complex, and accelerated to 5 GeV using
superconducting L-band RF. Before injection into the damping ring,
superconducting solenoids rotate the spin vector into the vertical,
and a separate superconducting RF structure is used for energy
compression.

The baseline design is for unpolarized positrons, although the beam
has a polarization of 30\%, and beamline space has been reserved for
an eventual upgrade to 60\% polarization.

To allow commissioning and tuning of the positron systems while the
high-energy electron beam is not available, a low-intensity auxiliary
(or ``keep-alive'') positron source is provided. This is a
conventional positron source, which uses a 500 MeV electron beam
impinging on a heavy-metal target to produce $\sim$10\% of the nominal
positron beam.  The keep-alive and primary sources use the same linac
to accelerate from 400 MeV to 5 GeV.

\subparagraph{Challenges} ~\\
The most challenging elements of the positron source are:

\begin{itemize}

  \item the 150 m long superconducting helical undulator, which has a
    period of 1.15 cm and a K-value of 0.92, and a 6 mm inner diameter
    vacuum chamber; \itemspace 
  \item the Ti-alloy target, which is a cylindrical wheel 1.4 cm thick
    and 1 m in diameter, which must rotate at 100 m/s in vacuum to
    limit damage by the photon beam; \itemspace 
  \item the normal-conducting RF system which captures the positron
    beam, which must sustain high accelerator gradients during
    millisecond-long pulses in a strong magnetic field, while
    providing adequate cooling in spite of high RF and particle-loss
    heating. \itemspace 

\end{itemize}

The target and capture sections are also high-radiation areas which
present remote handling challenges.

\subsection{Damping Rings}

\subparagraph{Functional requirements} ~\\
The damping rings must perform four critical functions:

\begin{itemize}

  \item accept $e^-$ and $e^+$ beams with large transverse and
    longitudinal emittances and damp to the low emittance beam
    required for luminosity production (by five orders of magnitude
    for the positron vertical emittance), within the 200 ms between
    machine pulses; \itemspace 
  \item inject and extract individual bunches without affecting the
    emittance or stability of the remaining stored bunches; \itemspace 
  \item damp incoming beam jitter (transverse and longitudinal) and
    provide highly stable beams for downstream systems; \itemspace 
  \item delay bunches from the source to allow feed-forward systems to
    compensate for pulse-to-pulse variations in parameters such as the
    bunch charge. \itemspace 

\end{itemize}

\subparagraph{System Description} ~\\
The ILC damping rings include one electron and one positron ring, each
6.7 km long, operating at a beam energy of 5 GeV. The two rings are
housed in a single tunnel near the center of the site, with one ring
positioned directly above the other. The plane of the DR tunnel is
located $\sim$10 m higher than that of the beam delivery system.  This
elevation difference gives adequate shielding to allow operation of
the injector system while other systems are open to human access.

The damping ring lattice is divided into six arcs and six straight
sections. The arcs are composed of TME cells; the straight sections
use a FODO lattice. Four of the straight sections contain the RF
systems and the superconducting wigglers. The remaining two sections
are used for beam injection and extraction. Except for the wigglers,
all of the magnets in the ring, are normal-conducting. Approximately
200 m of superferric wigglers are used in each damping ring. The
wigglers are 2.5 m long devices, operating at 4.5K, with a peak field
of 1.67 T.

The superconducting RF system is operated CW at 650 MHz, and provides
24 MV for each ring. The frequency is chosen to be half the linac RF frequency to easily accommodate different bunch patterns.  The single-cell cavities
operate at 4.5 K and are housed in eighteen 3.5 m long
cryomodules. Although a number of 500 MHz CW RF systems are currently
in operation, development work is required for this 650 MHz system,
both for cavities and power sources.

The momentum compaction of the lattice is relatively large, which
helps to maintain single bunch stability, but requires a relatively
high RF voltage to achieve the design RMS bunch length (9 mm). The
dynamic aperture of the lattice is sufficient to allow the large
emittance injected beam to be captured with minimal loss.

\subparagraph{Challenges} ~\\
The principal challenges in the damping ring are:
\begin{itemize}

  \item control of the electron cloud effect in the positron damping
    ring. This effect, which can cause instability, tune spread, and
    emittance growth, has been seen in a number of other rings and is
    relatively well understood. Simulations indicate that it can be
    controlled by proper surface treatment of the vacuum chamber to
    suppress secondary emission, and by the use of solenoids and
    clearing electrodes to suppress the buildup of the cloud. \itemspace 
  \item control of the fast ion instability in the electron damping
    ring. This effect can be controlled by limiting the pressure in
    the electron damping ring to below 1 nTorr, and by the use of
    short gaps in the ring fill pattern. \itemspace 
  \item development of a very fast rise and fall time kicker for single
    bunch injection and extraction in the ring. For the most demanding
    region of the beam parameter range, the bunch spacing in the
    damping ring is $\sim$3 ns, and the kicker must have a rise plus fall
    time no more than twice this. Short stripline kicker structures
    can achieve this, but the drive pulser technology still needs 
    development. \itemspace 

\end{itemize}

\subsection{Ring to Main Linac (RTML)}
\subparagraph{Functional requirements} ~\\
The RTML must perform several critical functions for each beam:

\begin{itemize}

  \item transport the beam from the damping ring to the upstream end
    of the linac; \itemspace 
  \item collimate the beam halo generated in the damping ring; \itemspace 
  \item rotate the polarization from the vertical to any arbitrary
    angle required at the IP; \itemspace 
  \item compress the long Damping Ring bunch length by a factor of
    $30\sim45$ to provide the short bunches required by the Main Linac
    and the IP; \itemspace 

\end{itemize}

\begin{figure}[htb] \vbabove 
\begin{center}
  \includegraphics[width=0.95\textwidth]{\picturefolder 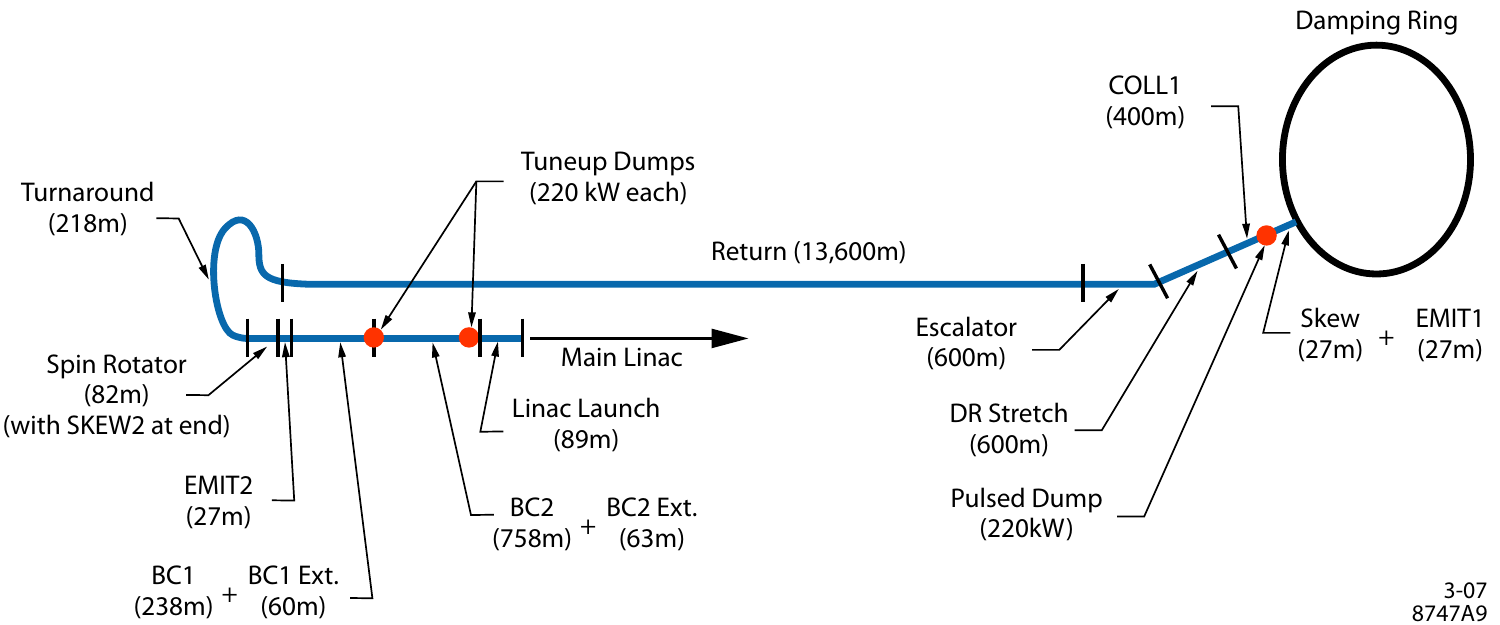}
 \vbabovecaption \caption{Schematic of the RTML.}
  \label{fig:OVRTML}
\end{center} \vbbelow 
\end{figure}

\subparagraph{System Description} ~\\
The layout of the RTML is identical for both electrons and positrons,
and is shown in Figure~\ref{fig:OVRTML}. The RTML consists of the
following subsystems:

\begin{itemize}

  \item an $\sim$15 km long 5 GeV transport line; \itemspace 
  \item betatron and energy collimation systems; \itemspace 
  \item a $180^{\circ}$ turn-around, which enables feed-forward beam
    stabilization; \itemspace 
  \item spin rotators to orient the beam polarization to the desired
    direction; \itemspace 
  \item a 2-stage bunch compressor to compress the beam bunch length
    from several millimeters to a few hundred microns as required at
    the IP. \itemspace 

\end{itemize}

The bunch compressor includes acceleration from 5 GeV to 13-15 GeV in
order to limit the increase in fractional energy spread associated
with bunch compression.

\subparagraph{Challenges} ~\\
The principal challenges in the RTML are:
\begin{itemize}

  \item control of emittance growth due to static misalignments,
    resulting in dispersion and coupling. Simulations indicate that
    the baseline design for beam-based alignment can limit the
    emittance growth to tolerable levels. \itemspace 
  \item suppression of phase and amplitude jitter in the bunch
    compressor RF, which can lead to timing errors at the IP. RMS
    phase jitter of 0.24$^{\circ}$ between the electron and positron
    RF systems results in a 2\% loss of luminosity. Feedback loops in
    the bunch compressor low-level RF system should be able to limit
    the phase jitter to this level. \itemspace 

\end{itemize}


\subsection{Main Linacs}

\subparagraph{Functional requirements} ~\\
The two main linacs accelerate the electron and positron beams from
their injected energy of 15 GeV to the final beam energy of 250 GeV,
over a combined length of 23 km. The main linacs must:

\begin{itemize}

  \item accelerate the beam while preserving the small bunch
    emittances, which requires precise orbit control based on data
    from high resolution beam position monitors, and also requires
    control of higher-order modes in the accelerating cavities; \itemspace 
  \item maintain the beam energy spread within the design requirement
    of $\sim$0.1~\% at the IP; \itemspace 
  \item not introduce significant transverse or longitudinal jitter,
    which could cause the beams to miss at the collision point. \itemspace 

\end{itemize}

\begin{figure}[htb] \vbabove 
\begin{center}
  \includegraphics[width=0.95\textwidth]{\picturefolder 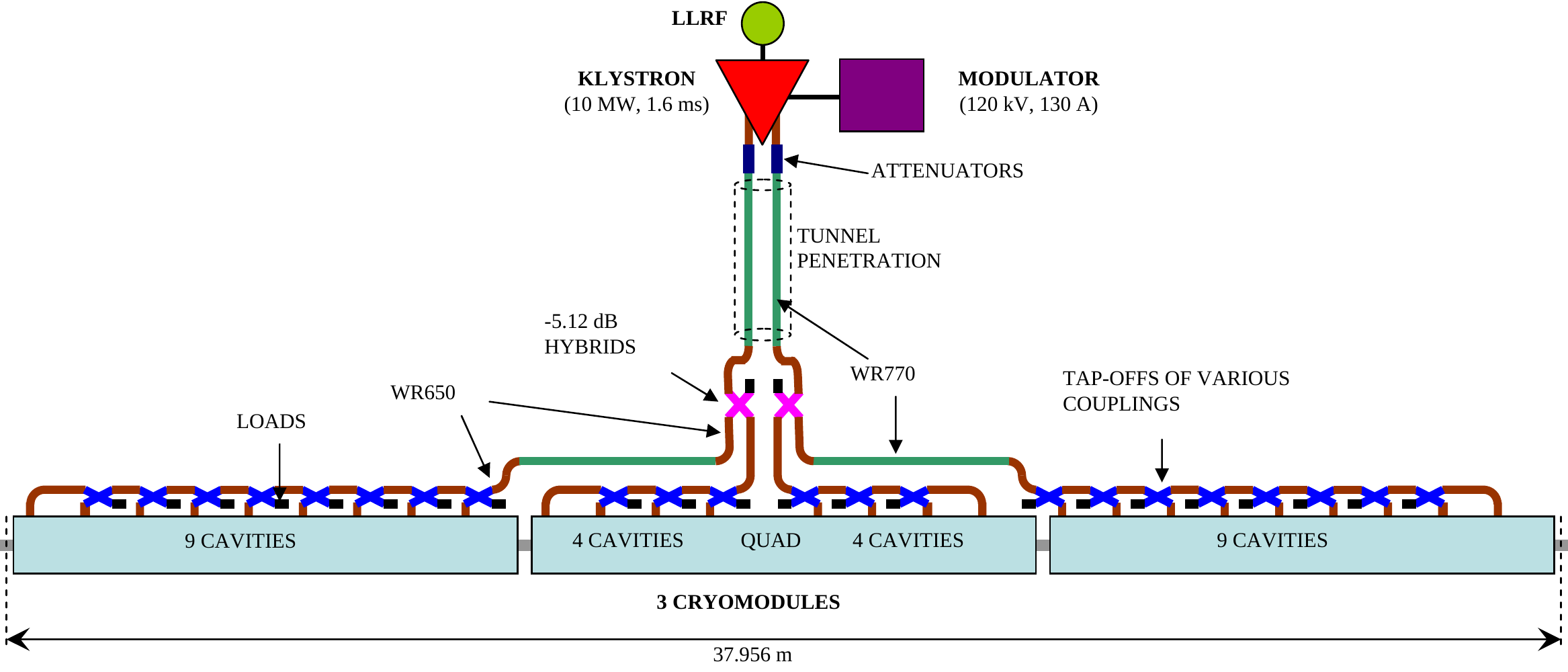}
 \vbabovecaption  \caption{RF unit layout.}
  \label{fig:OVRFunit}
\end{center} \vbbelow 
\end{figure}

\subparagraph{System description} ~\\
The ILC Main Linacs accelerate the beam from 15 GeV to a maximum
energy of 250 GeV at an average accelerating gradient of 31.5
MV/m. The linacs are composed of RF units, each of which are formed by
three contiguous SCRF cryomodules containing 26 nine-cell
cavities. The layout of one unit is illustrated in
Figure~\ref{fig:OVRFunit}. The positron linac contains 278 RF units,
and the electron linac has 282 RF units\footnote{Approximately 3 GeV of
  extra energy is required in the electron linac to compensate for
  positron production.}.

Each RF unit has a stand-alone RF source, which includes a
conventional pulse-transformer type high-voltage (120 kV) modulator, a
10 MW multi-beam klystron, and a waveguide system that distributes the
RF power to the cavities (see Figure~\ref{fig:OVRFunit}). It also
includes the low-level RF (LLRF) system to regulate the cavity field
levels, interlock systems to protect the source components, and the
power supplies and support electronics associated with the operation
of the source.

The cryomodule design is a modification of the Type-3 version
(Figure~\ref{fig:OVcryomodule}) developed and used at DESY. Within the
cryomodules, a 300 mm diameter helium gas return pipe serves as a
strongback to support the cavities and other beam line components.
The middle cryomodule in each RF unit contains a quad package that
includes a superconducting quadrupole magnet at the center, a cavity
BPM, and superconducting horizontal and vertical corrector
magnets. The quadrupoles establish the main linac magnetic lattice,
which is a weak focusing FODO optics with an average beta function of
$\sim$80 m. All cryomodules are 12.652 m long, so the active-length to
actual-length ratio in a nine-cavity cryomodule is 73.8\%. Every
cryomodule also contains a 300 mm long high-order mode beam absorber
assembly that removes energy through the 40-80 K cooling system from
beam-induced higher-order modes above the cavity cutoff frequency.

To operate the cavities at 2 K, they are immersed in a saturated He II
bath, and helium gas-cooled shields intercept thermal radiation and
thermal conduction at 5-8 K and at 40-80 K. The estimated static and
dynamic cryogenic heat loads per RF unit at 2 K are 5.1 W and 29 W,
respectively. Liquid helium for the main linacs and the RTML is
supplied from 10 large cryogenic plants, each of which has an
installed equivalent cooling power of $\sim$20 kW at 4.5 K. The main
linacs follow the average Earth's curvature to simplify the liquid
helium transport.

\begin{figure}[htb] \vbabove 
\begin{center}
  \includegraphics[width=0.95\textwidth]{\picturefolder 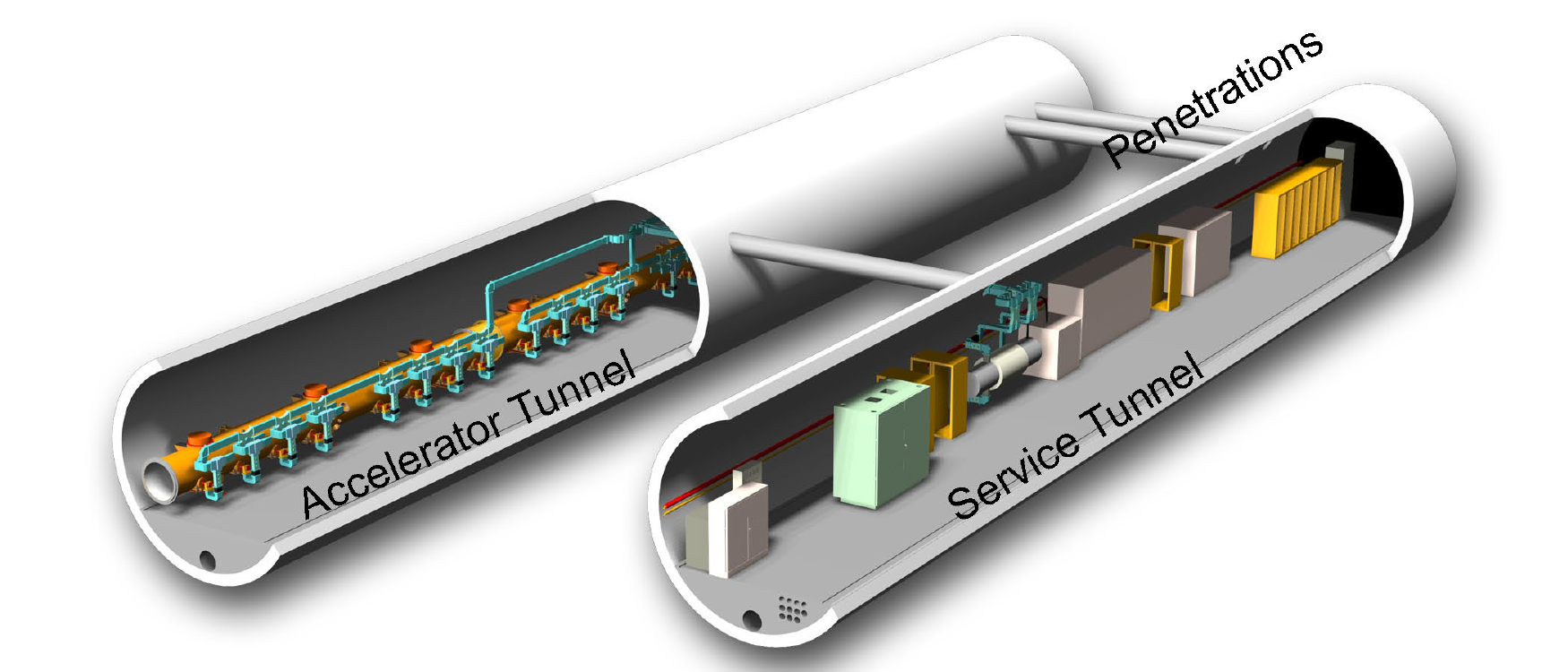}
  \vbabovecaption \caption{Cutaway view of the linac dual-tunnel configuration.}
  \label{fig:OVtunnel}
\end{center} \vbbelow 
\end{figure}

The Main Linac components are housed in two tunnels, an accelerator
tunnel and a service tunnel, each of which has an interior diameter of
4.5 meters. To facilitate maintenance and limit radiation exposure,
the RF source is housed mainly in the service tunnel as illustrated in
Figure~\ref{fig:OVtunnel}.

The tunnels are typically hundreds of meters underground and are
connected to the surface through vertical shafts\footnote{Except for
  the Asian sample site: see Section~\ref{sect:ESss}.}. Each of the main linacs
includes three shafts, roughly 5 km apart as dictated by the cryogenic
system. The upstream shafts in each linac have diameters of 14 m to
accommodate lowering cryomodules horizontally, and the downstream
shaft in each linac is 9 m in diameter, which is the minimum size
required to accommodate tunnel boring machines. At the base of each
shaft is a 14,100 cubic meter cavern for staging installation; it also houses utilities and parts of the cryoplant, most of which are
located on the surface.

\subparagraph{Challenges} ~\\
The principal challenges in the main linac are:
\begin{itemize}

  \item achieving the design average accelerating gradient of 31.5
    MV/m. This operating gradient is higher than that typically
    achievable today and assumes further progress will be made during
    the next few years in the aggressive program that is being pursued
    to improve cavity performance. \itemspace 
  \item control of emittance growth due to static misalignments,
    resulting in dispersion and coupling. Beam-based alignment
    techniques should be able to limit the single-bunch emittance
    growth. Long-range multibunch effects are mitigated via HOM
    damping ports on the cavities, HOM absorbers at the quadrupoles,
    and HOM detuning. Coupling from mode-rotation HOMs is limited by
    splitting the horizontal and vertical betatron tunes. \itemspace 
  \item control of the beam energy spread. The LLRF system monitors
    the vector sum of the fields in the 26 cavities of each RF unit
    and makes adjustments to flatten the energy gain along the bunch
    train and maintain the beam-to-RF phase constant. Experience from
    FLASH and simulations indicate that the baseline system should
    perform to specifications. \itemspace 

\end{itemize}


\subsection{Beam Delivery System}

\subparagraph{Functional requirements} ~\\
The ILC Beam Delivery System (BDS) is responsible for transporting the
$e^+e^-$ beams from the exit of the high energy linacs, focusing them to
the sizes required to meet the ILC luminosity goals, bringing them
into collision, and then transporting the spent beams to the main beam
dumps. In addition, the BDS must perform several other critical
functions:

\begin{itemize}

  \item measure the linac beam and match it into the final focus; \itemspace 
  \item protect the beamline and detector against mis-steered beams
    from the main linacs; \itemspace 
  \item remove any large amplitude particles (beam-halo) from the
    linac to minimize background in the detectors; \itemspace 
  \item measure and monitor the key physics parameters such as energy
    and polarization before and after the collisions. \itemspace 

\end{itemize}

\begin{figure}[htb] \vbabove 
\begin{center}
  \includegraphics[width=\textwidth]{\picturefolder 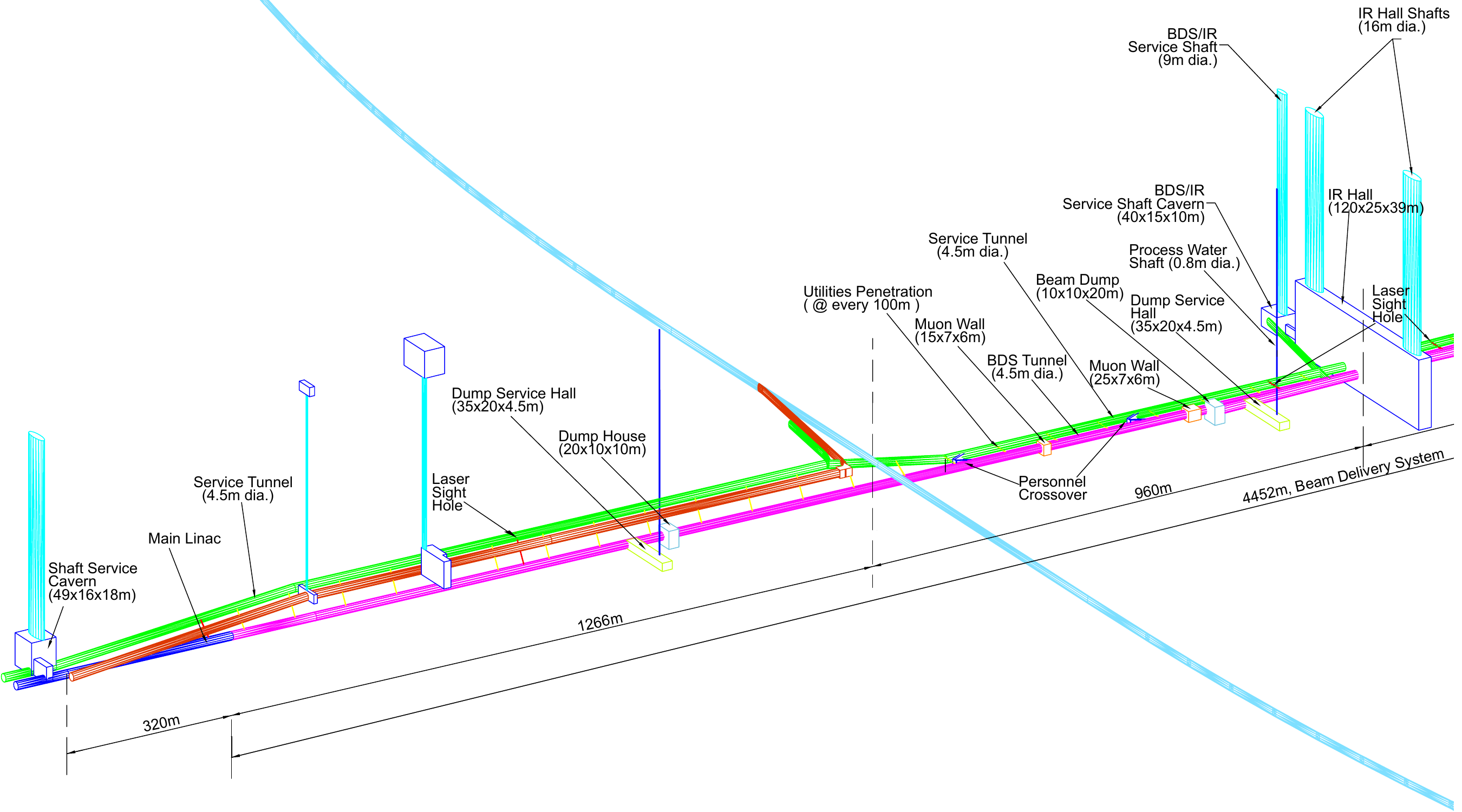}
  \vbabovecaption \caption[BDS layout, beam and service tunnels.] {BDS layout, 	beam and service tunnels (shown in magenta and
    green), shafts, experimental hall.  The line crossing the BDS
    beamline at right angles is the damping ring, located 10 m above
    the BDS tunnels.}
  \label{fig:OVBDS}
\end{center} \vbbelow 
\end{figure}

\subparagraph{System Description} ~\\
The layout of the beam delivery system is shown in
Figure~\ref{fig:OVBDS}. There is a single collision point with a 14
mrad total crossing angle. The 14 mrad geometry provides space for
separate extraction lines but requires crab cavities to rotate the
bunches in the horizontal plane for effective head-on
collisions. There are two detectors in a common interaction region
(IR) hall in a so-called ``push-pull'' configuration. The detectors
are pre-assembled on the surface and then lowered into the IR hall
when the hall is ready for occupancy.

The BDS is designed for 500 GeV center-of-mass energy but can be
upgraded to 1 TeV with additional magnets.

The main subsystems of the beam delivery, starting from the exit of
the main linacs, are:

\begin{itemize}

  \item a section containing post-linac emittance measurement and
    matching (correction) sections, trajectory feedback, polarimetry
    and energy diagnostics; \itemspace 
  \item a fast pulsed extraction system used to extract beams in case
    of a fault, or to dump the beam when not needed at the IP; \itemspace 
  \item a collimation section which removes beam halo particles that
    would otherwise generate unacceptable background in the detector,
    and also contains magnetized iron shielding to deflect muons; \itemspace 
  \item the final focus (FF) which uses strong compact superconducting
    quadrupoles to focus the beam at the IP, with sextupoles providing
    local chromaticity correction; \itemspace 
  \item the interaction region, containing the experimental
    detectors. The final focus quadrupoles closest to the IP are
    integrated into the detector to facilitate detector ``push-pull''; \itemspace 
   \item the extraction line, which has a large enough bandwidth to
     cleanly transport the heavily disrupted beam to a high-powered
     water-cooled dump. The extraction line also contains important
     polarization and energy diagnostics. \itemspace 

\end{itemize}

\subparagraph{Challenges} ~\\
The principal challenges in the beam delivery system are:
\begin{itemize}

  \item tight tolerances on magnet motion (down to tens of
    nanometers), which make the use of fast beam-based feedback
    systems mandatory, and may well require mechanical stabilization
    of critical components (e.g. final doublets). \itemspace 
  \item uncorrelated relative phase jitter between the crab cavity
    systems, which must be limited to the level of tens of
    femtoseconds. \itemspace 
  \item control of emittance growth due to static misalignments, which
    requires beam-based alignment and tuning techniques similar to the
    RTML. \itemspace 
  \item control of backgrounds at the IP via careful tuning and
    optimization of the collimation systems and the use of the
    tail-folding octupoles. \itemspace 
  \item clean extraction of the high-powered disrupted beam to the
    dump. Simulations indicate that the current design is adequate
    over the full range of beam parameters. \itemspace 

\end{itemize}


\setcounter{secnumdepth}{4}
\setcounter{subsection}{0} 
\setcounter{subsubsection}{0}
\usecounter{subsubsection}
\section{Sample Sites}\label{sect:ESss}

Conventional Facilities and Siting (CFS) is responsible for civil
engineering, power distribution, water cooling and air conditioning
systems. The value estimate (see Section~\ref{sec:ValueEstimate})
for the CFS is approximately 38\% of the total estimated
project value.

In the absence of a single agreed-upon location for the ILC, a sample
site in each region was developed. Each site was designed to support
the baseline design described in Section~\ref{Ovr:baseline}. Although many of the basic requirements are identical, differences in geology, topography
and local standards and regulations lead to different construction
approaches, resulting in a slight variance in value estimates across
the three regions. Although many aspects of the CFS (and indeed
machine design) will ultimately depend on the specific host site
chosen, the approach taken here is considered sufficient for the
current design phase, while giving a good indication of the influence
of site-specific issues on the project as a whole.

Early in the RDR process, the regional CFS groups agreed upon a matrix of criteria for any sample site. All three sites satisfied these criteria, including the mandatory requirement that the site can support the extension to the 1 TeV
center-of-mass machine.

\begin{figure}[htb] \vbabove 
\begin{center}
  \includegraphics[width=0.95\textwidth]{\picturefolder 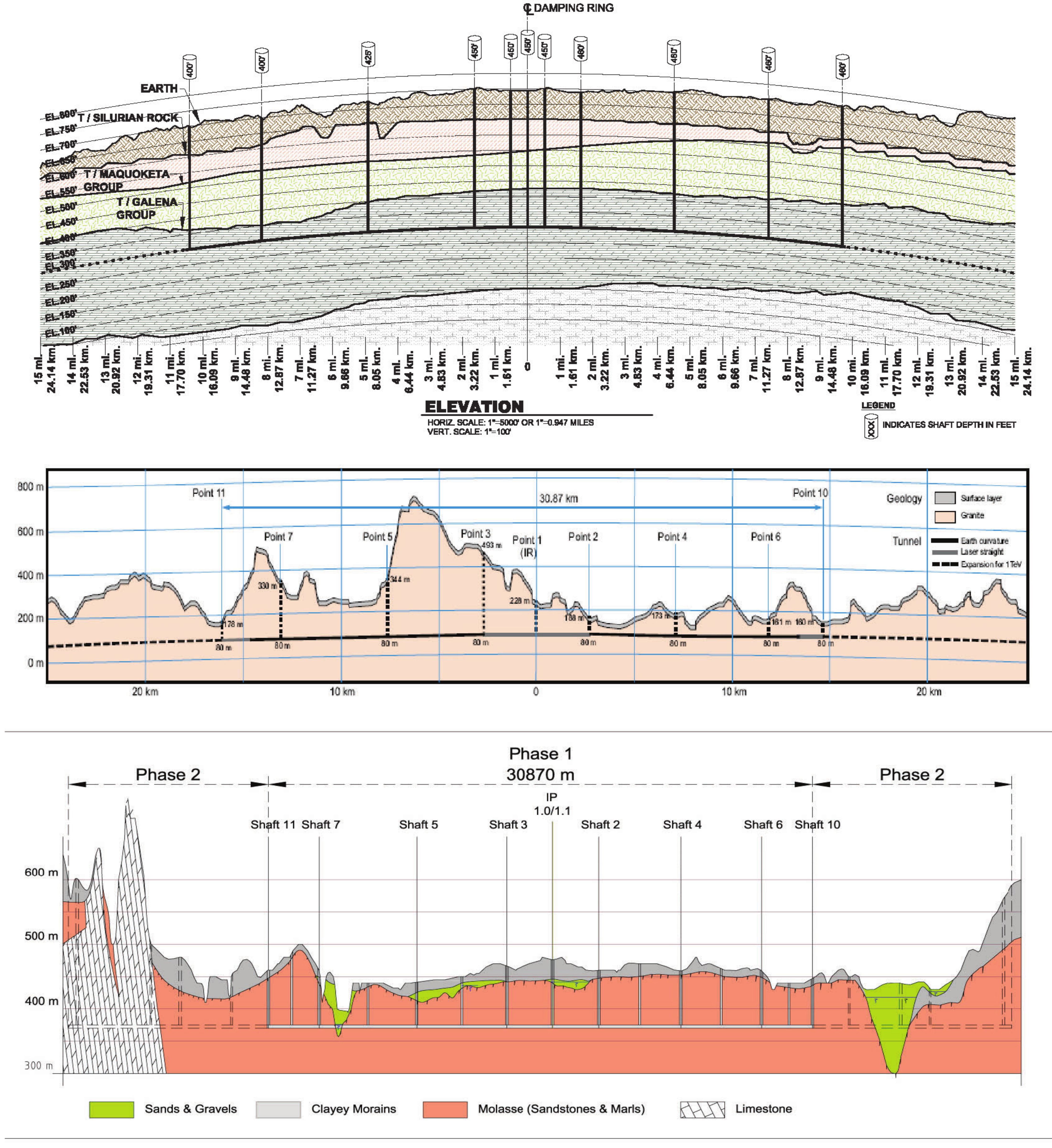}
 \vbabovecaption  \caption[Geology and tunnel profiles for the three regional sites.] {Geology and tunnel profiles for the three regional sites,
    showing the location of the major access shafts (tunnels for the
    Asian site). Top: the Americas site close to Fermilab. Middle: the
    Asian site in Japan. Bottom: the European site close to CERN.}
  \label{fig:OVsitesec}
\end{center} \vbbelow 
\end{figure}

The three sample sites have the following characteristics:

\begin{itemize}

  \item The Americas sample site lies in Northern Illinois near Fermilab.
    The site provides a range of locations to
    position the ILC in a north-south orientation. The site chosen has
    approximately one-quarter of the machine on the Fermilab site. The
    surface is primarily flat. The long tunnels are bored in a
    contiguous dolomite rock strata (¡ÆGalena Platteville¡Ç), at a
    typical depth of 30-100 m below the surface. \itemspace 
  \item The Asian site has been chosen from several possible ILC
    candidate sites in Japan. The sample site has a uniform terrain
    located along a mountain range, with a tunnel depth ranging from
    40 m to 600 m. The chosen geology is uniform granite highly suited
    to modern tunneling methods. One specific difference for the Asian
    site is the use of long sloping access tunnels instead of vertical
    shafts, the exception being the experimental hall at the
    Interaction Region, which is accessed via two 112 m deep vertical
    shafts. The sloping access tunnels take advantage of the
    mountainous location. \itemspace 
  \item The European site is located at CERN, Geneva, Switzerland, and
    runs parallel to the Jura mountain range, close to the CERN
    site. The majority of the machine is located in the `Molasse' (a
    local impermeable sedimentary rock), at a typical depth of 370 m. \itemspace 

\end{itemize}

The elevations of the three sample sites are shown in
Figure~\ref{fig:OVsitesec}. The tunnels for all three sites would be
predominantly constructed using Tunnel Boring Machines (TBM), at
typical rates of 20--30 m per day. The Molasse of the European site
near CERN requires a reinforced concrete lining for the entire tunnel
length. The Asian site (granite) requires rock bolts and a 5 cm
`shotcrete' lining. The US site is expected to require a concrete
lining for only approximately 20\% of its length, with rock-bolts
being sufficient for permanent structural support.

A second European sample site near DESY, Hamburg, Germany, has also
been developed. This site is significantly different from the three
reported sites, both in geology and depth (25~m deep), and requires
further study.

In addition, the Joint Institute for Nuclear Research has submitted a
proposal to site the ILC in the neighborhood of Dubna, Russian
Federation.

The three sites reported in detail here are all `deep-tunnel'
solutions. The DESY and Dubna sites are examples of `shallow' sites. A
more complete study of shallow sites -- shallow tunnel or
cut-and-cover -- will be made in the future as part of the Engineering
Design phase.

\clearpage
\setcounter{figure}{0}
\setcounter{table}{0}
\setcounter{secnumdepth}{4}
\setcounter{subsection}{0} 
\setcounter{subsubsection}{0}
\usecounter{subsubsection}
\chapter{Detectors}
The challenge for the ILC detectors is to optimize the scientific
results from a broad experimental program aimed at understanding the
mechanism of mass generation and electroweak symmetry breaking. This
includes the search for supersymmetric particles, and their detailed
study if they are found, and the hunt for signs of extra space-time
dimensions and quantum gravity. Precision measurements of Standard
Model processes can reveal new physics at energy scales beyond
direct reach. The detectors must also be prepared for the
unexpected.

Experimental conditions at the ILC provide an ideal environment for
the precision study of particle production and decay, and offer the
unparalleled cleanliness and well-defined initial conditions
conducive to recognizing new phenomena. Events are recorded without
trigger bias, with detectors designed for optimal physics
performance. The physics poses challenges, pushing the limits of jet
energy resolution, tracker momentum resolution, and vertex impact
parameter resolution. Multi-jet final states and supersymmetry
(SUSY) searches put a premium on hermeticity and full solid angle
coverage. Although benign by LHC standards, the ILC environment
poses challenges of its own.

The World Wide Study of Physics and Detectors for Future Linear
Colliders has wrestled with these challenges for more than a decade,
advancing the technologies needed for ILC detectors. Different
concepts for detectors have evolved\cite{tdr,Detectors:behnke}, as the
rapid collider progress has spurred the experimental community. Four
concept 
reports\cite{Detectors:sid,Detectors:ldc,Detectors:gld,Detectors:fourth} were presented in Spring,
2006. All of these detectors meet the ILC physics demands, and
can be built with technologies that are within reach today. There is
a growing community involved in refining and optimizing these
designs, and advancing the technologies. Full detector engineering
designs and proof of principle technology demonstrations can be
completed on the timetable proposed for the ILC Engineering Design
Report as long as there is adequate support for detector R\&D and
integrated detector studies.

\section{Challenges for Detector Design and Technology}

The relatively low radiation environment of the ILC allows detector
designs and technologies not possible at the LHC, but the demanding
physics goals still challenge the state of the art, particularly in
readout and sensor technologies.

Many interesting ILC physics processes appear in multi-jet final
states, often accompanied by charged leptons or missing energy.
Precision mass measurements require a jet energy resolution of
$\frac{\sigma_{E_{jet}}}{E_{jet}} = \frac{30\%}{\sqrt{E_{jet}}}$ for $E_{jet}$ 
up to approximately 100 GeV, and
$\frac{\sigma_{E_{jet}}}{E_{jet}}\le3\%$ beyond, more
than a factor of 2 better than achieved at LEP/SLC.

Detailed studies of leptons from W and Z decays require efficient
electron and muon ID and accurate momentum measurements over the
largest possible solid angle. Excellent identification of electrons
and muons within jets is critical because they indicate the presence
of neutrinos from heavy quark decays, and tag the jet flavor and
quark charge.

The jet mass resolution appears achievable if the detector has an
excellent, highly efficient, nearly hermetic tracking system and a
finely segmented calorimeter. Charged tracks reconstructed in the
tracker can be isolated in the calorimeter, and their contributions
removed from the calorimeter energy measurement. This ``particle
flow'' concept has motivated the development of high granularity
calorimeters, and highly efficient tracking systems. The main
challenge is the separation of neutral and charged contributions
within a dense jet environment. 

It is possible to satisfy the calorimeter granularity required for
the particle flow concept with electromagnetic cell sizes of about
$1\times1$ cm$^2$, and comparable or somewhat larger hadronic cells.
An electromagnetic energy resolution of $\sim15\%/\sqrt{E}$ and a
hadronic resolution of $\sim40\%/\sqrt{E}$ is sufficient.  

The momentum resolution required to satisfy the demands of particle
flow calorimetry and missing energy measurements is particularly
challenging and exceeds the current state of the art. Good momentum
resolution from the beam energy down to very low momentum is needed
over the full solid angle. Pattern recognition must be robust and
highly efficient even in the presence of backgrounds. This requires
minimal material to preserve lepton ID and permit high performance
calorimetry.

\begin{figure}[htb]
   \begin{center} 
\begin{tabular}{cc}
\includegraphics[width=0.483\textwidth]{\picturefolder 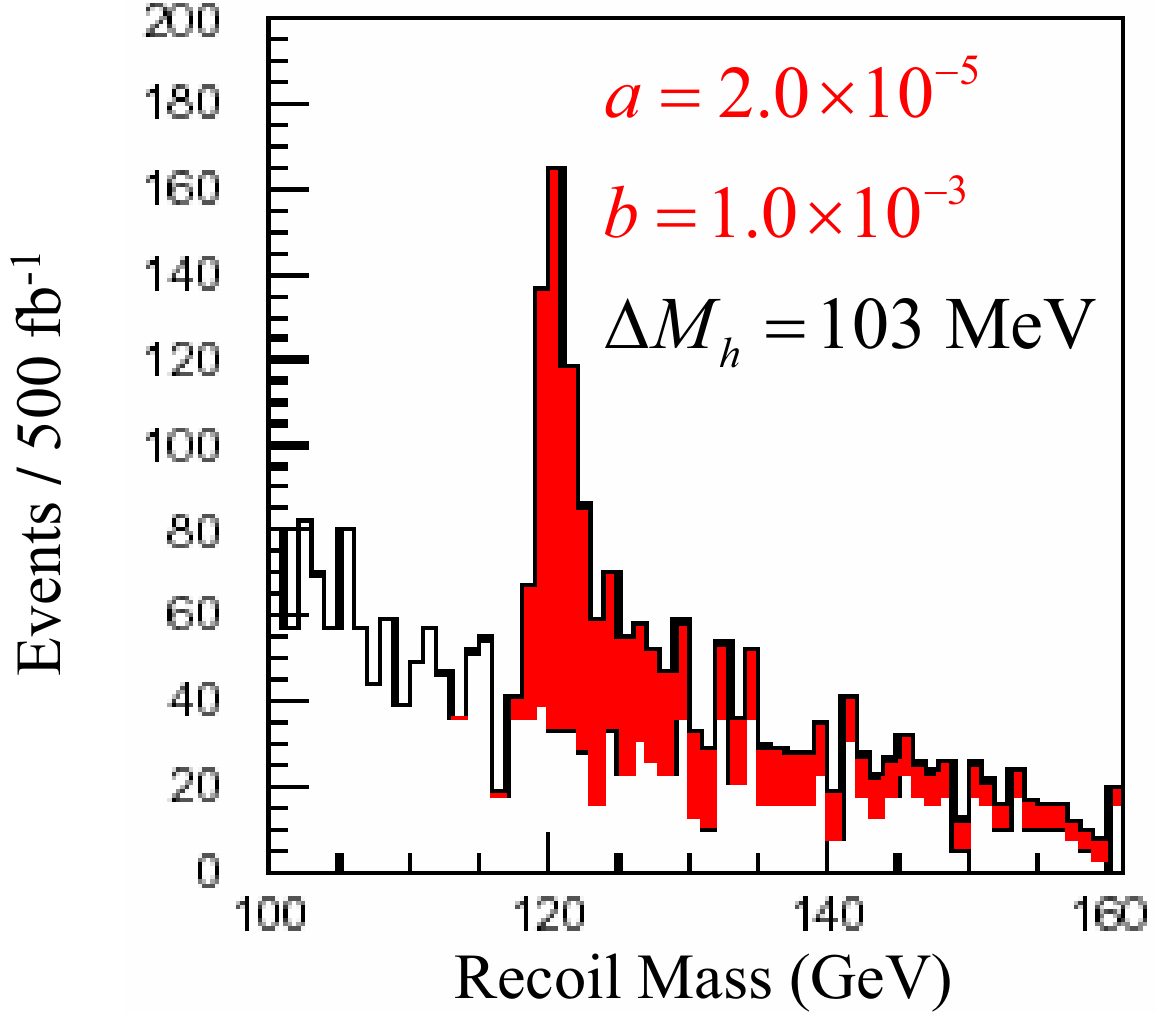} &
\includegraphics[width=0.44\textwidth]{\picturefolder 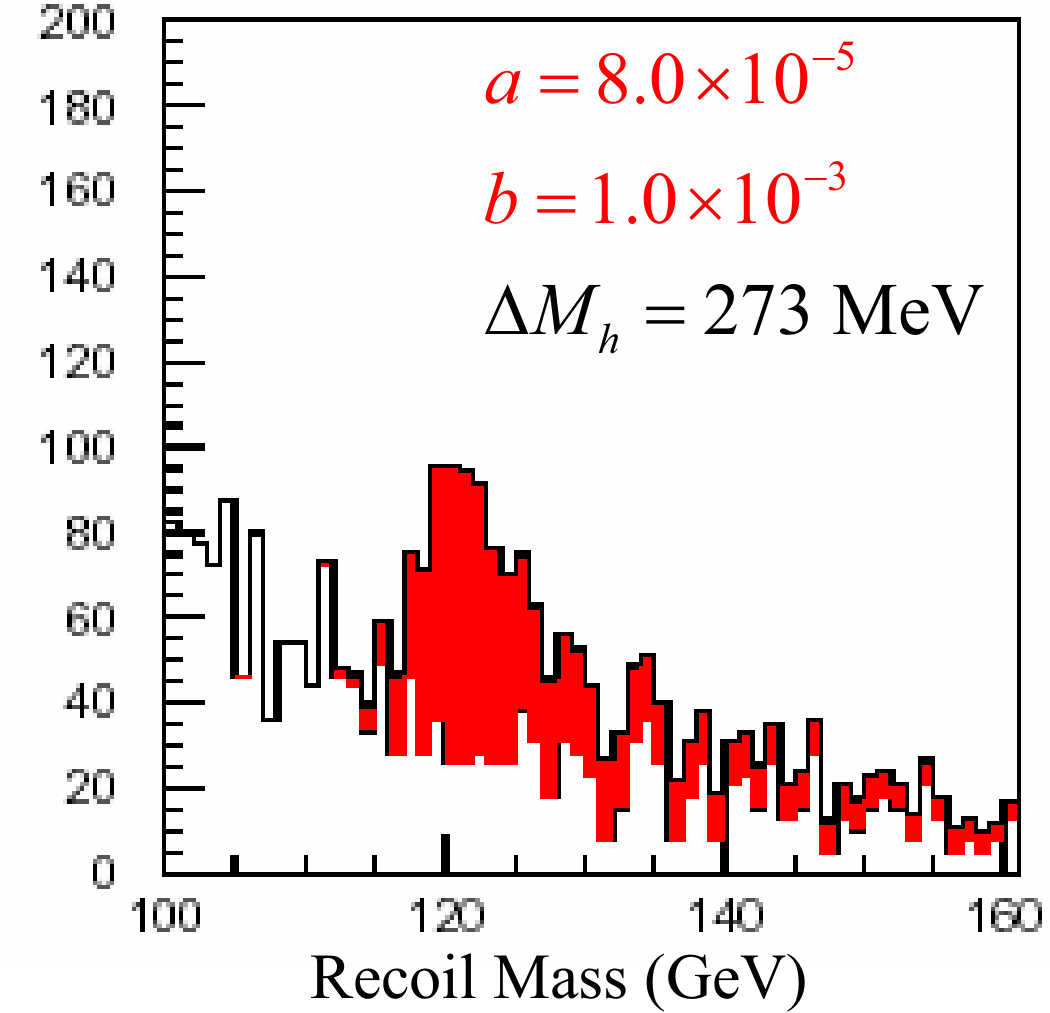} 
\end{tabular}
      \caption[Higgs recoil mass spectra]{Higgs recoil mass spectra for tracker 	momentum resolution,%
    $\frac{\delta p_{t}}{p_{t}^{2}} = a \oplus\frac{b}{p_{t}\sin\theta},$ for
    120 GeV Higgs mass, $\sqrt{s}$ = 350 GeV, and 500 fb$^{-1}$.}
      \label{fig:recoil}
   \end{center} \vbbelow
\end{figure}

``Higgs-strahlung'' production in association with a Z is a
particularly powerful physics channel. It allows precision Higgs
mass determination, precision studies of the Higgs branching
fractions, measurement of the production cross section and
accompanying tests of SM couplings, and searches for invisible Higgs
decays.  The resolution of the recoil mass from a Z decaying to
leptons depends on beam energy accuracy, beam energy spread and
tracking precision.  Figure~\ref{fig:recoil} shows an example of the
recoil mass distribution\cite{Detectors:schreiber} opposite the Z.
The tracker is also critical to mass determination of kinematically
accessible sleptons and neutralinos, and accurate measurements of
the center of mass energy.

Vertex detection identifies heavy particle decay vertices, enabling
flavor and charge tagging. Multilayer vertex detection also provides
efficient stand-alone pattern recognition, momentum measurement for
soft tracks, and seeds for tracks in outer trackers. The ILC physics
goals push vertex detector efficiency, angular coverage, and impact
parameter resolution beyond the current state of the art, even
surpassing the SLD CCD vertex detector\cite{Detectors:abe}.  The ILC
beamstrahlung e$^+$e$^-$ pairs present a background of up to 100
hits/mm$^2$/train for the innermost detector elements.
It is essential to reduce the number of
background hits, either by time-slicing the bunch train into pieces
of less than 150 bunch crossings, or by discriminating charged tracks from
background. The simultaneous challenges of rapid readout,
constrained power budget, transparency and high resolution are being
actively addressed by several efforts. The ILCs low data rates and
low radiation loads allow consideration of new technologies that
reach beyond LHC capabilities.

The very forward region of the ILC detector is instrumented with a
calorimeter (BeamCal) that extends calorimeter hermeticity to small
angles.  To search for new particles, this instrument must veto
electrons in a high radiation and high background environment.
Measurement of the energy deposited by beamstrahlung pairs and
photons in the BeamCal and associated photon calorimeter (GamCal)
provides a bunch-by-bunch luminosity measurement that can be used
for intra-train luminosity optimization. Beam parameters can also be
determined from the shapes of the observed energy depositions given
sufficiently fast readout electronics and adequate high bandwidth
resolution. Near the beampipe the absorbed radiation dose is up to
10 MGy per year.

Polarimetry and beam energy spectrometry must be able to achieve
very low systematic errors, with beam energy measured to 200 ppm,
and polarization to 0.1\%. High-field superconducting solenoid
designs must be refined, with development of new conductors. The
solenoid design must also accommodate dipole and solenoid
compensation, have high field uniformity, and support push-pull.
Muon detectors must be developed.

Detector system integration depends on engineering and design
work in several areas.
Stable, adjustable, vibration free support of the final quadrupoles
is needed. Support of the fragile beampipe with its massive masking
is also a concern. The detectors are required to move on and off
beamline quickly and reproducibly (``push-pull'').  The detectors
must be calibrated, aligned, and accessed, without compromising
performance.

Research and development on all of these detector issues must be
expanded in order to achieve the needed advances.

\section{Detector Concepts}

Four detector concepts are being studied as candidate detectors for
the ILC experimental program.  These represent complementary
approaches and technology choices.  Each concept is designed with an
inner vertex detector, a tracking system based on either a gaseous Time
Projection Chamber or silicon detectors, 
a calorimeter to reconstruct jets, a muon system, and a
forward system of tracking and calorimetry. Table~\ref{Detectors:concepts} 
presents some of the key parameters of each
of the four detector concepts.  GLD, LDC and SiD employ particle
flow for jet energy measurements.  SiD has the strongest magnetic
field and the smallest radius, while LDC and GLD rely on smaller
fields with larger tracking radii. Each approach uses different
emphasis to address the optimization.  The 4$^{\rm th}$ concept employs a 
dual-readout fiber calorimeter and a novel outer muon system.

\begin{table}[hbt]
\begin{center}
\caption{Some key parameters of the four detector concepts.
\label{Detectors:concepts}}
\begin{tabular}{|l|c|c|c|c|c|c|}\hline%
Concept & Tracking & Solenoidal & Solenoid & Vertex & ECAL & Overall 
\vspace{-6pt} \\%
  & Technology & Field & Radius, & Inner & Barrel & Detector 
\vspace{-6pt} \\%
  & & Strength & Length & Radius & Inner & Outer \vspace{-6pt} \\%
  & & (Tesla) & (m) & (mm) & Radius, & Radius, \vspace{-6pt} \\%
  & & & & & Half- & Half-  \vspace{-6pt}\\
  & & & & & Length & Length \vspace{-6pt}\\
  & & & & & (m) & (m) \\ \hline  & & & & \vbdlspacing \hline%
GLD & TPC/Si & 3 & 4 & 20 & 2.1 & 7.20 \\%
        &         &    &    9.5        &     & 2.8 & 7.50 \\ \hline%
LDC & TPC/Si & 4 & 3& 16 & 1.60 & 6.00 \\%
        &         &   &       6.6      &         & 2.3 & 6.20 \\ \hline%
SiD & Silicon & 5 & 2.5 & 14 & 1.27 & 6.45 \\%
        &           &   &       5.5          &    & 1.27 & 5.89 \\ \hline%
4$^{\rm th}$  & TPC & 3.5 & 3 & 15 & 1.5 & 5.50 \\
                    &   or drift      &      & 8   &  &  1.8 & 6.50 \\ 
\hline%
\end{tabular}
\end{center}
\end{table} 

Software models of the detectors have produced realistic simulations
of the physics performance, making it clear that the detectors can
do the physics.  The community is also preparing for the evolution
to collaborations.

\subsection{The Silicon Detector (SiD) Concept}

\begin{figure}[htb]
   \begin{center} \vbabove
\includegraphics[width=0.7\textwidth]{\picturefolder 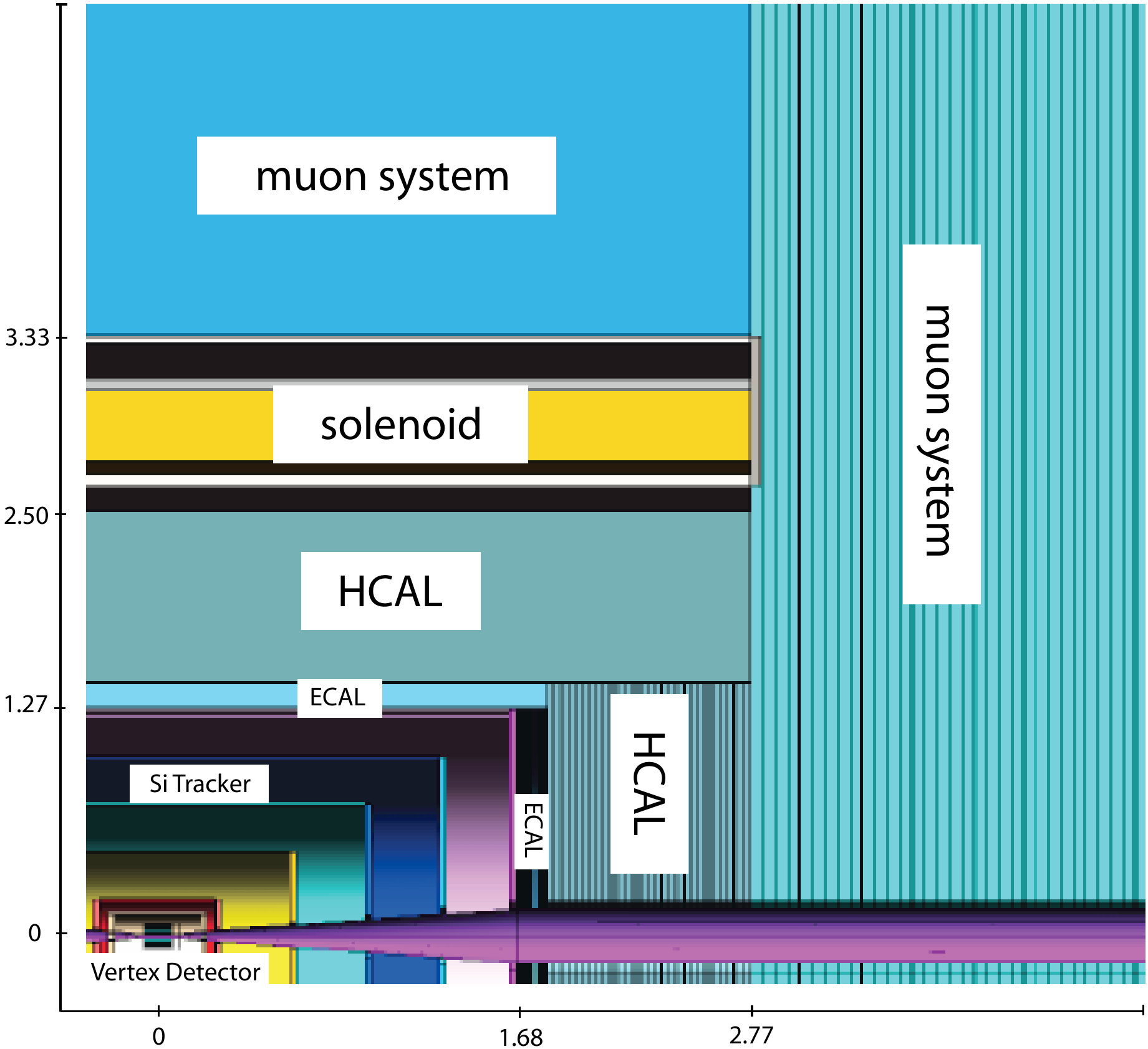}
\vbabovecaption
      \caption{Illustration of a quadrant of SiD.}
      \label{fig:sid}
   \end{center} \vbbelow
\end{figure}

The SiD concept is based on silicon tracking and a silicon-tungsten
sampling calorimeter, complemented by a powerful pixel vertex
detector, outer hadronic calorimeter, and muon system. Silicon
detectors are fast and robust, and can be finely segmented. Most SiD
systems can record backgrounds from a single bunch crossing
accompanying a physics event, maximizing event cleanliness. The
vertex detector, the tracker and the calorimeter can all absorb
significant radiation bursts without ``tripping'' or sustaining
damage, maximizing running efficiency.  The SiD Starting
Point\cite{Detectors:sid} is illustrated in Figure~\ref{fig:sid}.

A highly pixellated silicon-tungsten electromagnetic calorimeter and
a multilayer, highly segmented hadron calorimeter, inside the
solenoid, are chosen to optimize particle flow calorimetry.  Cost
and performance considerations dictate a 5 Tesla solenoid, at
relatively small radius.

SiD tracking works as an integrated system, incorporating the
pixellated vertex detector (5 barrels and 4 endcap layers), the
central silicon microstrip tracker (5 layers, barrels and endcaps),
and the electromagnetic calorimeter.  The vertex detector plays a
key role in pattern recognition; tracks produced by decays beyond
the second layer of the central tracker, but within the ECAL, are
captured with a calorimeter-assisted tracking algorithm. The
resolution of the combined system is $\frac{\sigma_p}{p^2}<2\times
10^{-5}$~$GeV^{-1}$ at high momentum.

The SiD electromagnetic calorimeter consists of layers of tungsten
and large-area silicon diode detectors in one mm gaps. The hadronic
calorimeter sandwich employs steel absorber plates and resistive
plate chambers (RPCs). Options include tungsten absorber, glass
RPCs, GEM foils, Micromegas, and scintillating tiles with silicon
photomultipliers. Muon detectors (following 6$\lambda$ at 3.5 m
radius) fill some gaps between iron plates of the flux return. Two
technologies are under consideration for the muon system,
strip-scintillator detectors and RPCs.

\subsection{The Large Detector Concept (LDC)}

\begin{figure}[htb]
   \begin{center} \vbabove
\begin{tabular}{cc}
\includegraphics[width=0.5\textwidth]{\picturefolder 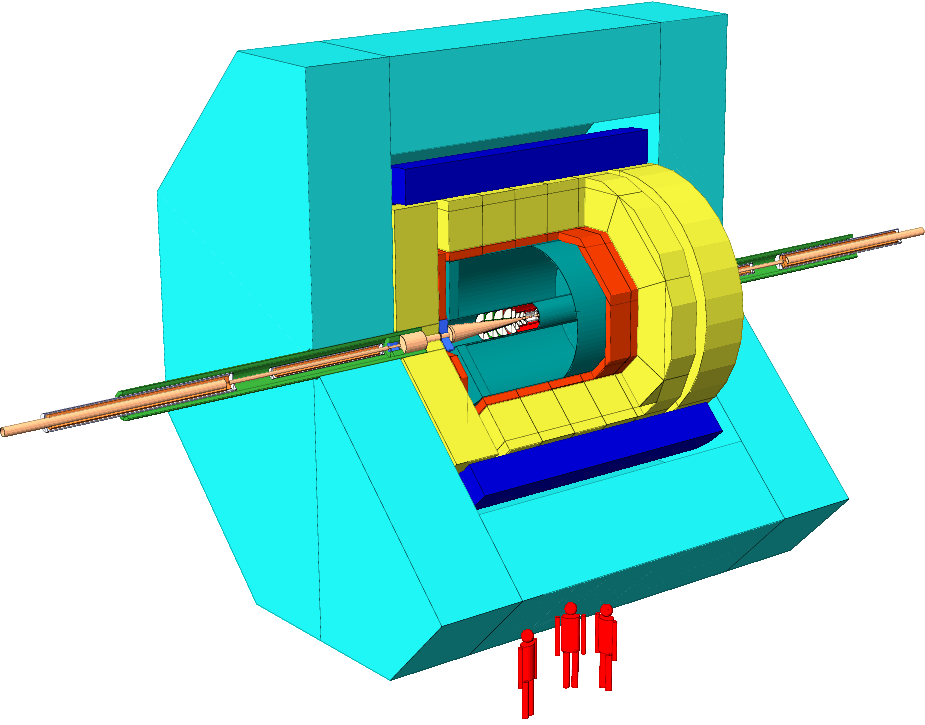} &
\includegraphics[width=0.42\textwidth]{\picturefolder 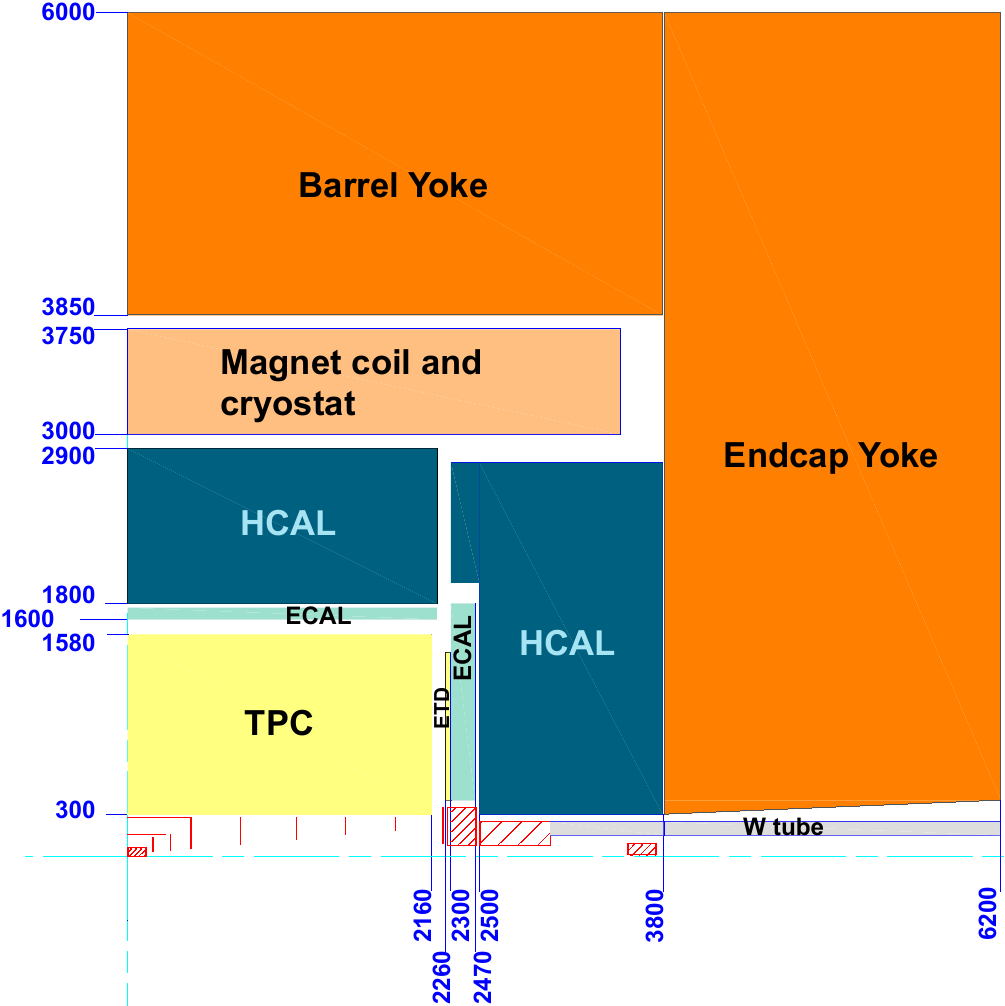} 
\end{tabular}
\vbabovecaption
      \caption[View of the LDC detector concept] {View of the LDC detector concept, as simulated with the
      MOKKA simulation package (left). 1/4 view of the LDC detector concept (right).}
      \label{fig:ldc}
   \end{center} \vbbelow
\end{figure}

The LDC\cite{Detectors:ldc} is based on a precision, highly
redundant and reliable Time Projection Chamber (TPC) tracking
system, and particle flow as a means to complete event
reconstruction, all inside a large volume magnetic field of up to 4
Tesla, completed by a precision muon system covering nearly the
complete solid angle outside the coil. A view of the simulated
detector is shown in Figure~\ref{fig:ldc} (left).

The TPC provides up to 200 precise measurements along a track,
supplemented by Si-based tracking detectors. A silicon vertex
detector gives unprecedented precision in the reconstruction of long
lived particles.

The proposed LDC detector has the following components:%
\begin{itemize}

\item a five layer pixel-vertex detector \itemspace %
\item a system of silicon strip and pixel detectors extending the
vertex detector \itemspace 
\item a large volume TPC \itemspace 
\item a system of ``linking'' detectors behind the endplate of the TPC
and in between the TPC outer radius and the ECAL inner radius \itemspace 
\item a granular Si-W electromagnetic calorimeter \itemspace 
\item a granular Fe-Scintillator hadronic calorimeter, gas hadronic
calorimeter is an option \itemspace 
\item a system of high precision extremely radiation hard calorimetric
detectors in the very forward region, to measure luminosity and to
monitor collision quality \itemspace 
\item a large volume superconducting coil, with longitudinal B-field
of 4 Tesla \itemspace 
\item an iron return yoke, instrumented to serve as a muon filter and
detector.\itemspace 

\end{itemize}

A schematic view of one quarter of this detector is shown in 
Figure~\ref{fig:ldc} (right).

\subsection{The GLD Concept}

\begin{figure}[htb]
   \begin{center} \vbabove
\includegraphics[width=\textwidth]{\picturefolder 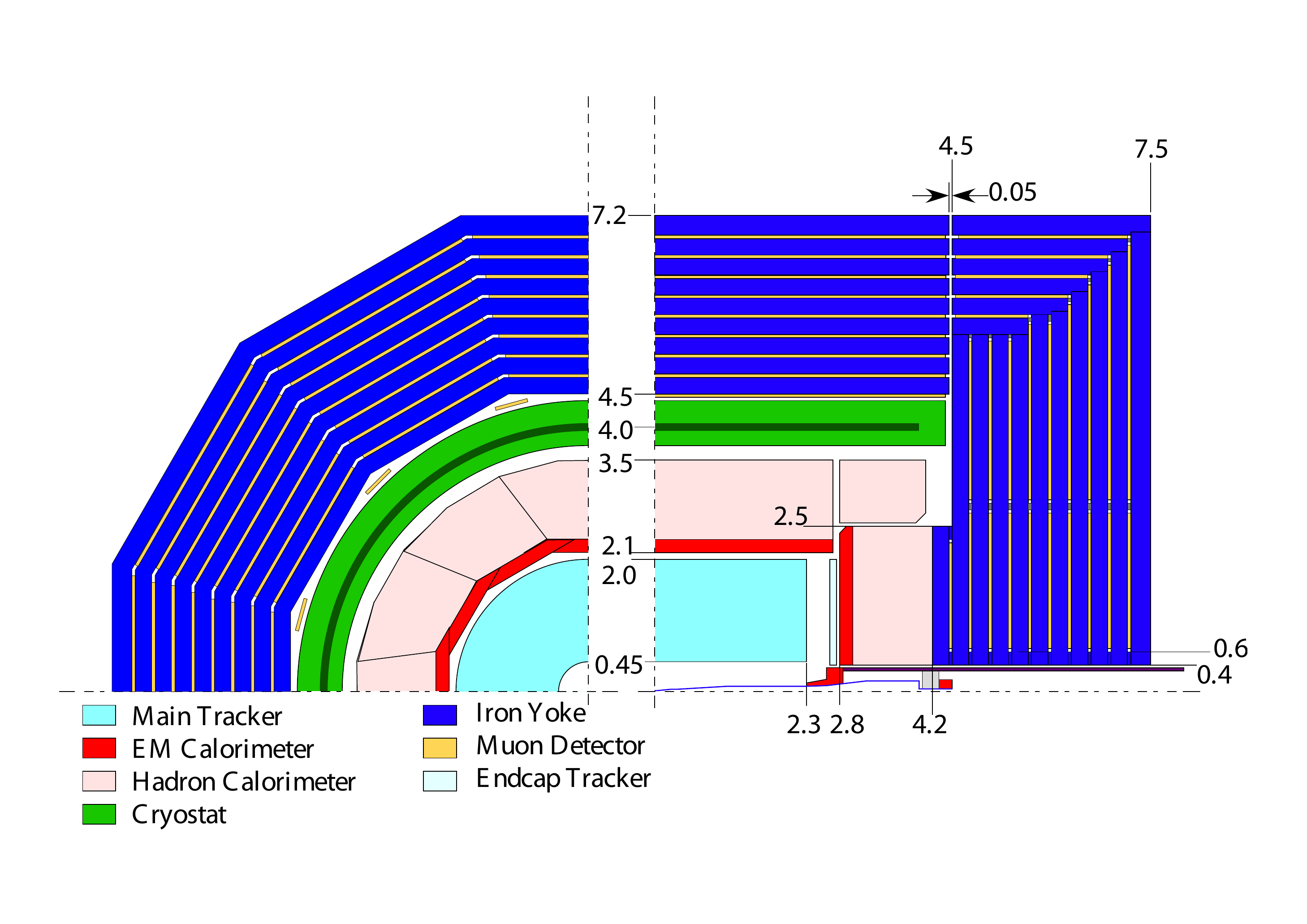}
\vbabovecaption
      \caption[Schematic view of two different quadrants of the GLD Detector.]
	{Schematic view of two different quadrants of the GLD Detector.
      The left figure shows the r$\phi$ view and the right shows the
      rz view. Dimensions are given in meters. The vertex detector and
      the silicon inner tracker are not shown.}
      \label{fig:gld}
   \end{center} \vbbelow
\end{figure}

The GLD detector\cite{Detectors:gld} concept has a large gaseous
tracker and finely granulated calorimeter within a large bore 3
Tesla solenoid. Figure~\ref{fig:gld} shows a schematic view of two
different quadrants of the baseline design of GLD.

The baseline design has the following sub-detectors:%
\begin{itemize}

\item a Time Projection Chamber as a large gaseous central tracker \itemspace 
\item a highly segmented electromagnetic calorimeter placed at large
radius and based on a tungsten-scintillator sandwich structure \itemspace 
\item a highly segmented hadron calorimeter with a lead-scintillator
sandwich structure and radial thickness of $\sim6\lambda$ \itemspace 
\item forward electromagnetic calorimeters which provide nearly full
solid angle coverage down to very forward angles \itemspace 
\item a precision silicon (FPCCD) micro-vertex detector \itemspace 
\item silicon inner and endcap trackers \itemspace 
\item a beam profile monitor in front of a forward electromagnetic
calorimeter \itemspace 
\item a scintillator strip muon detector interleaved with the iron
plates of the return yoke \itemspace 
\item a solenoidal magnet to generate the 3 Tesla magnetic field. \itemspace %

\end{itemize}

\subsection{Fourth Concept (``4$^{\rm th}$'') Detector}

The Fourth Concept detector\cite{Detectors:fourth} consists of four
essential detector systems. The calorimeter is a spatially
fine-grained dual-readout fiber sampling calorimeter augmented with
the ability to measure the neutron content of a shower.  The dual
fibers are scintillation and Cerenkov for separation of hadronic and
electromagnetic components of hadronic
showers\cite{Detectors:akchurin}. A separate crystal calorimeter
with dual readout in front of the fiber calorimeter is being
studied.

The muon system is a dual-solenoid magnetic field configuration in
which the flux from the inner solenoid is returned through the
annulus between this inner solenoid and an outer solenoid. The
magnetic field between the two solenoids back-bends the muons for a
second measurement of the momentum (with drift tubes after the
calorimeter).

\begin{figure}[htb]
   \begin{center} \vbabove
\includegraphics[width=\textwidth]{\picturefolder 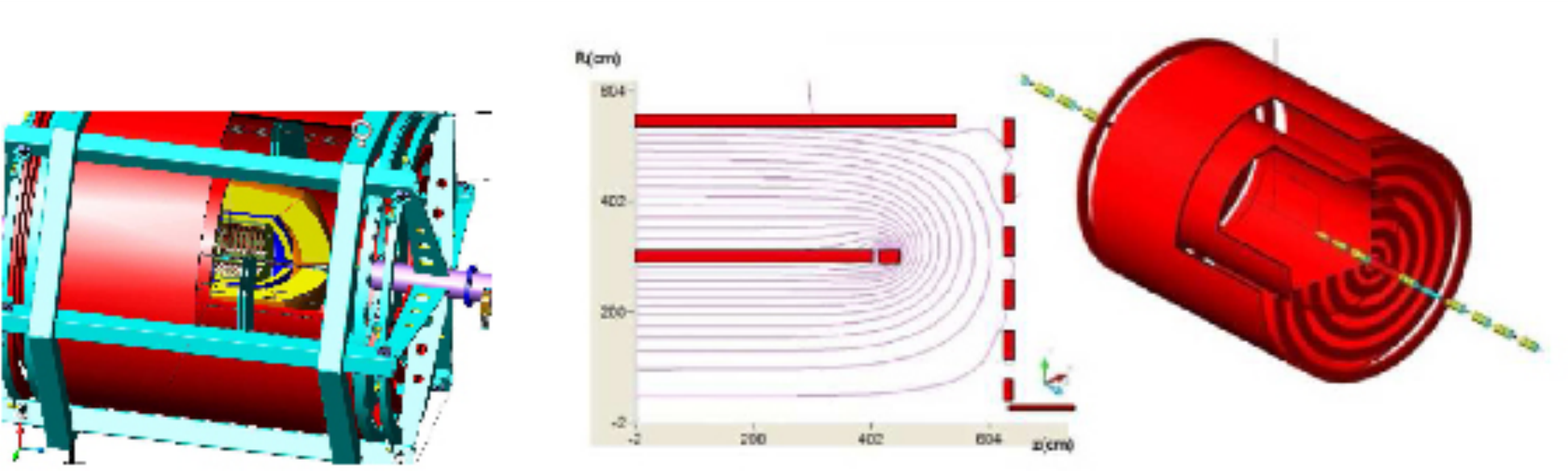}
\vbabovecaption
      \caption[Cut-away view of the 4th Detector]
	{Cut-away view of the 4th Detector (left).
      Drawings showing the two solenoids and the ``wall of coils''
      and resulting field lines in an r-z view (right).
      This field is uniform to 1\% at 3.5 T in the TPC tracking
      region, and also uniform and smooth at -1.5 T in the muon
      tracking annulus between the solenoids.}
      \label{fig:fourth}
   \end{center} \vbbelow
\end{figure}

The iron-free magnetic field is confined to a cylinder with
negligible fringe fields and with the capability to control the
fields at the beam.  The twist compensation solenoid just outside
the wall of coils is shown in Figure~\ref{fig:fourth} (right).  The
iron-free configuration may allow mounting of all beam line elements
on a single support, which could reduce the effect of vibrations at
the final focus (FF) as the beams move coherently up and down
together. In addition, the FF elements can be brought close to the
vertex chamber for better control of the beam crossing.   The
iron-free magnetic field configuration allows any crossing angle.

The pixel vertex detector is the SiD detector design. The Time
Projection Chamber (TPC) is very similar to those being developed by
the GLD and LDC concepts.

\section{Detector and Physics Performance}

Significant progress has been made in the development of complete
simulation and reconstruction software systems for the ILC
detectors, lending reality and credibility to studies of detector
performance and physics studies. These are available in software
repositories\cite{Detectors:gaede}.

The detectors have tracking systems composed of a number of
different sub-systems. Using realistic algorithms, and including a
simulation of the expected background rates, track reconstruction
efficiencies close to 99\% have been demonstrated, with momentum
resolutions of $\frac{\sigma_{p_t}}{p_t^2}<1\times10^{-4}\;{\rm
GeV^{-1}}$.

Below 1 TeV the best event reconstruction resolution is believed to
result from a particle flow algorithm. Simulations have shown
jet-energy resolutions are near the goal of 
$\frac{\sigma_{E_{jet}}}{E_{jet}} = \frac{30\%}{\sqrt{E_{jet}}}$ for $E_{jet}$ 
up to approximately 100 GeV, and $\frac{\sigma_{E_{jet}}}{E_{jet}}\le3\%$ 
beyond. Table~\ref{Detectors:partflow} presents some recent results for jet energy
resolution using particle flow in detailed, realistic simulations \cite{Detectors:thomson}.

\begin{table}[hbt]
\begin{center}
\caption{Jet energy resolutions based on simulations of LDC.
\label{Detectors:partflow}}
\begin{tabular}{|r|r|}
\hline%
$E_{jet}$ & $\sigma_{E_{jet}}$ \\ \hline   & \vbdlspacing \hline%
45 GeV & 4.4\% \\ \hline%
100 GeV & 3.0\% \\ \hline%
180 GeV & 3.1\% \\ \hline%
250 GeV & 3.4\% \\ \hline%
\end{tabular}
\end{center}
\end{table}

\begin{figure}[htb]
   \begin{center} \vbabove
\includegraphics[width=0.6\textwidth]{\picturefolder 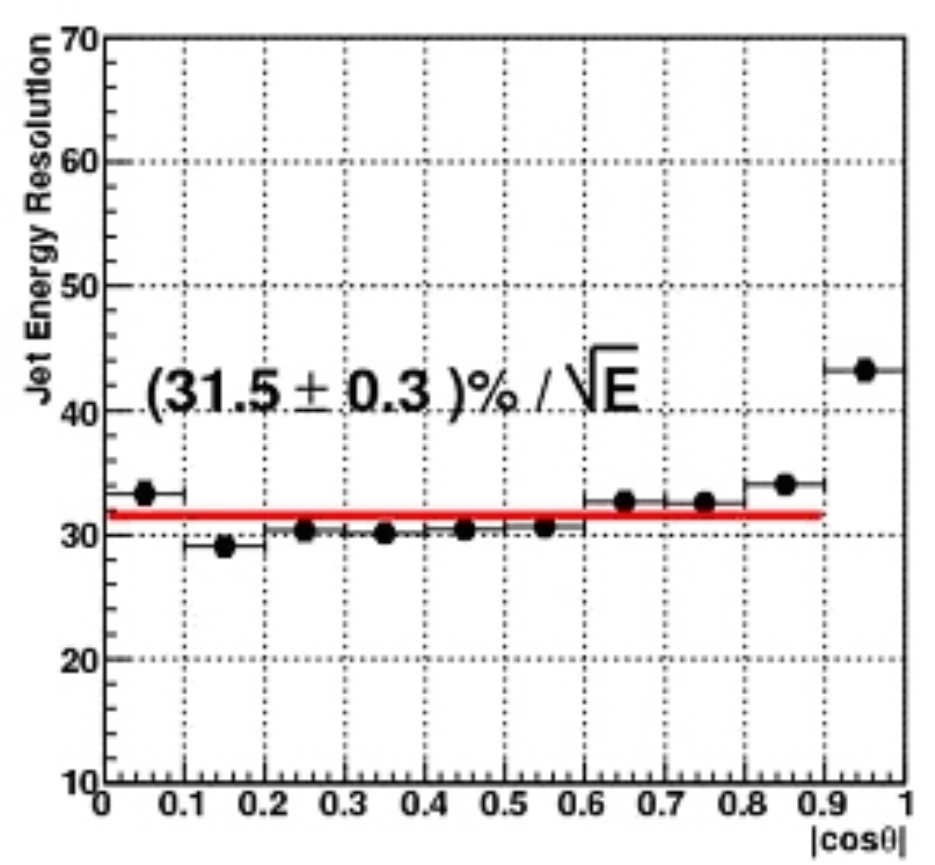}
\vbabovecaption
      \caption[RMS90 as a function of $|\cos\theta_q|$]
	{Energy resolution for 90\%
      of events
      (RMS90) as a function of $|\cos\theta_q|$ for
      $e^+e^-\rightarrow q\bar{q}$ (light quarks) events at $\sqrt{s}$ =
      91.2 GeV in the GLD detector.}
      \label{fig:rms90}
   \end{center} \vbbelow
\end{figure}

Figure~\ref{fig:rms90} presents a
calculation of the energy rms for 90\% of $\sqrt{s}$ = 91.2 GeV
events (RMS90) as a function of the production angle of the jets for
GLD. In the barrel the averaged energy resolution is 2.97 GeV,
which corresponds to 3.3\%.

\begin{figure}[h]
   \begin{center} \vbabove
\includegraphics[width=0.75\textwidth]{\picturefolder 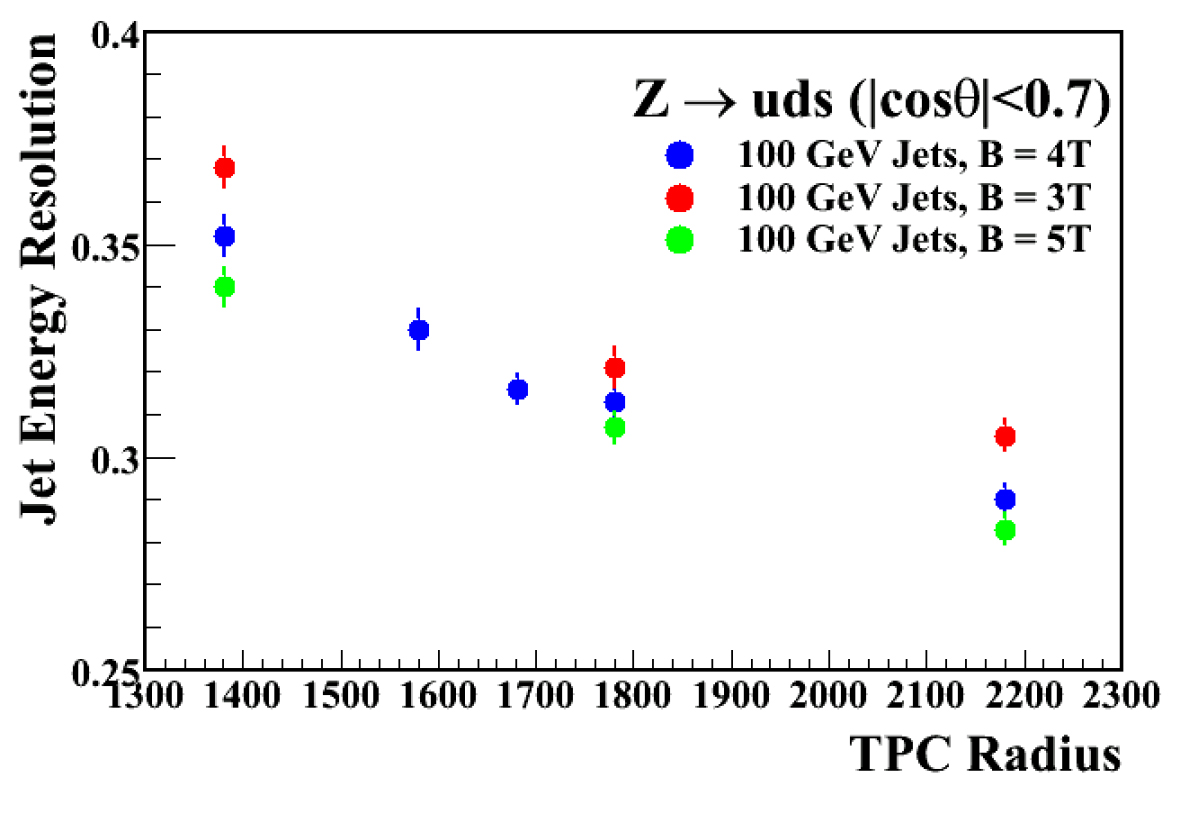}
\vbabovecaption
      \caption[The jet energy resolution] 
	{The jet energy resolution expressed as $\sigma_{jet}
      \sqrt{E_{jet}}$, as a function of of the TPC radius for a
      few magnetic field values. The TPC radius is equivalent to the
      inner radius of calorimeter.}
      \label{fig:jetres}
   \end{center} \vbbelow
\end{figure}

Particle Flow Algorithm (PFA) resolution is expected to improve as
the calorimeter radius and magnetic field increase. In order to
achieve the PFA performance goal with an acceptable detector cost,
SiD adopts the strongest magnetic field with the smallest radius,
GLD the weakest magnetic field but the largest radius, with LDC in
between. The performances as a function of TPC radius for a few
magnetic field values are shown in Figure~\ref{fig:jetres}.
As expected, the jet energy resolution improves with increasing
calorimeter radius when the magnetic field is fixed.

Study of Higgs boson properties could be a major focus of the ILC
physics program.
The challenging measurement of the Higgs mass using the recoil mass
method is presented in Figure~\ref{fig:recoil}.
In a related study, the precision on the mass measurement
for a 120 GeV Higgs boson at $\sqrt{s}$ = 350 GeV is shown
to be 135 MeV with SiD.

\section{Interfacing the Detector to the Machine}

The interaction region is the interface between the detector and the
accelerator. Its complexity motivates integration of the beam delivery 
system with the detector design.

The beams are delivered through the largest possible apertures en
route to the collision point, but are constrained to pass through a
beampipe of minimal radius at the IP to optimize vertex detector
performance. A series of detectors record the remnants from the beam
interactions in the very forward direction, monitor the beam
properties, and measure the delivered luminosity. Tungsten masks
shield most of the detector from the backgrounds produced in the
collision.

Several beam processes create backgrounds which are potentially
problematic for the detector. The main background is large numbers
of very forward going photons and electron-positron pairs produced
by ``beamstrahlung''. Other backgrounds include  synchrotron
radiation, muons produced upstream of the IR when beam tails impinge
on collimators, and neutrons created by the absorption of
beamstrahlung photons and pairs on beamline elements.

To guide the disrupted beam and charged background particles out of
the detector and to minimize backgrounds, the detector magnetic
field is perturbed to point in the direction of the outgoing beam.
This is done by superposing  a small dipole field on the detector's
main solenoidal field. This Detector Integrated Dipole (DID) is
beneficial once the crossing angle increases beyond a few mrad.

The detectors most sensitive to pair backgrounds are the vertex
detector and the beamcal. The innermost layer of the vertex detector
sits between 1.3 and 1.55 cm from the interaction point, and must
contend with $\sim100$ particles/mm$^2$/bunch train, which generates
high occupancies. The beamcal must contend with an energy deposition
of 100 TeV/ beam crossing, which results in a high radiation dose.
The number of particles passing outside the vertex detector, at
radii beyond 10 cm, is rather small. For a silicon based tracking
system it is not a real concern. In a TPC based tracking system,
where many bunches are integrated into one image of the tracker, the
total occupancy is expected to be below one percent, and is not a
problem.

The ILC reference design has one interaction region with beams
crossing at 14 mrad, and is equipped with two detectors which can be
moved quickly into and out of the interaction region (push-pull
operation) to share luminosity. The option with two beam delivery
systems continues to be investigated. Push-pull is being engineered
to proceed efficiently, allowing for quick vacuum and cryogenic
disconnects, signal and power umbilicals, and the means to
reestablish alignment and calibration quickly. The two detectors
provide redundancy, cross-checks and insurance against mishaps.

Precise knowledge of the beam energy and polarization is critical to
the physics program, and they can be measured both upstream and
downstream of the detector, using  energy spectrometers and
polarimeters.

\clearpage
\setcounter{figure}{0}
\setcounter{table}{0}
\setcounter{secnumdepth}{4}
\setcounter{subsection}{0} 
\setcounter{subsubsection}{0}
\usecounter{subsubsection}
\chapter{Value Estimates}\label{sec:ValueEstimate}
\section{The Accelerator}

A preliminary cost analysis has been performed for the ILC Reference
Design.  A primary goal of the estimate was to allow
cost-to-performance optimization in the Reference Design, before
entering into the engineering design phase.  Over the past year, the
component costs were estimated, various options compared and the
design evolved through about ten significant cost-driven changes,
resulting in a cost reduction of about 25\%, while still maintaining
the physics performance goals.

The ILC cost estimates have been performed using a ``value'' costing
system, which provides basic agreed-to value costs for components in
ILC Units\footnote{For this value estimate, 1 ILC Unit = 1 US 2007\$
  (= 0.83 Euro = 117 Yen).}, and an estimate of the explicit labor (in
person hours) that is required to support the project.  The estimates
are based on making world-wide tenders (major industrialized nations),
using the lowest reasonable price for the required quality.  There are
three classes of costs:

\begin{itemize}

  \item site-specific costs, where a separate estimate was made in
    each of the three regions; \itemspace 
  \item conventional costs for items where there is global capability
    -- here a single cost was determined; \itemspace 
  \item costs for specialized high-tech components (e.g. the SCRF linac
    technology), where industrial studies and engineering estimates
    were used. \itemspace 

\end{itemize}

The total estimated value for the shared ILC costs for the Reference
Design is 4.79 Billion (ILC Units). An important outcome of the value
costing has been to provide a sound basis for determining the relative
value of the various components or work packages. This will enable
equitable division of the commitments of the world-wide collaboration.

In addition, the site specific costs, which are related to the direct
costs to provide the infrastructure required to site the machine, are
estimated to be 1.83 Billion (ILC Units). These costs include the
underground civil facilities, water and electricity distribution and
buildings directly supporting ILC operations and construction on the
surface.  The costs were determined to be almost identical for the
Americas, Asian, and European sample sites. It should be noted that
the actual site-specific costs will depend on where the machine is
constructed, and the facilities that already exist at that location.

Finally, the explicit labor required to support the construction
project is estimated at 24 million person-hours; this includes
administration and project management, installation and testing.  This
labor may be provided in different ways, with some being contracted
and some coming from existing labor in collaborating institutions.

The ILC Reference Design cost estimates and the tools that have been
developed will play a crucial role in the engineering design effort,
both in terms of studying options for reducing costs or improving
performance, and in guiding value engineering studies, as well as
supporting the continued development of a prioritized R\&D program.

The total estimated value cost for the ILC, defined by the Reference
Design, including shared value costs, site specific costs and explicit
labor, is comparable to other recent major international projects,
e.g. ITER, and the CERN LHC when the cost of pre-existing facilities
are taken into account. The GDE is confident that the overall scale of
the project has been reliably estimated and that cost growth can be
contained in the engineering phase, leading to a final project cost
consistent with that determined at this early stage in the design.

\section{The Detectors}

Three detector concepts, GLD, LDC, and SiD, estimated the costs of their respective detector designs. Each used a complete work breakdown structure, and identified the significant costs associated with subsystems, and costs associated with assembly and installation. Estimates were guided by the GDE costing rules, and included approximately 35\% contingency.  The three estimates are reasonably consistent, but are divided differently between M\&S and labor, a result of regional accounting differences.

The cost drivers for the M\&S budgets are the calorimeters and the solenoidal magnet and flux return iron. Integration, transportation, and computing have been included, as have indirect costs associated with both M\&S and labor.

The coil costs for each of the concepts were consistent with the costs for the BaBar, Aleph, and CMS coils when compared as a function of stored energy. The cost breakdowns across detector subsystems for each of three concepts differ concept to concept.  This is to be expected, as SiD has costed electronics, installation, and management as separate items whereas LDC and GLD have embedded these in the subdetectors. In another example, GLD chooses to cost both hadron and electromagnetic calorimeters as a single item, since the detectors used are similar. LDC and SiD have separated these expenses, because the detection techniques are quite different.

Based on the SiD and LDC estimates, the value (M\&S) cost is in the range 360-420 Million (ILC Units) each. GLD does not estimate M\&S separately. Manpower for SiD and LDC (including contingency) is estimated at 1250-1550 person-years. Combining M\&S and person-years, the total detector cost lies in the range of 460-560 Million (ILC Units) for any of the detector concepts. The cost scale for the two detectors envisioned for the ILC is about 10\% of the cost of the machine.

\clearpage
\setcounter{figure}{0}
\setcounter{table}{0}
\setcounter{secnumdepth}{4}
\setcounter{subsection}{0} 
\setcounter{subsubsection}{0}
\usecounter{subsubsection}
\chapter{Next Steps: R\&D and the Engineering Design Phase}
\section{Accelerator R\&D}

For the last year, the focus of the core GDE activity has been on
producing the RDR and value estimate. In parallel, ILC R\&D programs
around the world have been ramping up to face the considerable
challenges ahead. The GDE Global R\&D Board -- a group of twelve GDE
members from the three regions -- has evaluated existing programs, and
has convened task forces of relevant experts to produce an
internationally agreed-upon prioritized R\&D plan for the critical
items. The highest-priority task force (S0/S1) addresses the SCRF
accelerating gradient:

\begin{itemize}

  \item S0: high-gradient cavity -- aiming to achieve 35 MV/m
    nine-cell cavity performance with an 80\% production yield; \itemspace 
  \item S1: high-gradient cryomodule -- the development of one or more
    high-gradient ILC cryomodules with an average operational gradient
    of 31.5 MV/m. \itemspace 

\end{itemize}

The S0/S1 task force has already produced focused and comprehensive
R\&D plans. Other task forces (S2: test linac; S3: Damping Ring; S4:
Beam Delivery System, etc.) are in the process of either completing
their reports, or just beginning their work.

\begin{figure}[htb]
\begin{center}
  \includegraphics[width=\textwidth]{\picturefolder 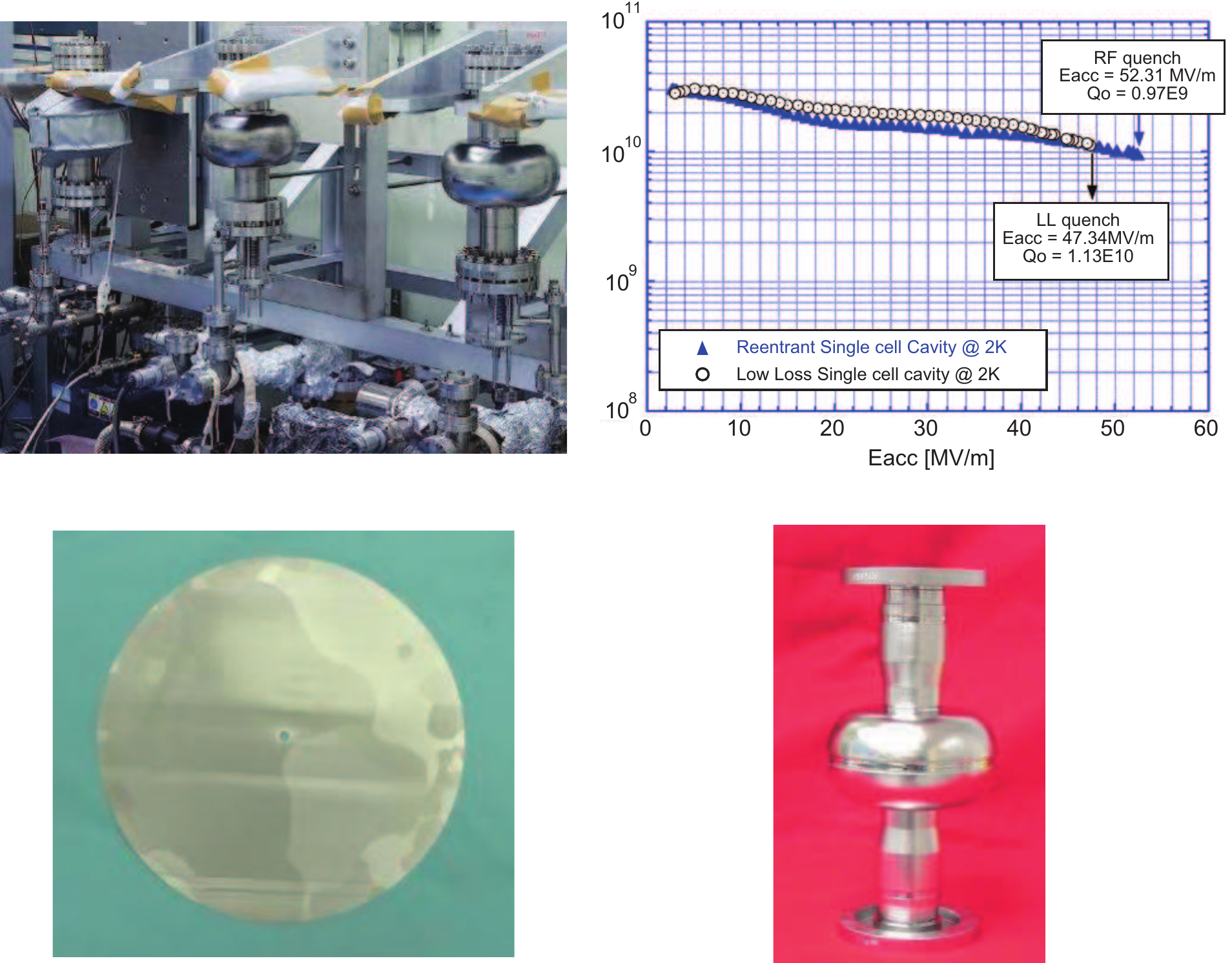}
  \caption[Cutting-edge SCRF R\&D.] {Cutting-edge SCRF R\&D. Top-left: ICHIRO single-cells being
    prepared for testing at KEK. Top-right: world-record performance
    from novel shape single-cells (ICHIRO and Cornell's reentrant
    cavity). Bottom-left: large-grain niobium disk (Jefferson
    Lab). Bottom-right: single-cell cavity produced from large-grain
    niobium material (Jefferson Lab).}
  \label{fig:OVcavRD}
\end{center}
\end{figure}

For the cost- and performance-critical SCRF, the primary focus of
S0/S1 remains the baseline choice, the relatively mature TESLA
nine-cell elliptical cavity. However, additional research into
alternative cavity shapes and materials continues in parallel. One
promising technique is the use of `large-grain'
niobium~\cite{large-grain}, as opposed to the small-grain material
that has been used in the past (Figure~\ref{fig:OVcavRD}).  Use of
large grain material may remove the need for electropolishing, since
the same surface finish can potentially be achieved with Buffered
Chemical Polishing (BCP) -- a possible cost saving. Several
single-cells have achieved gradients in excess of 35 MV/m (without
electropolishing) and more recent nine-cell cavity tests have shown
very promising results.

Various new and promising cavity shapes are also being investigated,
primarily at KEK and Cornell. While the basic nine-cell form remains,
the exact shape of the `cells' is modified to reduce the peak magnetic
field at the niobium surface. In principle these new shapes can
achieve higher gradients, or higher quality factors
($Q_0$). Single-cells at KEK (ICHIRO) and Cornell (reentrant) have
achieved the highest gradients to date ($\sim$50 MV/m, see
Figure~\ref{fig:OVcavRD}). R\&D towards making high-performance
nine-cell cavities using these designs continues as future possible
alternatives to the ILC baseline cavity.

The GDE formally supports R\&D on alternative designs for components other than the cavities, where the new designs promise potential cost and/or performance benefits. Some key examples are alternative RF power
source components, of which the Marx modulator is currently the most
promising.  In addition, R\&D on critical technologies will
continue through the EDR. Topics include items such as the damping
ring kickers and electron-cloud mitigation techniques, the positron
target and undulator, the magnets around the beam interaction point,
and global issues that require very high availability such as the
control system, the low-level RF, and the magnet power supplies.

\section{The Detector Roadmap: R\&D and Engineering Designs}

The detector R\&D and integrated detector design efforts must keep pace with progress on the ILC.
The detector R\&D program, which has already developed over many years, includes efforts in all regions, with inter-regional collaboration in some cases, and inter-regional coordination in all cases. The R\&D is reviewed within the global context by the World Wide Study. This R\&D is critical to the success of the ILC experimental program.

To focus integrated detector design efforts over the next few years, the current studies for four distinct concepts will be concentrated into two engineering design efforts, in time for the submission of two detector EDRs at the same time as the ILC machine EDR. The ILCSC will issue a call for Letters of Intent to the ILC detector community during Summer 2007 to initiate this process. The next steps are still being developed by the ILCSC, but will include appointing a Research Director, who will be responsible for developing the ILC experimental program, and establishing an International Detector Advisory Group, which will help define the two experiments suitable for engineering design. The resulting two detectors are expected to have complementary and contrasting strengths, as well as broad international participation. 
The two detector concepts should be defined by early 2009, and their engineering designs will then be completed over the next two or three years.

\section{Towards the Engineering Design Report (EDR)}

While investment into the critical R\&D remains a priority, a
significant ramping-up of global engineering resources will now be
required to produce an engineered technical design by 2010. An
important aspect of this work will be the refinement and control of
the published cost estimate by value engineering. The EDR phase will
also require a restructuring of the GDE to support the expanded
scope. A more traditional project structure will be adopted based on
the definition of a discrete set of Work Packages. The responsibility
for achieving the milestones and deliverables of each Work Package
will be assigned to either a single institute, or consortium of
institutes, under the overall coordination of a central project
management team. The Work Packages need to be carefully constructed to
accommodate the direct needs of the Engineering Design phase,
while at the same time reflecting the global nature of the project. An
important goal of the current planning is to integrate the engineering
design and fundamental R\&D efforts, since these two aspects of the
project are clearly not independent. The new project structure will be
in place by mid 2007.

The GDE remains committed to the technically-driven schedule of supplying the EDR in 2010, making start of construction possible as early as 2012 – consistent with expected early results from the LHC. The critical path and cost drivers have been clearly identified during the RDR phase, and they define the priorities for the next three years of the Engineering Design phase. The R\&D program will be fine-tuned to mitigate the remaining identified technical risks of the design. A key element of the engineering activity will be the formation of a qualified industrial base in each region for the SCRF linac technology. A equally critical focus will be on the civil construction and conventional facilities – the second primary cost driver – where an early site selection would clearly be advantageous.

Finally, the GDE also remains committed to completing these challenging goals as a truly international organization, by building on and consolidating the successful collaboration which produced the RDR. The support of the world-wide funding agencies is critical in this endeavor. The GDE – together with the leaders of the particle physics community – will continue to work with the regional funding agencies and governments to begin construction of this project in the early part of the next decade.

\backmatter

\clearpage
\listoffigures 

\clearpage
\listoftables

\typeout{Lastpage = \arabic{page}}

\clearpage
\begin{thebibliography}{99}
\addcontentsline{toc}{chapter}{Bibliography}


\bibitem{PhysCase:QuantumUniverse} A.~Albrecht, {\it et al.},
Report of the DOE/NSF High Energy Physics Advisory Panel (2004).
%
\bibitem{PhysCase:ilc}H.~Shapiro {\it et al.}, ``Report of the
Committee on Elementary Particle Physics in the 21st Century,'' Board
of Physics and Astronomy, National Research Council, National
Academies Press, Washington D.C. (2006); ``The European Strategy for
Particle Physics,'' Report of CERN Council Strategy Group (2006); GLC
Project: ``Linear Collider for TeV Physics'' ,KEK-REPORT-2003-7; I.
Corbett {\it et al.}, ``Report of the Consultative Group on
High-Energy Physics,'' OECD Global Science Forum (2002); S. Yamada
{\it et al.},``Report of the JLC Globalization Committee'' (2002);
Aguilar-Saavedra, J. A. {\it et al.}, ``TESLA Technical Design Report
Part III: Physics at an e+e- Linear Collider,'' hep-ph/0106315; ACFA
Linear Collider Working Group, ``Particle Physics Experiments at JLC,''
hep-ph/0109166; Abe, T. and {\it et al.},  ``Linear Collider Physics Resource Book for Snowmass,''
American Linear Collider Working Group (2001).
%
\bibitem{PhysCase:hepap06} J.~Bagger {\it et al.}, ``Discovering the
Quantum Universe: The Role of Particle Colliders,'' Report of the
DOE/NSF High Energy Physics Advisory Panel (2006).
%
%
\bibitem{PhysCase:atlas} ATLAS Collaboration, ``ATLAS Physics Technical
Design Report,'' CERN-LHCC-99-14 and CERN-LHCC-99-15,
\url{http://atlas.web.cern.ch/Atlas/GROUPS/PHYSICS/TDR/TDR.html};
``CMS Physics TDR,'' CERN/LHCC/2006-021 (2006).
%
\bibitem{PhysCase:cousins} R.~Cousins, J.~Mumford and V.~Valuev, 
``Forward-Backward Asymmetry of Simulated and Reconstructed Z-prime $\rightarrow \mu^+ \mu^-$ Events in CMS''
CERN-CMS-NOTE-2005-022 (2005)
%
%
%
\bibitem{PhysCase:ilcpars} ``Parameters for the Linear Collider,''
\url{http://www.fnal.gov/directorate/icfa/LC_parameters.pdf} (2003) and
\url{http://www.fnal.gov/directorate/icfa/para-Nov20-final.pdf} (2006).
%
\bibitem{PhysCase:gudi} G.~Moortgat-Pick {\it et al.},
``The Role of Polarized Positrons and Electrons in Revealing Fundamental Interactions at the Linear Collider,''
hep-ph/0507011 (2005).
%
\bibitem{PhysCase:hawkings} R.~Hawkings and K.~Moenig, 
`` Electroweak and CP violation physics at a Linear Collider Z-factory,'' hep-ex/9910022 (1999).
%
\bibitem{PhysCase:heusch} C.~A.~Heusch, 
``The International Linear Collider in its Electron-Electron Version,''
Int. J. Mod. Phys.A20 (2005).
%
\bibitem{PhysCase:ginzburg} I.~F.~Ginzburg, G.~L.~Kotkin, V.~G.~Serbo
and V.~I.~Telnov, JETP Lett. 34 (1981) 491; Badelek, B. {\it
et al.}, hep-ex/0108012.

\bibitem{itrp} ITRP Recommendation,
\url{http://www.fnal.gov/directorate/icfa/ITRP\_Report\_Final.pdf} (2004).

\bibitem{tdr} R. Brinkmann {\it et al.}, eds., ``TESLA Technical Design Report,''
DESY-2001-011 (March, 2001).

\bibitem{nlc} T. O. Raubenheimer {\it et al.}, eds., ``Zeroth Order Design Report for
the Next Linear Collider,'' SLAC-R-474 (1996); N. Phinney, ed., ``2001
Report on the Next Linear Collider: A report submitted to Snowmass
'01,'' SLAC-R-571 (2001).

\bibitem{glc} ``GLC project: Linear Collider for TeV Physics,''
KEK-Report-2003-7, \url{http://lcdev.kek.jp/Roadmap/} (2003)

\bibitem{euro-xfel} M. Altarelli {\it et al.}, ``The European X-Ray Free-Electron Laser
Technical Design Report,'' DESY 2006-097 (2006).

\bibitem{high-g} F. Furuta {\it et al.}, ``Experimental Comparison at KEK of High Gradient
Performance of Different Single-Cell Superconducting Cavity Designs,''
EPAC06 (2006); R. L. Geng {\it et al.},
``High-Gradient Activities at Cornell: Re-entrant Cavities,'' SRF 2005 (2005).

\bibitem{Detectors:behnke}
K. Abe {\it et al.}, ``Particle Physics
Experiments at JLC,'' KEK-Report-2001-11 and hep-ph/0109166 (2001);%
T. Abe {\it et al.}, ``Linear Collider Physics Resource Book for
Snowmass 2001,'' BNL-52627, CLNS01/1279,
FERMILAB-Pub-01/058-E,LBNL-47813, SLAC-R-570, UCRL-ID-143810-DR
(2001).
%
\bibitem{Detectors:sid} The SiD Concept Group, ``SiD Detector Outline Document,
\url{http://physics.uoregon.edu/~lc/wwstudy/concepts/} (2006).
%
\bibitem{Detectors:ldc} The LDC Concept Group, ``LDC Detector Outline
Document,'' \url{http://www.ilcldc.org} (2006).
%
\bibitem{Detectors:gld} GLD Concept Study Group, ``GLD Detector Outline
Document,'' physics/0607154 (2006).
%
\bibitem{Detectors:fourth} The 4th Concept Group, ``Detector Outline Document,''
\url{http://physics.uoregon.edu/~lc/wwstudy/concepts/} (2006).
%
\bibitem{Detectors:schreiber} H.~J.~Schreiber, ``Branching fraction measurements of the SM higgs with a mass of 160 GeV at future linear colliders,'' LC-PHSM-2000-035 (2000);
T. Barklow, ``Physics impact of detector performance,'' 2005 ILC Workshop,
\url{http://www-conf.slac.stanford.edu/lcws05/program/talks/18mar2005.ppt}
(2005);
H.-J.~Yang and K.~Riles, ``Impact of tracker design on higgs and
slepton measurements,'' physics/0506198 (2005).
%
\bibitem{Detectors:abe} T.~Abe, ``A Study of Topological Vertexing for 
Heavy Quark Tagging'' , SLAC-PUB-8775 (2001).
%
\bibitem{Detectors:akchurin} N.~Akchurin {\it et al.}, NIM
{\bf A537}, 537-561 (2005).
%
\bibitem{Detectors:gaede}  F.~Gaede {\it et al.},
\url{http://ilcsoft.desy.de/marlin}; T.~Johnson {\it et al.}, ``lcorg.sim: A
Java-based Reconstruction and Analysis Toolkit,''
\url{http://www.lcsim.org}; C.~Gatto {\it et al.},
\url{http://www.fisica.unile.it/~danieleb/IlcRoot}.
%
\bibitem{Detectors:thomson} M.~Thomson, Paris ILC Software
Meeting (2007).
%

\bibitem{large-grain} P.~Kneisel {\it et al.}, ``Preliminary Results from
  Single Crystal and Very Large Crystal Niobium Cavities,'' PAC05 (2005).


\end{thebibliography}
\end{document}